\documentclass[preprint,floatfix,aps,prd,showpacs,amsmath,amssymb,amsfonts,superscriptaddress]{revtex4-1}
\usepackage{graphicx}
\usepackage{color}

\usepackage{dsfont}
\usepackage[normalem]{ulem}

\usepackage{mathtools}

\usepackage{refcount}
\usepackage{comment}
\usepackage{tensor}
\usepackage{xcolor}
\usepackage{soul}
\usepackage{mathtools}
\usepackage[utf8]{inputenc}
\usepackage{multirow} 
\usepackage{subfigure}

\usepackage{adjustbox}
\usepackage{rotating}
\usepackage[active]{srcltx}

\expandafter\ifx\csname package@font\endcsname\relax\else
\expandafter\expandafter
\expandafter\usepackage
\expandafter\expandafter
\expandafter{\csname package@font\endcsname}%
\fi
\allowdisplaybreaks
\usepackage{float}
\usepackage{hyperref}
\hypersetup{
	colorlinks=true,       
	linkcolor=blue,          
	citecolor=blue,        
}
\usepackage{varioref}
\usepackage{cleveref}

\usepackage{colortbl}      

\usepackage[section]{placeins}

\usepackage{fancybox}
\labelformat{figure}{Fig.~#1}

\newcommand{\bea}{\begin{aligned}}
\newcommand{\eea}{\end{aligned}}
\def\bea{\begin{eqnarray}}
\def\eea{\end{eqnarray}}
\def\beq{\begin{equation}}
\def\eeq{\end{equation}}
\def\bse{\begin{subequations}}
\def\ese{\end{subequations}}
\def\ov{\overline}

\begin{document}

\title{Interacting dark sector from Horndeski theories and beyond: Mapping fields and fluids}

\author{Pulkit Bansal} 
\email{pulkit.l.bansal@gmail.com}
\affiliation{Department of Physics, Indian Institute of Technology Bombay, Mumbai 400076, India}
\author{Joseph P. Johnson}
\email{josephpj@iisermohali.ac.in}
\affiliation{Department of Physical Sciences, Indian Institute of Science Education and Research Mohali, SAS Nagar, Punjab 140306, India}
\author{S. Shankaranarayanan}
\email{shanki@iitb.ac.in}
\affiliation{Department of Physics, Indian Institute of Technology Bombay, Mumbai 400076, India}
%
\begin{abstract}
In Cosmology, when dissipative effects are minimal, the energy content of the Universe can be effectively described as a sum of perfect fluids. Perfect fluid descriptions ensure thermal equilibrium since they equilibrate immediately. However, interactions among different energy content of the Universe might prevent such rapid equilibration. This limitation calls for a more fundamental framework that incorporates these interactions directly at the level of the action. In earlier work, two of the authors demonstrated that an interacting dark energy (DE)–dark matter (DM) field theory action could be derived from a modified gravity action via a conformal transformation, establishing a one-to-one correspondence between the field theory action and fluid for a unique interaction term~\cite{2021-Johnson.Shankaranarayanan-Phys.Rev.D}. 
In this work, we extend that analysis by considering quadratic order Horndeski gravity, identifying two classes of models --- field coupling and field-kinetic coupling.
Our approach generalizes the coupling function for DE-DM interactions by incorporating an additional dependence on kinetic terms. We establish a field-to-fluid mapping for dark matter and find that this mapping only holds for a specific form of the interaction strength. Interestingly, we show that this interaction strength \emph{excludes} non-gravitational interactions between dark energy and dark radiation. Numerical analysis reveals that purely kinetic interactions within the dark sector can significantly alter cosmological evolution compared to non-interacting scenarios, highlighting the strong dependence of cosmological dynamics on coupling strength. A preliminary examination of linear scalar perturbations indicates that the field-kinetic coupling results in a non-zero gravitational slip parameter and momentum exchange.
\end{abstract}
\maketitle

\section{Introduction}
\label{Section:Introduction}

Cosmological evolution of the Universe is most successfully explained by the $\Lambda$CDM model within the framework of General Relativity (GR)~\cite{2000-Padmanabhan-TheoreticalAstrophysicsVolume, 2005-Mukhanov-PhysicalFoundationsCosmology, 2008-Weinberg-Cosmology}, in which majority of the matter-energy content in the Universe is made up of dark matter (DM) and dark energy (DE) described by non-interacting perfect fluids~\cite{Peebles:2002gy,Padmanabhan:2002ji,Tsujikawa:2010zza,Bertone:2016nfn,Joyce:2016vqv,Sofue:2020rnl,2022-Shanki.Joseph-GRG}. From the outset, the cosmological constant has faced several theoretical challenges, such as the fine-tuning problem and the coincidence problem~\cite{Weinberg:1988cp,2006-Copeland.etal-Int.J.Mod.Phys.}. Despite these issues, it has provided a better fit for cosmological observations such as the accelerated expansion of the late Universe~\cite{1998-Riess.Others-Astron.J.,1999-Perlmutter.Others-Astrophys.J.} and the Cosmic Microwave Background (CMB) observations from the early Universe~\cite{2007-Spergel.Others-Astrophys.J.Suppl., 2018-Akrami.Others-Astron.Astrophys., 2020-Aghanim.others-Astron.Astrophys.} compared to any alternatives. However, as observational precision increases and stringent constraints are placed on various cosmological and model parameters, several cosmological tensions have emerged, most notably the $H_0$ tension and the $\sigma_8$ tension between the early and late Universe observations~\cite{2022-Brout.others-Astrophys.J.,2022-Perivolaropoulos.Skara-NewAstron.Rev., 2020-Aghanim.others-Astron.Astrophys., 2013-Marra.etal-Phys.Rev.Lett., 2014-Bennett.etal-Astrophys.J.,2016-Riess.others-Astrophys.J., 2019-Riess.etal-Astrophys.J.,2023-Gupta-Mon.Not.Roy.Astron.Soc.}. Although the $\Lambda$CDM model fits these data sets well, the estimated parameter values show a conflict of approximately $3\sigma$~\cite{2021-DiValentino.etal-CQG,Kamionkowski:2022pkx}. Aiming to solve these issues, alternatives to the $\Lambda$CDM model had been proposed and widely studied~\cite{2021-DiValentino.etal-CQG,Kamionkowski:2022pkx}. 
These dynamic DE models come in a wide variety, ranging from simple parameterizations of the DE equation of state, such as CPL model~\cite{2001-Chevallier.Polarski-Int.J.Mod.Phys.D, 2003-Linder-Phys.Rev.Lett.,2017-Tripathi.etal-JCAP}, to more complex field theory models like quintessence and k-essence~\cite{1988-Ratra.Peebles-Phys.Rev.D, 2000-Armendariz-Picon.etal-Phys.Rev.Lett., 2001-Armendariz-Picon.etal-Phys.Rev.D, 2021-Rajvanshi.etal-Class.Quant.Grav., Bamba:2012cp}. However, most of these models restrict the interaction between DE and other matter-energy components to purely gravitational interactions and rely on differences in cosmological evolution to address cosmological tensions. While some of these models succeeded in reducing one of the tensions, in most cases, they exacerbated the others~\cite{Schoneberg:2021qvd,Abdalla:2022yfr}.

While efforts to uncover their true nature are ongoing, this ambiguity provides an opportunity to explore potential non-gravitational interactions between DE and DM~\cite{Wang:2016lxa,Wang:2024vmw}. Several models of the dark sector interaction have been proposed in the literature to alleviate the Hubble tension. It has been shown that, while $H_0$ tension can be eased by considering energy transfer between DE and DM, it still does not solve $\sigma_8$ tension~\cite{2022-Schoeneberg.etal-Phys.Rept.,2021-DiValentino.etal-Class.Quant.Grav., 2024-Wang-Rept.Prog.Phys., 2024-Giare-}. In most of these models, the interaction between the fluid terms in the dark sector is proposed phenomenologically. Due to the little knowledge we possess regarding the dark sector, in numerous models, the interaction intensity $Q_{\nu}$ within the dark sector is introduced by hand.  Recently, two current authors examined the interactions in the dark sector from a classical field theoretic action for an interacting canonical scalar field description for DE and DM with an arbitrary coupling function~\cite{2021-Johnson.Shankaranarayanan-Phys.Rev.D,2022-Johnson.etal-JCAP}. The authors showed a unique field-fluid mapping for DM and constructed an autonomous system of equations for interacting dark sector. 

More specifically, two of the current authors showed that interacting dark sector action can be obtained from an $f(R)$ gravity action via a conformal transformation, which introduces interaction terms in the Einstein frame action~\cite{2021-Johnson.Shankaranarayanan-Phys.Rev.D,2022-Johnson.etal-JCAP}. This has been later generalized to $f(R,\chi)$ models ($\chi$ is a scalar field) and shown that for a specific form of interaction, it is possible to map the field theory description of the interacting dark sector to a fluid description. These models had the action with interaction terms that are functions of the DE scalar field. This mapping of the field to fluid description raises a few questions: Is the interaction unique, or does it depend on the form of the scalar field action? If yes, can we obtain a general class of interaction that can help us establish the fluid-to-field correspondence? Is it possible to completely capture the interacting dark sector using field theoretic action?

In this work, we extend the analysis of Ref.~\cite{2021-Johnson.Shankaranarayanan-Phys.Rev.D} by considering Horndeski gravity~\cite{1974-Horndeski-IJTP,2019-Kobayashi-Rept.Prog.Phys., 2010-DeFelice-Phys.Rev.Lett.}. 
Horndeski gravity has the advantage of being the most general metric theory of gravity with a scalar in four dimensions, resulting in second-order equations of motion. Horndeski's theory encompasses a wide range of gravity theories, including General Relativity, Brans-Dicke theory, Quintessence, Dilaton, Chameleon, kinetic gravity braiding, and covariant Galileon, as special cases~\cite{Deffayet:2009wt}. This enables us to simultaneously analyze several gravity theories and choices of interaction in a single framework. 

Using the familiar conformal and extended conformal transformations~\cite{1993-Bekenstein-Phys.Rev.D,2013-Zumalacarregui.etal-PRD,2015-vandeBruck.etal-JCAP}, we obtain interaction terms for a more general class of models. To keep things transparent, we focus on two types of coupling: (i) conformal coupling that depends only on the scalar field and (ii) extended conformal coupling that depends on the scalar field and the kinetic term. Interestingly, we show that starting with a k-essence description of the DM scalar field, the interaction term, after the field-fluid mapping, is identical to the earlier work~\cite{2021-Johnson.Shankaranarayanan-Phys.Rev.D}. However, for the extended conformal coupling, we obtain general interaction terms. We show that the new terms in the interaction have interesting physical consequences relating to energy and momentum transfer in the dark sector. 
Recently, it has been suggested that the momentum transfer in the dark sector can overcome the $H_0$ and $\sigma_8$ tensions simultaneously~\cite{Chamings:2019kcl,2020-Amendola.Tsujikawa-JCAP}. While the momentum transfer in these works is phenomenological, we derive this from a field theory action in this work. This allows us to study energy and momentum transfer and their effects on the background and perturbed evolution without any ambiguities. We show that this model can successfully describe the late Universe cosmology. By looking at the dynamical system analysis of the background evolution, we show that the model has stable solutions consistent with observations without fine-tuning.

From a detailed analysis of the background dynamics and a preliminary analysis of the linear scalar perturbations, we show that, for the extended conformal couplings with an additional kinetic dependence, there is a greater freedom and hence the possibility of non-trivial deviations from simpler field-coupling models as a consequence of energy and momentum transfer. This opens up the possibility of having more control over the structure formation in the model without affecting the background expansion rate, which could help to alleviate $H_0$ tension and $\sigma_8$ tension simultaneously.
We also see that the field-kinetic coupling leads to a non-zero gravitational slip parameter, which is a useful tool to detect departures from GR~\cite{2006-Bertschinger-Astrophys.J., 2007-Caldwell.etal-Phys.Rev.D,2019-Joseph.Shanki-PRD}.

This work is organized as follows. In Sec.~\eqref{Section:Field-Coupling DE-DM}, we consider a subclass of Horndeski theories with a field-dependent non-minimal coupling to gravity and perform a conformal transformation. We then establish a field-fluid mapping of the DM field for the resulting DE-DM interactions in the Einstein frame. In Sec.~\eqref{Section:Field-Kinetic Coupling DE-DM}, we extend the analysis of Sec.~\eqref{Section:Field-Coupling DE-DM} to include quadratic-order second derivative terms of $\phi$ in the Horndeski Lagrangian. In Sec.~\eqref{Section:Cosmological background evolution}, we investigate the cosmological evolution of the two classes of coupling models. We establish the equations of motion for the cosmological evolution of the two models for an arbitrary form of field coupling $\alpha(\phi)$ and a specific form of a purely kinetic coupling $\alpha(X)$, respectively. Assuming a k-essence DE field $\phi$, we construct an autonomous system of equations for the two classes of coupling models and analyze the fixed points and the system stability. In Sec.~\eqref{Section:Numerical Evolution}, we consider a quintessence DE field and a specific k-essence DE field, respectively, and numerically study the background evolution. In Sec.~\eqref{Section:Scalar perturbations}, we perform a preliminary analysis of the linear scalar perturbations. Furthermore, we investigate the impact of
the field-kinetic coupling on energy and momentum transfer between DE and DM. We summarize our results and discuss future directions in Sec.~\eqref{Section:Conclusions}.
The four appendices contain the details of the calculations and the plots from numerical evolution.

\begin{table}[!htb]
    \centering
\begin{tabular}{||l|l||} 
\hline \hline  Special Cases  & Description of the theory \\
\hline \hline $\mathcal{\Tilde{L}} = \Tilde{G}_{4}(\phi)\Tilde{R} + \Tilde{G}_{2}(\phi, \Tilde{X})$ & K-essence coupled non-minimally to gravitation
\\
& (includes $f(\Tilde{R}, \phi)$, DBI, BD, Chameleon, Dilaton)
\\
\hline $\mathcal{\Tilde{L}} = \Tilde{G}_{3}(\phi, \Tilde{X})\Tilde{\Box}\phi$ & Kinetic-gravity braiding/G-inflation
\\
\hline $\Tilde{G}_{5} \neq 0$ & Non-minimal coupling to Einstein tensor
\\
\hline $\Tilde{G}_{2} = 8\xi^{(4)}\Tilde{X}^2(3 - \mathrm{ln}\Tilde{X}), \Tilde{G}_{3} = 4\xi^{(3)}\Tilde{X}(7 - 3\, \mathrm{ln}\Tilde{X}),$ & Non-minimal field coupling to Gauss-Bonnet term
\\
$\Tilde{G}_{4} = 4\xi^{(2)}\Tilde{X}(2 - \mathrm{ln}\Tilde{X}), \Tilde{G}_{5} = -4\xi^{(1)}\mathrm{ln}\Tilde{X}.$ & $(\xi ^{(n)} := \partial^n\xi/\partial\phi^n)$
\\
\hline $\mathcal{\Tilde{L}} = \sum_{i = 1}^{5} c_{i}\mathcal{\Tilde{L}}_{i},$ & Covariant Galileon (covariant scalar field
\\
$\mathcal{\Tilde{L}}_{1} = M^3\phi, \mathcal{\Tilde{L}}_{2} = (\Tilde{\nabla}\phi)^2, \mathcal{\Tilde{L}}_{3} = (\Tilde{\Box}\phi)(\Tilde{\nabla}\phi)^2/M^3,$ & respecting Galilean symmetry in flat spacetime)
\\
$\mathcal{\Tilde{L}}_{4} = (\Tilde{\nabla}\phi)^2[2(\Tilde{\Box}\phi)^2 - 2\Tilde{\phi}_{\mu\nu}\Tilde{\phi}^{\mu\nu} - \Tilde{R}(\Tilde{\nabla}\phi)^2/2]/M^6,$ & 
\\
$\mathcal{\Tilde{L}}_{5} = (\Tilde{\nabla}\phi)^2 [(\Tilde{\Box}\phi)^3 - 3(\Tilde{\Box}\phi)\Tilde{\phi}_{\mu\nu}\Tilde{\phi}^{\mu\nu} + 2\tensor{\Tilde{\phi}}{_\mu^\nu}\tensor{\Tilde{\phi}}{_\nu^\rho}\tensor{\Tilde{\phi}}{_\rho^\mu}$ & 
\\
\ \ \ \ \ \ \ $ -6\Tilde{\phi}_{\mu}\Tilde{\phi}^{\mu\nu}\Tilde{\phi}^{\rho}\Tilde{G}_{\nu\rho}]/M^9.$ &
\\
\hline \hline
\end{tabular}
    \caption{Scalar-tensor theories contained in the Horndeski action \eqref{Eq:Horndeski action} \cite{Deffayet:2009wt, 2010-Kobayashi-Phys.Rev.Lett., 2019-Kobayashi-Rept.Prog.Phys.,Unnikrishnan:2013rka,2010-DeFelice-Phys.Rev.Lett.}. }
    \label{Table: Horndeski special cases}
\end{table}
In this work we use the natural units where $c=1, M_{\mathrm{Pl}}^2 = (8 \pi G)^{-1}$, and the metric signature $(-,+,+,+)$. Greek alphabets denote the 4-dimensional space-time coordinates, and Latin alphabets denote the 3-dimensional spatial coordinates. Overbarred quantities (like $\overline{\rho}(t), \overline{p}(t)$) are evaluated for the FLRW background, and a \textit{overdot} represents the derivative with respect to cosmic time $t$. Unless otherwise specified, subscript ``$\phi$" denotes derivative with respect to $\phi$, subscript ``$\chi$" denotes derivative with respect to $\chi$, and subscript ``$X$" denotes derivative with respect to the kinetic term of the scalar field $\phi$. Moreover, subscript $DM$ denotes dark matter, and subscript $DE$ denotes dark energy. Couplings of the form $\alpha(\phi)$ are referred to as \emph{field couplings} while those of the form $\alpha(\phi, X)$ are referred to as \emph{field-kinetic couplings}.

\section{Dark sector interaction from a Horndeski action with $G_{4}(\phi)$ type non-minimal coupling} \label{Section:Field-Coupling DE-DM}

As discussed in the introduction, we focus on a general scalar-tensor theory that leads to second-order equations of motion and avoids ghosts or negative energy states. The Horndeski theory meets these criteria for describing the scalar field $\phi$ coupled to gravity. To streamline the analysis, we adopt a k-essence framework for the second scalar field $\chi$, which is minimally coupled to the Horndeski action. Thus, the Horndeski action, describing the dynamics of the scalar field $\phi$ and gravity minimally coupled to another scalar field $\chi$, is expressed as follows~\cite{2019-Kobayashi-Rept.Prog.Phys.}:
\bea \label{Eq:Horndeski action}
S = \int \mathrm{d}^{4}x\sqrt{-\Tilde{g}}
& &
\left[\frac{M^{2}_{\mathrm{Pl}}}{2}\Tilde{G}_{4}(\phi, \Tilde{X})\Tilde{R} + \Tilde{G}_{2}(\phi, \Tilde{X}) - \Tilde{G}_{3}(\phi, \Tilde{X})\Tilde{\Box}\phi
\nonumber
\right. \\
& & \left. 
 + \frac{M^{2}_{\mathrm{Pl}}}{2}\Tilde{G}_{4\Tilde{X}}[(\Tilde{\Box}\phi)^{2} - \Tilde{\nabla}_{\mu}\Tilde{\nabla}_{\nu}\phi\Tilde{\nabla}^{\mu}\Tilde{\nabla}^{\nu}\phi] + \Tilde{G}_{5}(\phi, \Tilde{X})\Tilde{G}_{\mu\nu}\Tilde{\nabla}^{\mu}\Tilde{\nabla}^{\nu}\phi 
 \right. \\
& & \left.
- \frac{1}{6}\Tilde{G}_{5\Tilde{X}}(\phi, \Tilde{X})\left\{ (\Tilde{\Box}\phi)^{3} + 2\Tilde{\nabla}_{\mu}\Tilde{\nabla}^{\nu}\phi\Tilde{\nabla}_{\nu}\Tilde{\nabla}^{\alpha}\phi\Tilde{\nabla}_{\alpha}\Tilde{\nabla}^{\mu}\phi - 3\Tilde{\nabla}_{\mu}\Tilde{\nabla}_{\nu}\phi\Tilde{\nabla}^{\mu}\Tilde{\nabla}^{\nu}\phi\Tilde{\Box}\phi\right\}
\nonumber
\right. \\
& & \left.
+ P_{1}(\chi, \Tilde{Y})\right] \, .
\nonumber
\eea
where $\Tilde{X} = -\Tilde{g}^{\mu\nu}\Tilde{\nabla}_{\mu}\phi\Tilde{\nabla}_{\nu}\phi\,/2$, $\Tilde{\Box}\phi = \Tilde{g}^{\mu\nu}\Tilde{\nabla}_{\mu}\Tilde{\nabla}_{\nu}\phi$ and $\Tilde{Y} = -\Tilde{g}^{\mu\nu}\Tilde{\nabla}_{\mu}\chi\Tilde{\nabla}_{\nu}\chi\, /2$. $\Tilde{G}_{2}, \Tilde{G}_{3}, \Tilde{G}_{4}$ and $\Tilde{G}_{5}$ are arbitrary functions of $\phi$ and $\Tilde{X}$. Different subclasses of Horndeski theories exist, depending on the form of these functions. Some of these cases are outlined in Table~\eqref{Table: Horndeski special cases}. In the above action, the terms containing $\Tilde{G}_{4}$ and $\Tilde{G}_{5}$ have quadratic and cubic order second derivative terms, respectively.
For our analysis in this section, we consider a class of kinetic braiding theories and their extensions to describe the dynamics of the scalar field $\phi$, non-minimally coupled to gravity. We also assume the non-minimally coupling to be independent of the kinetic term $\Tilde{X}$. This combines the first two special cases of Table~\eqref{Table: Horndeski special cases}, and corresponds to setting $\Tilde{G}_{4\Tilde{X}} = \Tilde{G}_{5} = 0$, i. e., we don't take into account the quadratic and cubic order second derivative terms of the Horndeski action. Therefore, the resulting special class of the Horndeski action describes Brans-Dicke, k-essence, kinetic gravity braiding (G-inflation), and theories of scalar field non-minimally coupled to gravity for the $\phi$ field. Such choices are motivated by the constraints from the joint electromagnetic and gravitational wave observations~\cite{2017-Creminelli.Vernizzi-Phys.Rev.Lett., 2017-Baker.etal-Phys.Rev.Lett., 2017-Ezquiaga.Zumalacarregui-Phys.Rev.Lett., Amendola:2018ltt, TerenteDiaz:2023iqk}. 

The non-minimal coupling of scalar field $\phi$ can be removed using the following conformal transformation~\cite{Tsujikawa:2010zza}:
\begin{equation} \label{Eq:Conformal transformation}
    g_{\mu\nu} = \Tilde{G}_{4}(\phi)\Tilde{g}_{\mu\nu} = \mathrm{e}^{-2\alpha(\phi)}\Tilde{g}_{\mu\nu} \, .
\end{equation}
With this metric redefinition and defining the functions $G_2, G_3$ for a simpler representation, we arrive at the following form of the action in the Einstein frame:
\begin{equation} \label{Eq:Field coupled:Final action}
    S = \int \mathrm{d}^{4}x\sqrt{-g} \left[ \frac{M^{2}_{\mathrm{Pl}}}{2}R + G_{2}(\phi, X) -G_{3}(\phi, X)\Box\phi  + \mathrm{e}^{4 \alpha(\phi)} 
    P_{1}(\chi, Z(\phi, Y))  \right] \, ,
\end{equation}
where $G_2(\phi, X), G_3(\phi, X)$ and $\alpha(\phi)$ are related to the original functions $\Tilde{G}_2, \Tilde{G}_3$ and $\Tilde{G}_4$ via the relations \eqref{Appendix:Eq:Field coupled:Redefinitions},  $X = -g^{\mu\nu}\nabla_{\mu}\phi\nabla_{\nu}\phi\,/2 \equiv -\phi_{\mu}\phi^{\mu}\,/2$, $\Box\phi = g^{\mu\nu}\nabla_{\mu}\nabla_{\nu}\phi \equiv g^{\mu\nu}\phi_{\mu\nu}$, $Y = -g^{\mu\nu}\nabla_{\mu}\chi\nabla_{\nu}\chi\,/2 \equiv -\chi_{\mu}\chi^{\mu}\,/2$ and $Z(\phi, Y) = \mathrm{e}^{-2 \alpha(\phi)}Y$. The details for the derivation of the action \eqref{Eq:Field coupled:Final action} are outlined in Appendix \ref{Appendix:Field-Coupling Derivation}. The above action corresponds to interacting scalar fields $(\phi, \chi)$ minimally coupled to gravity. 

The variation of action \eqref{Eq:Field coupled:Final action} with respect to metric $g_{\mu\nu}$ gives the Einstein equations:
\begin{equation} \label{Eq:Einstein equation}
    G_{\mu\nu} = \frac{1}{M^{2}_{\mathrm{Pl}}}T_{\mu\nu} \, ,
\end{equation}
where the energy-momentum tensor is given by:
\begin{eqnarray} \label{Eq:Field coupled:Total EM-tensor}
\begin{aligned}
T_{\mu\nu} =
& \
g_{\mu\nu} \, G_{2} + \phi_{\mu}\phi_{\nu} \, G_{2X} + 
 g_{\mu\nu}\phi^{\sigma}\nabla_{\sigma} \, G_{3} - \phi_{\mu}\phi_{\nu}\Box\phi \, G_{3X} - \phi_{\nu}\nabla_{\mu}G_{3} - \phi_{\mu}\nabla_{\nu}G_{3}
\\
&
+ g_{\mu\nu}\mathrm{e}^{4 \alpha(\phi)}P_{1} + \mathrm{e}^{2 \alpha(\phi)}\chi_{\mu}\chi_{\nu}P_{1Z} \, .
\end{aligned}
\end{eqnarray}
Due to direct interactions between the two scalar fields $\phi$ and $\chi$, there is no unique way to write the energy-momentum tensor of the individual scalar fields. For obvious reasons, we split the total tensor in the following way:
\begin{subequations} \label{Eq:Field coupled:Splitting EM-tensor}
\begin{align}
T_{\mu\nu} =
&
\ T^{(\phi)}_{\mu\nu} + T^{(\chi)}_{\mu\nu} \, ,  \label{SubEq:Field coupled:Total EM-tensor} \\
\begin{split}
T^{(\phi)}_{\mu\nu} = & \
g_{\mu\nu}G_{2} + \phi_{\mu}\phi_{\nu} \, G_{2X} \, + 
g_{\mu\nu}\phi^{\sigma} \nabla_{\sigma}G_{3} - \phi_{\mu}\phi_{\nu}\Box\phi \,  G_{3X} \label{SubEq:Field coupled:phi EM-tensor} - \phi_\nu \, \nabla_{\mu}G_{3} - 
\phi_{\mu} \nabla_{\nu}G_{3} \, ,
\end{split}
\\
T^{(\chi)}_{\mu\nu} = & \ g_{\mu\nu}\mathrm{e}^{4 \alpha(\phi)}P_{1} + \mathrm{e}^{2 \alpha(\phi)} \chi_{\mu}\chi_{\nu}P_{1Z} \, .
\label{SubEq:Field coupled:chi EM-tensor}
\end{align}
\end{subequations}
From action \eqref{Eq:Field coupled:Final action}, the equations of motion describing the fields $\phi$ and $\chi$, respectively take the form:
\bse \label{Eq:Field coupled:Scalar field EOMs}
\begin{eqnarray}
& & \mathrm{e}^{4\alpha(\phi)}P_{1\chi} + \mathrm{e}^{2 \alpha(\phi)}\Box\chi \, P_{1Z} + 2\mathrm{e}^{2 \alpha(\phi)}\chi^{\mu}\nabla_{\mu}\alpha(\phi) \, P_{1Z} + \mathrm{e}^{2\alpha(\phi)}\chi^{\mu}\nabla_{\mu}P_{1Z} = 0 \, ,  \label{Subeq:Field coupled:chi EOM} \\[5pt] 
& & 
G_{2\phi} + \Box\phi \, G_{2X} - 2XG_{2X\phi} - \phi_{\mu\sigma}\phi^{\sigma}\phi^{\mu} \, G_{2XX}
- 2\,\Box\phi \, G_{3\phi} + 2XG_{3\phi\phi} \nonumber \\
&+ &
2[\phi_{\mu\nu}\phi^{\mu}\phi^{\nu} + X\Box\phi] \, G_{3\phi X} - [\phi_{\mu\sigma}\phi^{\sigma}\phi^{\mu\nu}\phi_{\nu} - \phi_{\mu\nu}\phi^{\mu}\phi^{\nu}\Box\phi] \, G_{3XX} + [\phi_{\mu\nu}\phi^{\mu\nu} - (\Box\phi)^{2}] \, G_{3X} \nonumber \\
&+ &
R_{\mu\nu}\phi^{\mu}\phi^{\nu} \, G_{3X} + 4\alpha_{\phi}\mathrm{e}^{4 \alpha(\phi)} \, P_{1} - 2Y\alpha_{\phi}\mathrm{e}^{2 \alpha(\phi)} \, P_{1Z} = 0 \, .
\end{eqnarray}
\ese
The set of equations \eqref{Eq:Einstein equation}, \eqref{Eq:Field coupled:Splitting EM-tensor} and \eqref{Eq:Field coupled:Scalar field EOMs} completely describe our system. Since the two scalar fields $\phi$ and $\chi$ interact, the conservation of the energy-momentum tensor of the individual components is violated. The interaction between the two scalar fields can be described as~\cite{2021-Johnson.Shankaranarayanan-Phys.Rev.D}:
\begin{equation} \label{Eq:Interaction strength definition}
    Q_{\nu} = \nabla^{\mu}T^{(\chi)}_{\mu\nu} = -\nabla^{\mu}T^{(\phi)}_{\mu\nu} \, .
\end{equation}
Substituting $T^{(\chi)}_{\mu\nu}$ from Eq. \eqref{SubEq:Field coupled:chi EM-tensor} in the above expression, we get:
\begin{equation} \label{Eq:Field coupled:Interaction strength:Field representation}
Q_{\nu} = 2\mathrm{e}^{2\alpha(\phi)}\alpha_{\phi}\phi_{\nu}\left[ 2\mathrm{e}^{2\alpha(\phi)}P_{1} - 
    Y\, P_{1Z} \right] \, .
\end{equation}

From Eq.~\eqref{SubEq:Field coupled:chi EM-tensor}, we find the trace $T^{(\chi)}$ of the energy-momentum tensor $T^{(\chi)}_{\mu\nu}$:
\begin{equation} \label{Eq:Field coupled:Trace of chi EM-tensor}
\begin{aligned}
&   T^{(\chi)} = 2\mathrm{e}^{2\alpha(\phi)}\left[2\mathrm{e}^{2\alpha(\phi)}P_{1} -  YP_{1Z} \right] \, .   
\end{aligned}
\end{equation}
Substituting this in Eq.~\eqref{Eq:Field coupled:Interaction strength:Field representation}, we can now rewrite the interaction strength $Q_{\nu}$ in the following form:
\begin{equation} \label{Eq:Field coupled:Interaction strength:Trace representation}
\begin{aligned}
Q_{\nu}^{(I)} 
= \alpha_{\phi}\phi_{\nu}T^{(\chi)}
= T^{(\chi)}\nabla_{\nu}(\alpha(\phi)) \, .
\end{aligned}
\end{equation}
This is the first key result of this work; it relates the interaction strength with the trace of the energy-momentum tensor of the $\chi$ field and the coupling $\alpha(\phi)$. Hence, we shall refer to such models as \emph{field couplings}.
To begin with, the form of $Q_{\nu}$ is the same as the one derived in Ref.~\cite{2021-Johnson.Shankaranarayanan-Phys.Rev.D}. Therefore, the form of the interaction strength established in Eq.~\eqref{Eq:Field coupled:Interaction strength:Trace representation} is not unique to a canonical scalar field theory and can be extended to a wider class of scalar field descriptions. Furthermore, the above similarity with the results established in Ref.~\cite{2021-Johnson.Shankaranarayanan-Phys.Rev.D} provides a physical mapping of the two scalar fields $(\phi, \chi)$. Specifically, this allows us to associate the scalar fields $\phi$ and $\chi$ with DE and DM, respectively. Hence, we will now refer to the $\phi$ field as the DE field and $\chi$ as the DM field. 

\subsection{Field-Fluid mapping for field-coupling model}

While the field theory description is a fundamental representation, the fluid description is more advantageous for relating to cosmological observations. 
In the fluid description, the prevailing method is to represent the DM field 
as a fluid. To go about this, we need to rewrite the interaction strength in Eq.~\eqref{Eq:Field coupled:Interaction strength:Trace representation} in the fluid description of DM. To start with, we write the energy-momentum tensor for the DM perfect fluid as:
\begin{equation} \label{Eq:Perfect fluid EM-tensor}
    T^{(DM)}_{\mu\nu} = p_{_{DM}}g_{\mu\nu} + (\rho_{_{DM}} + p_{_{DM}})u_{\mu}u_\nu \, ,
\end{equation}
with the corresponding energy density $\rho_{_{DM}}$ and the pressure $p_{_{DM}}$ related to the scalar field description as:
\begin{subequations} \label{Eq:Field coupled:DM field-fluid mapping}
\begin{align}
&
\rho_{_{DM}} = \mathrm{e}^{2\alpha(\phi)}\left[ 2YP_{1Z} - \mathrm{e}^{2\alpha(\phi)}P_{1} \right] \, ,
\\
&
p_{_{DM}} = \mathrm{e}^{4\alpha(\phi)}P_{1} \, .
\end{align}
\end{subequations}
The four-velocity $u_{\mu}$ of the DM fluid is given by:
\begin{equation} \label{Eq:Field coupled:DM fluid 4-velocity}
    u_{\mu} = -[-g^{\alpha\beta}\chi_{\alpha}\chi_{\beta}]^{-\frac{1}{2}}\chi_{\mu} \, .
\end{equation}
Finally, the interaction strength $Q_{\nu}^{(I)}$ takes the form:
\begin{equation} \label{Eq:Field coupled:Interaction strength:Fluid representation}
    Q_{\nu}^{(I)} = (3p_{_{DM}} - \rho_{_{DM}})\nabla_{\nu}\alpha(\phi) \, .
\end{equation}
If the interaction strength $Q_{\nu}^{(I)}$ takes the form described above, we can represent the corresponding DE-DM interaction theory using a classical field theoretic action \eqref{Eq:Field coupled:Final action}, with the fluid-to-field mapping given by the relations in Eqs.~\eqref{Eq:Field coupled:DM field-fluid mapping} and \eqref{Eq:Field coupled:DM fluid 4-velocity}. This demonstrates that the field-fluid mapping of DM for the DE-DM interaction theory established in Ref. \cite{2021-Johnson.Shankaranarayanan-Phys.Rev.D} can be extended to more generalized descriptions of the dark sector components with an arbitrary coupling function $\alpha(\phi)$ and is not limited to the canonical field description. Furthermore, this implies that the form of interaction strength described in Eq. \eqref{Eq:Field coupled:Interaction strength:Fluid representation} within the fluid description of DM cannot distinguish between canonical and non-canonical descriptions of the DM scalar field. Therefore, the form of interaction strength $Q_{\nu}$ in the fluid picture, as established in Eq. \eqref{Eq:Field coupled:Interaction strength:Fluid representation} and Ref. \cite{2021-Johnson.Shankaranarayanan-Phys.Rev.D}, is not unique for a fluid-field mapping of DM. This is evident from the mapping described in Eq. \eqref{Eq:Field coupled:DM field-fluid mapping}, where $P_{1}(\chi, Z)$ is an arbitrary function describing the dynamics of the scalar field $\chi$. Note that unless otherwise specified, we do not make any assumption regarding the DM fluid. All the evolution equations in this work are derived for an arbitrary, non-zero energy density and pressure for the DM fluid, satisfying energy conservation. We shall assume a pressureless DM fluid only when we discuss the stability of the system in Sec.~\eqref{Section:Cosmological background evolution} (Tables \eqref{Table:Field-coupling:Fixed points} and \eqref{Table:Kinetic-coupling:Fixed points}, to be more specific) and perform a numerical analysis in Sec.~\eqref{Section:Numerical Evolution}\,.

\section{Dark sector interaction from a Horndeski action with $G_{4}(\phi, X)$ type non-minimal field-kinetic coupling} \label{Section:Field-Kinetic Coupling DE-DM}

Building on the insights from the previous section, we now expand the analysis by considering the Horndeski action \eqref{Eq:Horndeski action}, relaxing the form of $\Tilde{G}_{4\Tilde{X}}$ while setting $\Tilde{G}_{5} = 0$. 
Considering non-zero $\Tilde{G}_{4\Tilde{X}}$ extends the scalar-tensor theory discussed in Sec.~\eqref{Section:Field-Coupling DE-DM} to include quadratic-order second derivative terms from the Horndeski Lagrangian, along with a non-minimal and kinetic type coupling of the field $\phi$ to gravity. Consequently, we ignore the cubic order and coupling terms to the Einstein tensor. As we show, the additional kinetic dependence of the coupling $\Tilde{G}_{4}(\phi, \Tilde{X})$ modifies the theory and impacts the results from the previous section. Consistent with our earlier discussion, we assume that the scalar fields $\phi$ and $\chi$, correspond to DE and DM, respectively.

One of the critical changes in the action introduced by non-zero $\Tilde{G}_{4\Tilde{X}}$ is that the derivatives of the scalar field $\phi$ are non-minimally coupled to gravity. These non-minimal coupling terms differ from those discussed in the previous section. Specifically, such a coupling leads to conformal coupling \eqref{Eq:Conformal transformation} where the scalar field ($\chi$) moves along the geodesics of a metric ${g}_{\mu \nu}$. When $\Tilde{G}_{4\Tilde{X}}$ is non-zero, a more general form of coupling, referred to as disformal coupling, needs to be considered~\cite{1993-Bekenstein-Phys.Rev.D,2013-Zumalacarregui.etal-PRD,2015-vandeBruck.etal-JCAP}. In disformal coupling, matter fields move along geodesics of the metric ${g}_{\mu \nu}=\tilde{g}_{\mu \nu}+B(\phi) \partial_\mu \phi \partial_\nu \phi$, where $B$ is an arbitrary function of $\phi$. Disformal interactions have been shown to arise in theories that use an (approximate) shift symmetry for the scalar field to protect the mass of the DE scalar, ensuring that it remains light on cosmological scales~\cite{2013-Zumalacarregui.etal-PRD,2015-vandeBruck.etal-JCAP}.

To remove the non-minimal coupling between gravitation and the scalar field $\phi$, represented by the function $\Tilde{G}_{4}(\phi, \Tilde{X})$ in action \eqref{Eq:Horndeski action}, we perform the following extended conformal transformation~\cite{1993-Bekenstein-Phys.Rev.D,2013-Zumalacarregui.etal-PRD,2015-vandeBruck.etal-JCAP}:
\begin{equation} \label{Eq:Extended conformal transformation}
\Tilde{g}_{\mu\nu} = \Tilde{G}^{-1}_{4}\left( \phi, \tilde{X} \right)g_{\mu\nu} = \mathrm{e}^{2\alpha(\phi, X)} \, g_{\mu\nu}  \, ,
\end{equation}
where $\alpha(\phi, X)$ is an arbitrary function to be determined. Under this transformation and redefining few functions for a simpler representation, the action in the Einstein frame takes the following simplified form:
\bea \label{Eq:Field-kinetic coupled:Final action}
\begin{aligned}
S = \int \mathrm{d}^{4}x\sqrt{-g}
&
\left[ \frac{M^{2}_{\mathrm{Pl}}}{2}R + G_{2}(\phi, X) - G_{3}(\phi, X)\Box\phi + A_{1}(\phi, X)\phi_{\mu\nu}\phi^{\mu\nu} + A_{2}(\phi, X)(\Box\phi)^{2} \right. \\
& \left.
+ A_{3}(\phi, X)(\phi_{\mu\nu}\phi^{\mu}\phi^{\nu})^{2} + A_{4}(\phi, X)\Box\phi\phi_{\mu\nu}\phi^{\mu}\phi^{\nu} + A_{5}(\phi, X)\phi_{\mu\nu}\phi^{\nu}\phi^{\mu\sigma}\phi_{\sigma}\right. \\
& \left. 
 + \mathrm{e}^{4\alpha(\phi, X)}P_{1}\left(\chi, \mathrm{e}^{-2\alpha(\phi, X)}Y \right)\right] \, ,
\end{aligned}
\eea
where the functions $A_{1}(\phi, X) - A_{5}(\phi, X)$ are:
\begin{subequations} \label{Eq:Field-kinetic coupled:DHOST terms}
\begin{align}
&
A_{1}(\phi, X)
    = + M_{\mathrm{Pl}}^{2}\left( \frac{\alpha_{X}}{1 - 2X\alpha_{X}} \right) \, , \\
&
A_{2}(\phi, X)
    = - M_{\mathrm{Pl}}^{2}\left( \frac{\alpha_{X}}{1 - 2X\alpha_{X}} \right) \, , \\
&
A_{3}(\phi, X) 
    = - 2M^{2}_{\mathrm{Pl}}\left( \frac{(\alpha_{X})^3}{1 - 2X\alpha_{X}} \right) \, , \\
&
A_{4}(\phi, X)
     = + 2M^{2}_{\mathrm{Pl}}\left( 
    \frac{(\alpha_{X})^2}{1 - 2X\alpha_{X}} \right) \, , \\
&
A_{5}(\phi, X) 
    = + M^{2}_{\mathrm{Pl}}(\alpha_{X})^2\left(\frac{1 + 2X\alpha_{X}}{1 - 2X\alpha_{X}}\right) \, .
\end{align}
\end{subequations}
Appendix \ref{Appendix:Field-Kinetic Coupling Derivation} contains the details for the derivation of Eqs.~\eqref{Eq:Field-kinetic coupled:Final action} and \eqref{Eq:Field-kinetic coupled:DHOST terms}.  For invertibility, the extended conformal transformation must satisfy the condition:
\begin{equation} \label{Eq:Extended Conformal:Invertibility condition}
    1 - 2X\alpha_{X} \neq 0 \, .
\end{equation}
Due to the additional kinetic dependence of the coupling function $\alpha(\phi, X)$, we shall refer to these models as \emph{field-kinetic coupling} scalar field models. Note that the action \eqref{Eq:Field-kinetic coupled:Final action} does not describe a Horndeski theory anymore since the additional quadratic order second derivative terms lead to equations of motion with higher-order derivative terms (beyond second order)~\cite{2019-Kobayashi-Rept.Prog.Phys.}. Theories of the form established in Eq.~\eqref{Eq:Field-kinetic coupled:Final action} fall under a special class of theories referred to as \emph{Degenerate higher-order scalar-tensor theories} or DHOST theories \cite{2014-Zumalacarregui-Phys.Rev.D, 2015-Gleyzes-Phys.Rev.Lett., Langlois:2018dxi}. The higher-order derivative terms do not cause any instability in the system since an invertible field redefinition does not change the number of physical degrees of freedom. Therefore, the DHOST theory described by the Eq.~\eqref{Eq:Field-kinetic coupled:Final action} still possesses 2+1 physical degrees of freedom \cite{Langlois:2017mdk}. DHOST theories of the form described by Eqs.~\eqref{Eq:Field-kinetic coupled:Final action} and \eqref{Eq:Field-kinetic coupled:DHOST terms} are referred to as \emph{quadratic DHOST theories}~\cite{Langlois:2015cwa, BenAchour:2016cay, Crisostomi:2016czh}. Specifically, they belong to a certain subclass, the so called type $Ia$ which is characterized by certain constraints on the form of the functions $A_{1} - A_{5}$ \cite{BenAchour:2016cay, 2019-Kobayashi-Rept.Prog.Phys.}. It has been shown in the literature that the apparent higher-order equations of motion can be reduced to a second-order system for the scale factor and the scalar field $\phi$ in the context of cosmology \cite{Crisostomi:2017pjs, Crisostomi:2018bsp, 2019-Frusciante-Phys.Lett.B, Lazanu:2024mzj}. It is possible to determine a conformal and disformal factor so that all Lagrangians belonging to type Ia theories can be mapped back to quadratic Horndeski theories \cite{BenAchour:2016cay}.
The pictorial representation of the mapping of different field theories under conformal and disformal transformations is given below:
\begin{align}
\mathrm{Horndeski \ theory}
&
\xLeftrightarrow[]{\text{conformal}} \mathrm{Horndeski \ theory} \, ,
\\
\mathrm{Quadratic \ Horndeski \ theory} 
&
\xLeftrightarrow[\text{disformal}]{\text{extended conformal}} \mathrm{Quadratic \ DHOST \ type \ Ia \ theory} \, .
\end{align}
Since we have moved to the Einstein frame, the variation of the action \eqref{Eq:Field-kinetic coupled:Final action} with respect to the metric $g_{\mu\nu}$ gives us the Einstein equations \eqref{Eq:Einstein equation}, where 
\begin{equation} \label{Eq:Field-kinetic coupled:Total EM-tensor}
\begin{aligned}
T_{\mu\nu} &= 
T_{\mu\nu}^{(II,\phi)} + g_{\mu\nu}\mathrm{e}^{4\alpha(\phi,X)} \, P_{1} + \mathrm{e}^{2\alpha(\phi,X)}\chi_{\mu}\chi_{\nu} \, P_{1Z}
 + 4\,\alpha_X\mathrm{e}^{4\alpha(\phi,X)}\phi_{\mu}\phi_{\nu}\,P_{1}
- 2Y\alpha_X\mathrm{e}^{2\alpha(\phi,X)}\phi_{\mu}\phi_{\nu} \, P_{1Z} \, .
\end{aligned}
\end{equation}
We have defined $Z(\phi, X, Y) = \mathrm{e}^{-2 \alpha(\phi, X)}Y$. $T_{\mu\nu}^{(II,\phi)}$ comprises of the terms $G_2$, $G_3$ and $A_1-A_5$, which are explicitly made up of quantities related to the DE field $\phi$ only. 
Since there is no unique way of expressing the energy-momentum tensor for the individual components, we have split $T_{\mu\nu}$ in the following way:
\begin{subequations} \label{Eq:Field-kinetic coupled:EM-tensor breakup}
\begin{align}
&
T_{\mu\nu}^{(DM)} = g_{\mu\nu}\mathrm{e}^{4\alpha(\phi,X)} \, P_{1} + \mathrm{e}^{2\alpha(\phi,X)}\chi_{\mu}\chi_{\nu} \, P_{1Z} \, ,
\label{SubEq:Field-kinetic coupled:DM tensor}
\\
&
T_{\mu\nu}^{(DE)} = T_{\mu\nu}^{(II,\phi)} +\alpha_X\phi_{\mu}\phi_{\nu}\,[4\mathrm{e}^{4\alpha(\phi,X)} \, P_{1}
- 2Y\mathrm{e}^{2\alpha(\phi,X)}P_{1Z}] \, .
\label{SubEq:Field-kinetic coupled:DE tensor}
\end{align}
\end{subequations}
Expressing the individual components of the energy-momentum tensors for the DE and DM fields in this form will enable us to easily map the DM field to the perfect fluid description and write down its four-velocity.
Varying the action \eqref{Eq:Field-kinetic coupled:Final action} with respect to the fields $\chi$ and $\phi$, respectively gives us the corresponding equations of motion which are given by:
\begin{subequations}
\begin{align}
\begin{split}
&
\mathrm{e}^{4\alpha(\phi,X)}P_{1\chi} + 2\alpha_{\phi}\mathrm{e}^{2\alpha(\phi,X)}\phi_{\mu}\chi^{\mu} \, P_{1Z} - 2\alpha_X\mathrm{e}^{2\alpha(\phi,X)}\,\phi_{\mu\nu}\phi^{\mu}\chi^{\nu}\,P_{1Z} + \mathrm{e}^{2\alpha(\phi,X)}\Box\chi \, P_{1Z} 
\\
- & 
2Y\mathrm{e}^{2\alpha(\phi,X)} \, P_{1\chi Z} - \chi_{\mu\nu}\chi^{\mu}\chi^{\nu} \, P_{1ZZ} - 2Y\alpha_{\phi}\chi^{\mu}\phi_{\mu} \, P_{1ZZ} + 2Y\alpha_X\phi_{\mu\nu}\phi^{\mu}\chi^{\nu} \, P_{1ZZ}
= 0 \, , 
\label{Eq:Field-kinetic coupled:chi EOM}
\end{split}
\\[8pt]
\begin{split}
&
E(\phi, X, \phi_{\mu\nu}\phi^{\mu}\phi^{\nu}, \Box\phi,...) + \alpha_X\phi_{\mu}\chi^{\mu}\left[ 4\mathrm{e}^{4\alpha(\phi,X)}P_{1\chi}
- 2Y\mathrm{e}^{2\alpha(\phi,X)}P_{1\chi Z} \right]
\\ 
+ &
2\left[ -\alpha_X\chi_{\mu\nu}\chi^{\mu}\phi^{\nu} + 4X\alpha_{\phi}\alpha_X + 2\alpha_X^2\phi_{\mu\nu}\phi^{\mu}\phi^{\nu} \right]\left[ \mathrm{e}^{2\alpha(\phi,X)}P_{1Z} - YP_{1ZZ} \right]
\\
+ &
\left[ \alpha_{\phi} + \alpha_X\Box\phi - 2X\alpha_{\phi X} - 8X\alpha_{\phi}\alpha_X - \alpha_{XX}\phi_{\mu\nu}\phi^{\mu}\phi^{\nu} - 4\alpha_X^2\phi_{\mu\nu}\phi^{\mu}\phi^{\nu} \right]
\\
\times &
\left[ 4\mathrm{e}^{4\alpha(\phi,X)}P_{1}
- 2Y\mathrm{e}^{2\alpha(\phi,X)}P_{1Z} \right]
= 0 \, .
\label{Eq:Field-kinetic coupled:phi EOM}
\end{split}
\end{align}
\end{subequations}
$E(\phi, X, \phi_{\mu\nu}\phi^{\mu}\phi^{\nu}, \Box\phi,...)$ accounts for the contribution of the terms $G_2, G_3$ and $A_1-A_5$ to the equations of motion. Explicit form of this term is complex and not particularly insightful, so it is not provided here. \href{https://www.dropbox.com/scl/fo/485inatss9xb9cufted8a/AB8bt3gaNJRNxofLvjrnJIQ?rlkey=2939qlm09cw90rmlugcxgzaz4&st=ej8g8fz2&dl=0}{For further details, please refer to the online Mathematica file.}
Using the definition \eqref{Eq:Interaction strength definition}, the interaction strength $Q_{\nu}^{(II)} $ takes the following form:
\bea \label{Eq:Field-kinetic coupled:Interaction strength:Trace form}
\begin{aligned}
Q_{\nu}^{(II)}
&
= [4\mathrm{e}^{4\alpha(\phi,X)} \, P_{1}
- 2Y\mathrm{e}^{2\alpha(\phi,X)} \, P_{1Z}]\nabla_{\nu}\alpha(\phi, X) \, ,
\\
&
= T^{(DM)}\nabla_{\nu}\alpha(\phi, X) \, ,
\\
&
= -\nabla^{\mu}T^{(II,\phi)}_{\mu\nu} - T^{(DM)}\left[ \alpha_X(\Box\phi\phi_{\nu} + \phi_{\mu\nu}\phi^{\mu}) + \phi_\nu\phi^\mu\nabla_\mu\alpha_{X} \right] - \left[\alpha_X\phi_{\nu}\right]\phi^{\mu}T^{(DM)}_{\mu} \, ,
\end{aligned}
\eea
where $T^{(DM)}$ refers to the trace of the energy-momentum tensor $T_{\mu\nu}^{(DM)}$. 
This is the second key result of this work, regarding which we want to discuss the following points: To begin with, the interaction term is different from the previous section and in Ref.~\cite{2021-Johnson.Shankaranarayanan-Phys.Rev.D}. In other words, the dependence of $\tilde{G}_{4}$ on $\tilde{X}$ introduces non-trivial interaction in the dark sector. In addition, in the particular case where the coupling function $\alpha(\phi, X)$ is independent of the kinetic term $X$, the above expression reduces to Eq.~\eqref{Eq:Field coupled:Interaction strength:Trace representation} and the theory reduces to the one discussed in Sec.~\eqref{Section:Field-Coupling DE-DM}.

\subsection{Field-Fluid mapping for the field-kinetic coupling model}

As mentioned earlier, moving to the perfect fluid description of the DM component turns out to be useful for cosmological observations. To go about this, we first define the energy density $\rho_{_{DM}}$ and the pressure $p_{_{DM}}$ as follows:
\begin{subequations} \label{Eq:Field-kinetic coupled:rho & p}
\begin{align}
& 
\rho_{_{DM}} = \mathrm{e}^{2\alpha(\phi,X)}\left[ 2Y \, P_{1Z} - \mathrm{e}^{2\alpha(\phi,X)} \, P_{1} \right] \, , \\
&
p_{_{DM}} = \mathrm{e}^{4\alpha(\phi,X)} \, P_{1} \, .
\end{align}
\end{subequations}
The four-velocity $u_{\mu}$ for the DM fluid is given by Eq.~\eqref{Eq:Field coupled:DM fluid 4-velocity}. This reduces $T_{\mu\nu}^{(DM)}$ to the energy-momentum tensor describing a perfect fluid \eqref{Eq:Perfect fluid EM-tensor}.
Correspondingly, we can rewrite $T_{\mu\nu}^{(DE)}$ in Eq.~\eqref{SubEq:Field-kinetic coupled:DE tensor}, leading to the following:
\begin{eqnarray} \label{Eq:Field-kinetic coupled:DE tensor:DM fluid representation}
    T_{\mu\nu}^{(DE)} = T_{\mu\nu}^{(II,\phi)} + (3p_{_{DM}} - \rho_{_{DM}}) \, \alpha_X\phi_{\mu}\phi_{\nu} \, .
\end{eqnarray}
We can rewrite the interaction strength $Q_{\nu}^{(II)}$ in Eq.~\eqref{Eq:Field-kinetic coupled:Interaction strength:Trace form} as:
\begin{equation} \label{Eq:Field-kinetic coupled:Interaction strength:Fluid representation}
\begin{aligned}
Q_{\nu}^{(II)} =
(3p_{_{DM}} - \rho_{_{DM}})\nabla_{\nu}\alpha(\phi, X)\, .
\end{aligned}
\end{equation}
Let us now compare the above expression with the interaction strength $Q_{\nu}^{(I)}$ in Eq.~\eqref{Eq:Field coupled:Interaction strength:Fluid representation}. 
First, the general form of  $Q_{\nu}$ is the same in both cases; however, the conformal/extended conformal transformations are very different. Second, in both cases, a nonvanishing zeroth component of the interaction strength indicates an energy exchange between DE and DM, which is attributed solely to the non-gravitational direct interactions characterized by the couplings $\alpha(\phi)$ and $\alpha(\phi, X)$, respectively. The spatial components of the interaction strength indicate a momentum exchange between the dark sector components. For a homogeneous and isotropic background, the spatial components vanish, corresponding to a net zero momentum exchange. As we shall show in Sec.~\eqref{Subsection:Momentum Transfer}, the theory predicts a non-zero momentum transfer in first-order perturbations. 

Third, for both cases, the interaction term vanishes for an equation of state when $p = \rho/3$ (corresponding to radiation), regardless of the nature of the coupling. This implies that a field theoretic description of the dynamical DE and dark radiation type model, as discussed in this section, will not exhibit non-gravitational interactions between the two dark sector components. This result can provide a strong theoretical constraint on dark radiation models~\cite{Archidiacono:2022ich}. Fourth, for a pressureless DM fluid description, the expression for the interaction strength $Q_{\nu}^{(II)}$ established in Eq.~\eqref{Eq:Field-kinetic coupled:Interaction strength:Fluid representation} takes the exact form as the interaction term in Eqs. (2.46) and (2.47) established in Ref.~\cite{2020-Kase.Tsujikawa-JCAP} for $\mathrm{e}^{\alpha} = (1 + f_{1})$. Note that the fluid in Ref.~\cite{2020-Kase.Tsujikawa-JCAP} is not pressureless, hence, the equations of motion are not strictly identical. 

Lastly, assuming an equation of state different from radiation, we can write the interaction strength $Q_{\nu}$ in another form:
\bea \label{Eq:Field-kinetic coupled:Interaction strength:Alternate form}
\begin{aligned}
Q_{\nu}^{(II)} =
& 
\ Q_{\nu}^{(II,\phi)} - \nabla^{\mu}\left[ \alpha_X(3p_{_{DM}} - \rho_{_{DM}})\phi_{\mu}\phi_{\nu}  \right] \, ,
\\
= & 
\ Q_{\nu}^{(II,\phi)} - (3p_{_{DM}} - \rho_{_{DM}})\left[ \alpha_X(\Box\phi\phi_{\nu} + \phi_{\mu\nu}\phi^{\mu}) + \phi_\nu\phi^\mu( \alpha_{\phi X}\phi_{\mu} - \alpha_{XX}\phi_{\mu\sigma}\phi^{\sigma} ) \right]
\\
&
- \left[\alpha_X\phi_{\nu}\right]\phi^{\mu}\nabla_{\mu}(3p_{_{DM}} - \rho_{_{DM}}) \, ,
\end{aligned}
\eea
where $Q_{\nu}^{(II,\phi)} = -\nabla^{\mu}T^{(II,\phi)}_{\mu\nu}$. The above expression suggests that the additional field-kinetic coupling introduces non-trivial interactions in the dark sector, which we discuss in detail in the following section.

\section{Background evolution of the dark sector}
\label{Section:Cosmological background evolution}

Following our discussion of interacting dark sector models in the previous two sections, we now study the cosmological evolution of these two models in the spatially flat FLRW background:
\begin{eqnarray} \label{Eq:Flat FLRW metric}
    \mathrm{d}s^2 = -\mathrm{d}t^2 + a^2(t)\mathrm{d\textbf{x}^2} \, ,
\end{eqnarray}
where $a(t)$ is the scale factor.  Since the interaction strength $Q_{\nu}^{(I)}$ in Eq.~\eqref{Eq:Field coupled:Interaction strength:Fluid representation} is a limit of $Q_{\nu}^{(II)}$ in Eq.~\eqref{Eq:Field-kinetic coupled:Interaction strength:Fluid representation}, in the rest of this work we will only consider $Q_{\nu}^{(II)}$ and we will refer  $Q_{\nu}^{(II)}$ as $Q_{\nu}$.
We now examine the background evolution and the field-fluid mapping for the DM sector. In the above FLRW background, the DM field ($\chi$)  equation of motion is:
\begin{equation} \label{Eq:FLRW:chi EOM}
\begin{aligned}
&
\mathrm{e}^{4\alpha(\phi,X)} \, P_{1\chi} - \mathrm{e}^{2\alpha(\phi,X)}\Dot{\chi}^{2} \, P_{1\chi Z} - \mathrm{e}^{2\alpha(\phi,X)}(\Ddot{\chi} + 3H\Dot{\chi}) \, P_{1Z} -\Dot{\chi}^{2}\Ddot{\chi} \,  P_{1ZZ}
\\
- &
\Dot{\phi}\Dot{\chi}\left[ \alpha_{\phi} + \alpha_X\Ddot{\phi} \right]\left[ 2\mathrm{e}^{2\alpha(\phi,X)} \, P_{1Z} - \Dot{\chi}^{2} \, P_{1ZZ} \right] = 0 \, .
\end{aligned}
\end{equation}
and the interaction strength \eqref{Eq:Field-kinetic coupled:Interaction strength:Trace form} is:
\begin{equation} \label{Eq:FLRW:Interaction strength:Field representation}
\begin{aligned}
\overline{Q}
=
\Dot{\phi}\left[ \alpha_{\phi} + \alpha_{X}\Ddot{\phi} \right]\left[ 4\mathrm{e}^{4\alpha(\phi,X)} \, P_{1} - \mathrm{e}^{2\alpha(\phi,X)}\Dot{\chi}^2 \, P_{1Z} \right] \, ,
\end{aligned}
\end{equation}
In the fluid picture, the equation of motion describing the DM fluid in the above FLRW background is:
\begin{equation} \label{Eq:FLRW:DM fluid EOM}
    \Dot{\rho}_{_{DM}} + 3H(\rho_{_{DM}} + p_{_{DM}}) = -\overline{Q} \, ,
\end{equation}
where the interaction strength in Eqs.~\eqref{Eq:Field-kinetic coupled:Interaction strength:Fluid representation} and \eqref{Eq:Field-kinetic coupled:Interaction strength:Alternate form} takes the form:
\bea \label{Eq:FLRW:Interaction strength:Fluid representation}
\overline{Q} 
& &
= (3p_{_{DM}} - \rho_{_{DM}})(\alpha_{\phi} + \alpha_{X}\Ddot{\phi})\Dot{\phi} \, ,
\\
& &
= \overline{Q}^{(\phi)} + (3p_{_{DM}} - \rho_{_{DM}})\left[ \alpha_{X}\Dot{\phi}(2\Ddot{\phi} + 3H\Dot{\phi}) + (\alpha_{\phi X} + \alpha_{XX}\Ddot{\phi})\Dot{\phi}^{3} \right] + (3\Dot{p}_{_{DM}} - \Dot{\rho}_{_{DM}})\alpha_{X}\Dot{\phi}^{2} \, .
\nonumber
\eea
This form of $\overline{Q}$ generalizes the results established in Ref.~\cite{2021-Johnson.Shankaranarayanan-Phys.Rev.D} for the field-fluid mapping to a larger class of the arbitrary coupling functions since we have considered an additional kinetic dependence of the coupling function $\alpha(\phi, X)$. It must be noted that for any other form of interaction strength, the field-fluid correspondence \emph{may not} exist.

For further analysis, we consider a k-essence type description for the DE field $\phi$ for both models. This corresponds to setting the terms $G_2$ and $G_3$ in the two actions \eqref{Eq:Field coupled:Final action} and \eqref{Eq:Field-kinetic coupled:Final action} as follows:
\begin{eqnarray} \label{Eq:k-essence assumption}
\begin{aligned}
G_{2}(\phi, X) = \beta(\phi)X + \gamma(\phi)X^2 - V(\phi)\, ,
& \quad G_{3}(\phi, X) = 0 \, ,
\end{aligned}
\end{eqnarray}
where $\beta(\phi)$ and $\gamma(\phi)$ are arbitrary functions of $\phi$ and $V(\phi)$ denotes the scalar field potential. We now consider two cases --- field coupling and purely kinetic coupling --- to understand the effect of these couplings in the late-time evolution of the Universe.

In the rest of this section, we will look at the 
evolution of these two models in detail. Specifically, we shall establish an autonomous system of equations for the field-coupling and kinetic-coupling models separately. We shall assume specific forms of these functions when we do the numerical evolution of the coupled models in Sec.~\eqref{Section:Numerical Evolution}.

\subsection{Autonomous system of equations for field-coupling model}
\label{sec:Autonomous01}

 We now define the following dimensionless variables:
\begin{equation} \label{Eq:Field coupled:Dimensionless variables defined}
\begin{aligned}
&
x = \frac{\Dot{\phi}}{\sqrt{6}HM_{\mathrm{Pl}}} \, ,\ \ \ \ \ \ \ \ \ \ \ \ y = \beta(\phi) \, ,\ \ \ \ \ \ \ \ \ \ \ \ v = \gamma(\phi)H^2M_{\mathrm{Pl}}^2 \, ,\ \ \ \ \ \ \ \ \ \ \ \ w = \frac{V}{3H^{2}M_{\mathrm{Pl}}^{2}} \, ,
\\ 
&
b_{1} = M_{\mathrm{Pl}}\frac{V_{\phi}}{V} \, ,\ \ \ \ \ \ \ \ \ \ \ \ \ \ \ b_{2} = \frac{VV_{\phi\phi}}{V_{\phi}^{2}} \, ,\ \ \ \ \ \ \ \ \ \ \beta_{1} = M_{\mathrm{Pl}}\frac{\beta_{\phi}}{\beta} \, ,
\\
&
\beta_{2} = \frac{\beta\beta_{\phi\phi}}{\beta_{\phi}^{2}} \, ,\ \ \ \ \ \ \ \ \ \ \ \ \ \ \ \
\gamma_{1} = M_{\mathrm{Pl}}\frac{\gamma_{\phi}}{\gamma} \, , \ \ \ \ \ \ \ \ \ \gamma_{2} = \frac{\gamma\gamma_{\phi\phi}}{\gamma_{\phi}^{2}} \, ,
\\
&
\alpha_{1} = \alpha(\phi, X) \, , \ \ \ \ \ \ \ \ \ \ \ \ \ \alpha_{2} = M_{\mathrm{Pl}}\frac{\alpha_{\phi}}{\alpha} \, , \ \ \ \ \ \ \ \ \alpha_{3} = \frac{\alpha\alpha_{\phi\phi}}{\alpha_{\phi}^{2}} \, .
\end{aligned}
\end{equation}
Note that $w$ is the dimensionless (scaled) potential. The following set of equations give the autonomous system describing an arbitrary field coupling $\alpha(\phi)$ between the dark sector components:
\begin{subequations} \label{Eq:Field coupled:Autonomous system}
\begin{align}
&
x' - \frac{3x\mathcal{G}}{2} + \frac{\mathcal{F}}{\sqrt{6}} = 0 \, , \\
&
y' - \sqrt{6}xy\beta_{1} = 0 \, , \\
&
v' + v( 3\mathcal{G} - \sqrt{6}\gamma_{1}x) = 0 \, , \\
&
w' - w(3\mathcal{G} + \sqrt{6}b_{1}x) = 0 \, , \\
&
\Omega_{_{DM}}' + 3\Omega_{_{DM}}(\omega_{_{DM}} + 1 - \mathcal{G}) + \overline{q} = 0 \, , \\
&
\Omega_{r}' + \Omega_{r}(4 - 3\mathcal{G}) = 0 \, , \\
&
b_{1}' + \sqrt{6}xb_{1}^{2}(1 - b_{2}) = 0 \, , \\
&
\beta_{1}' + \sqrt{6}x\beta_{1}^{2}(1 - \beta_{2}) = 0 \, , \\
&
\gamma_{1}' + \sqrt{6}x\gamma_{1}^{2}(1 - \gamma_{2}) = 0 \, , \\
&
\alpha_{1}' - \sqrt{6}x\alpha_{1}\alpha_{2} = 0 \, , \\
&
\alpha_{2}' + \sqrt{6}x\alpha_{2}^{2}(1 - \alpha_{3}) = 0 \, .
\end{align}
\end{subequations}
Here, a prime denotes a derivative with respect to the number of e-foldings $N \equiv \mathrm{ln}(a)$. The parameters $\mathcal{F}$ and $\mathcal{G}$ are defined as:
\begin{subequations} \label{Eq:Field coupled:F and G}
\begin{align}
\mathcal{G} =
&
\ \frac{\Omega_{r}}{3} + \omega_{_{DM}}\Omega_{_{DM}} + x^{2}y + 3x^{4}v - w + 1 \, ,
\\
\begin{split}
\mathcal{F} =
&
\ 3\left[\frac{\sqrt{6}x(y + 6x^2v) + x^{2}(\beta_{1}y + 9x^2\gamma_{1}v) + wb_{1} - \alpha_{1}\alpha_{2}\Omega_{_{DM}}(3\omega_{_{DM}} - 1)}{(y + 18x^2v)}\right] \, ,
\end{split}
\end{align}
\end{subequations}
and the energy constraint is given by:
\begin{equation} \label{Eq:Field coupled:Dimensionless Energy constraint}
\Omega_{_{DM}} + \Omega_{r} + x^{2}(y + 9x^2v) + w - 1 = 0 \, .
\end{equation}
The scaled interaction term ($\overline{q}$) is defined as:
\begin{equation} \label{Eq:Field coupled:Scaled Interaction term}
    \overline{q}  = \frac{\overline{Q}}{3H^{3}M_{\mathrm{Pl}}^{2}} = (3\omega_{_{DM}} - 1)\sqrt{6}x\alpha_{1}\alpha_{2}\Omega_{_{DM}} \, .
\end{equation}
Note that various cosmological parameters can be expressed in terms of these variables as:
\begin{equation} \label{Eq:Field coupled:Dimensionless cosmological parameters}
\begin{aligned}
&
\rho_{i} = 3H^{2}M_{\mathrm{Pl}}^{2}\Omega_{i} \, ,
\\
&
\Omega_{_{DE}} = x^{2}(y + 9x^2v) + w \, ,
\\
&
\omega_{_{DE}} = \frac{x^{2}y + 3x^{4}v - w}{x^{2}y + 9x^{4}v + w} \, ,
\\
&
\epsilon = -\frac{\dot{H}}{H^2} = \frac{3}{2}\left(\frac{\Omega_{r}}{3} + \omega_{_{DM}}\Omega_{_{DM}} + x^{2}y + 3x^{4}v - w + 1 \right) = \frac{3\mathcal{G}}{2} \, ,
\end{aligned}
\end{equation}
where $\epsilon$ is the slow-roll parameter, $\omega_{_{DE}}$ and $\omega_{_{DM}}$ refer to the equation of state of DE and DM, respectively. For the stability analysis, we assume a pressureless DM fluid corresponding to $\omega_{_{DM}} = 0$.
We note that this assumption will put constraints on the evolution of the corresponding DM scalar field ($\chi$) related to the energy density and pressure of the DM fluid through Eqs.~\eqref{Eq:Field coupled:DM field-fluid mapping}, \eqref{Eq:Field-kinetic coupled:rho & p} for the two coupling models\,. Refer to Appendix \ref{Appendix:Pressureless fluid constraints} for further discussion\,.

\begin{sidewaystable}
\centering
\begin{tabular}{||>{\columncolor{gray!20}}c@{}|c|c|c|c|c|c|c|c|c|c|c|c|c||>{\columncolor{gray!20}}c@{}|@{}>{\columncolor{gray!20}}c@{}|@{}>{\columncolor{gray!20}}c@{}|@{}>{\columncolor{gray!20}}c@{}|@{}>{\columncolor{gray!20}}c@{}||}
\hline \hline
\textbf{F.P.} & $\mathbf{x^{*}}$ & $\mathbf{y^{*}}$ & $\mathbf{v^{*}}$ & $\mathbf{w^{*}}$ & $\mathbf{b^{*}_{1}}$ & $\mathbf{b^{*}_{2}}$ & $\mathbf{\beta^{*}_{1}}$ & $\mathbf{\beta^{*}_{2}}$ & $\mathbf{\gamma^{*}_{1}}$ & $\mathbf{\gamma^{*}_{2}}$ & $\mathbf{\alpha^{*}_{1}}$ & $\mathbf{\alpha^{*}_{2}}$ & $\mathbf{\alpha_3^{*}}$ & $\mathbf{\Omega^{*}_{_{DE}}}$ & $\mathbf{\Omega^{*}_{r}}$ & $\mathbf{\Omega^{*}_{_{DM}}}$ & $\mathbf{\epsilon^{*}}$ & \textbf{Behavior}$\mathbf{^{*}}$
\\
\hline\hline
1a & 0 & $\neq 0$ & 0 & 0 & - & - & - & - & - & - & - & - & - & 0 & 1 & 0 & 2 & Saddle point
\\
\hline
1b & 0 & $\neq 0$ & 0 & 0 & - & - & - & - & - & - & 0 & - & - & 0 & 0 & 1 & $\frac{3}{2}$ & Saddle point
\\
\hline \hline
1c & $\lambda$ & $\frac{1}{\lambda^2} - 9\lambda^2\lambda_1$ & $\lambda_1$ & 0 & 0 & - & 0 & - & $\frac{\sqrt{6}( 1 - 3\lambda^4\lambda_1)}{\lambda}$ & 1 & - & 0 & - & 1 & 0 & 0 & $3(1 - 3\lambda^4\lambda_1)$ & Attractor for 
\\
& & & & & & & & & & & & & & & & & & $\lambda^4\lambda_1 \geq 1/3$
\\
\hline 
1d & $\lambda$ & $\frac{1}{\lambda^2} - 9\lambda^2\lambda_1$ & $\lambda_1$ & 0 & 0 & - & 0 & - & $\frac{\sqrt{6}( 1 - 3\lambda^4\lambda_1)}{\lambda}$ & 1 & 0 & - & 1 & 1 & 0 & 0 & $3(1 - 3\lambda^4\lambda_1)$ & Attractor for 
\\
& & & & & & & & & & & & & & & & & & $\lambda^4\lambda_1 \geq 1/3, \alpha_2\lambda \leq 0$
\\
\hline 
1e & $\lambda$ & $\lambda_1$ & 0 & $1 - \lambda^2\lambda_1$ & $-\sqrt{6}\lambda\lambda_1$ & 1 & 0 & - & 0 & - & - & 0 & - & 1 & 0 & 0 & $3\lambda^2\lambda_1$ & Attractor for 
\\
& & & & & & & & & & & & & & & & & & $0 \leq \lambda^2\lambda_1 \leq 1/2$
\\
\hline 
1f & $\lambda$ & $\lambda_1$ & 0 & $1 - \lambda^2\lambda_1$ & $-\sqrt{6}\lambda\lambda_1$ & 1 & 0 & - & 0 & - & 0 & - & 1 & 1 & 0 & 0 & $3\lambda^2\lambda_1$ & Attractor for 
\\
& & & & & & & & & & & & & & & & & & $0 \leq \lambda^2\lambda_1 \leq 1/2, \alpha_2\lambda \leq 0$
\\
\hline 
1g & $\lambda$ & 0 & $\lambda_1$ & $1 - 9\lambda^4\lambda_1$ & $-6\sqrt{6}\lambda^3\lambda_1$ & 1 & 0 & - & $6\sqrt{6}\lambda^3\lambda_1$ & 1 & - & 0 & - & 1 & 0 & 0 & $18\lambda^4\lambda_1$ & Attractor for 
\\
& & & & & & & & & & & & & & & & & & $0 \leq \lambda^4\lambda_1 \leq 1/12$
\\
\hline 
1h & $\lambda$ & 0 & $\lambda_1$ & $1 - 9\lambda^4\lambda_1$ & $-6\sqrt{6}\lambda^3\lambda_1$ & 1 & 0 & - & $6\sqrt{6}\lambda^3\lambda_1$ & 1 & 0 & - & 1 & 1 & 0 & 0 & $18\lambda^4\lambda_1$ & Attractor for 
\\
& & & & & & & & & & & & & & & & & & $0 \leq \lambda^4\lambda_1 \leq 1/12, \alpha_2\lambda \leq 0$
\\
\hline \hline
1i & 0 & $ \neq 0$ & - & 1 & 0 & - & - & - & - & - & - & - & - & 1 & 0 & 0 & 0 & Attractor
\\
\hline 
1j & $\lambda$ & $\frac{1}{\lambda^2}$ & 0 & 0 & 0 & - & 0 & - & 0 & - & - & 0 & - & 1 & 0 & 0 & 3 & Saddle point
\\
\hline
1k & $\lambda$ & 0 & $\frac{1}{9\lambda^4}$ & 0 & 0 & - & 0 & - & $\frac{2}{\lambda}\sqrt{\frac{2}{3}}$ & 1 & - & 0 & - & 1 & 0 & 0 &  2 & Saddle point
\\
\hline \hline
1l & $\lambda$ & $ -\frac{2}{\lambda^2}$ & $\frac{1}{3\lambda^4}$ & 0 & 0 & - & 0 & - & 0 & - & - & 0 & - & 1 & 0 & 0 & 0 & Attractor
\\
\hline 
1m & $\lambda$ & $\frac{1}{6\lambda^2}$ & 0 & $\frac{5}{6}$ & $-\frac{1}{\sqrt{6}\lambda}$ & 1 & 0 & - & 0 & - & - & 0 & - & 1 & 0 & 0 & $\frac{1}{2}$ & Attractor
\\
\hline 
1n & $\lambda$ & 0 & $\frac{1}{36\lambda^4}$ & $\frac{3}{4}$ & $-\frac{1}{\sqrt{6}\lambda}$ & 1 & 0 & - & $\frac{1}{\sqrt{6}\lambda}$ & 1 & - & 0 & - & 1 & 0 & 0 & $\frac{1}{2}$ & Attractor 
\\
\hline 
\hline
\end{tabular}
\caption{Fixed points (F.P.) and stability analysis of the autonomous system established for field-coupled DE and DM.}
\label{Table:Field-coupling:Fixed points}
\end{sidewaystable}
We will now look at the fixed points of the autonomous system and discuss the stability of the system at these points.  Table~\eqref{Table:Field-coupling:Fixed points} summarizes the fixed points and the stability of the system at the corresponding points. Note that some fixed point conditions can be realized for any physical values of certain variables. These are denoted by cells with a hyphen (-) in the table. The fixed points have been categorized as the following: 
\begin{itemize}
    \item \emph{Radiation-dominated era}: Point $1a$ represents the radiation-dominated fixed point. From the eigenvalues of the Jacobian matrix of the system, we find that this is a saddle point.
    \item \emph{Matter-dominated era}: Point $1b$ represents the matter-dominated fixed point and is a saddle point.
    \item \emph{DE-dominated era}: Fixed points $1c-1n$ represent DE-dominated points. Note that certain non-zero combinations of either of the two variables $y, v$ and $w$, as established in Table \eqref{Table:Field-coupling:Fixed points}, lead to an accelerated expansion and an attractor behavior of the fixed points. 
\end{itemize}
Compared to the canonical description of the DE field $\phi$, the k-essence description gives a wider range of cases where an accelerated expansion of the Universe can be realized. For instance, cases $1c$, $1d$, and $1l$ represent attractor and accelerated expansion fixed points, which are purely kinetically driven in contrast to the canonical accelerated expansion fixed points, which are potential driven. The attractor nature of these cases also ensures that the universe remains in an accelerated phase. For instance, if $\lambda_1 = 1/(3\lambda^4)$ for points $1c$ and $1d$, we have $\epsilon^{*} = 0 \implies H^{*} = constant$, hence, leading to a de Sitter universe. Additionally, we can see that the interacting model can consistently lead to a radiation-dominated $\rightarrow$ matter-dominated $\rightarrow$ DE-dominated era. Some examples are: $1a \rightarrow 1b \rightarrow 1l$, $1a \rightarrow 1b \rightarrow 1m$, $1a \rightarrow 1b \rightarrow 1n$.

\subsection{Kinetic-coupling models: A specific case}
\label{Subsection:Kinetic model}

The action \eqref{Eq:Field-kinetic coupled:Final action} describes the dynamics of an interacting dark sector with a field-kinetic coupling in the Einstein frame.
Since the functions $A_{1}-A_{5}$, given by Eq.~\eqref{Eq:Field-kinetic coupled:DHOST terms}, are expressed in terms of the coupling $\alpha(\phi, X)$, to keep the equations tractable, we consider a specific form of $\alpha(\phi, X)$\,. From Eq.~\eqref{Eq:Field-kinetic coupled:DHOST terms}\,, it is evident that the numerator of the DHOST terms depends on higher powers of $\alpha_X$ while the denominator depends on $1 - 2X\alpha_X$\,. Since the invertibility condition, given by Eq.~\eqref{Eq:Extended Conformal:Invertibility condition}\,, ensures that the denominator of these terms does not vanish, $\alpha$ can take any form such that $1 - 2X\alpha_X \neq 0$. Besides this, there is no condition on $\alpha$. Given this, we shall assume a functional form of $\alpha$ that simplifies the denominator and hence, the overall form of all the DHOST terms $A_1-A_5$\,. To do this, we make the following choice for $\alpha(\phi, X)$:
\begin{eqnarray} \label{Eq:Kinetic-coupling:Coupling form assumption}
    \alpha_{X} = \frac{\alpha_0}{2X} \implies \alpha(\phi, X) = 
    \frac{\alpha_0}{2}\ln X + C(\phi) \, ,
\end{eqnarray}
where $\alpha_0$ is an arbitrary constant, denoting the kinetic coupling strength, and $C(\phi)$ is an arbitrary constant of integration. We note that for any other form of the coupling function $\alpha(\phi, X)\,,$ the equations of motion become highly complex and increasingly difficult to analyze. Since the coupling $\alpha$ must satisfy the invertibility condition \eqref{Eq:Extended Conformal:Invertibility condition}, we have $\alpha_0 \neq 1$\,. Furthermore, we set the function $C(\phi) = 0\,.$ Thus, we shall consider a purely kinetic coupling between the dark sector components and $\Dot{\phi} \neq 0$\,.

Substituting the form of the coupling function $\alpha(X)$ back in Eqs.~\eqref{Eq:FLRW:DM fluid EOM} and \eqref{Eq:FLRW:Interaction strength:Fluid representation} leads to the following fluid equation of motion describing DM:
\bea \label{Eq:Kinetic-coupling:FLRW:DM EOM}
& & 
\Dot{\rho}_{_{DM}} + 3H(\rho_{_{DM}} + p_{_{DM}}) = \alpha_0 
\left( \rho_{_{DM}} - 3p_{_{DM}} \right)\left(\frac{\Ddot{\phi}}{\Dot{\phi}}\right) \, .
\eea

Substituting the form of $G_2, G_3$, $\alpha(X)$ and fluid description of DM in Eq.~\eqref{Eq:Field-kinetic coupled:phi EOM} leads to the following equation of motion describing the scalar field $\phi$:
\bea 
\begin{aligned} \label{Eq:Kinetic-coupling:FLRW:Original phi EOM}
&
- \beta(\phi)\Dot{\phi}\Ddot{\phi} - 3\gamma(\phi)\Dot{\phi}^3\Ddot{\phi} - 3H\left( \beta(\phi)\Dot{\phi}^2 + \gamma(\phi)\Dot{\phi}^4 \right) - \frac{1}{2}\beta_{\phi}\Dot{\phi}^3 - \frac{3}{4}\gamma_{\phi}\Dot{\phi}^5 - V_{\phi}\Dot{\phi} 
\\
&
+ \ \frac{6\alpha_0(\alpha_0+1)M_{\mathrm{Pl}}^2}{(\alpha_0-1)}\left[ 9H^3 + 9H\Dot{H} + \Ddot{H} + (3\alpha_0-1)\left( \Dot{H} + 3H^2 \right)\left( \frac{\Ddot{\phi}}{\Dot{\phi}} \right)
\right. \\
& \left.
+ \ 3\alpha_0\left( \frac{\Ddot{\phi}}{\Dot{\phi}} - 3H \right)\left( \frac{\Ddot{\phi}}{\Dot{\phi}} \right)^2 + 2\alpha_0\left( -2\left( \frac{\Ddot{\phi}}{\Dot{\phi}} \right) + 3H \right)\left( \frac{\dddot{\phi}}{\Dot{\phi}} \right) + \alpha_0\left( \frac{\ddddot{\phi}}{\Dot{\phi}} \right)
\right]
\\
&
- \ \alpha_0^2\left( 3p_{_{DM}} - \rho_{_{DM}} \right)\left( \frac{\Ddot{\phi}}{\Dot{\phi}} \right) - 12\alpha_0Hp_{_{DM}} - 3\alpha_0\Dot{p}_{_{DM}}
= - \alpha_0\left( 3p_{_{DM}} - \rho_{_{DM}} \right)\left( \frac{\Ddot{\phi}}{\Dot{\phi}} \right) \, .
\end{aligned}
\eea 

For a spatially flat FLRW background, the total energy density and pressure of the matter fields can be written as:
\bea
\begin{aligned} \label{Eq:Kinetic-coupling:Total rho & P}
\rho_{_{\rm total}} = 
& \
\frac{3\alpha_0(\alpha_0+1)M_{\mathrm{Pl}}^2}{(\alpha_0-1)}\left[ -2\Dot{H} - \frac{2(3\alpha_0+2)}{(\alpha_0+1)}H^2 - 2\alpha_0 \left(\frac{\dddot{\phi}}{\Dot{\phi}}\right) + 3\alpha_0\left( \frac{\Ddot{\phi}}{\Dot{\phi}} \right)^2 - 2(3\alpha_0-1)H\left( \frac{\Ddot{\phi}}{\Dot{\phi}} \right) \right]
\\
& \
 + \frac{1}{2}\beta(\phi)\Dot{\phi}^2 + \frac{3}{4}\gamma(\phi)\Dot{\phi}^4 + V(\phi) + (1-\alpha_0)\rho_{_{DM}} + 3\alpha_0p_{_{DM}} + \rho_r \, ,
\\
p_{_{\rm total}} = 
& \
\frac{\alpha_0M_{\mathrm{Pl}}^2}{(\alpha_0-1)}\left[ -4\Dot{H} -6H^2 - 2(\alpha_0+1)\left( \frac{\dddot{\phi}}{\Dot{\phi}} \right) + (\alpha_0+1)\left( 3\alpha_0+2 \right)\left( \frac{\Ddot{\phi}}{\Dot{\phi}} \right)^2 \right]
\\
& \
 + \frac{1}{2}\beta(\phi)\Dot{\phi}^2 + \frac{1}{4}\gamma(\phi)\Dot{\phi}^4 - V(\phi) + p_{_{DM}} + p_r \, .
\end{aligned}
\eea
Since the radiation has no direct interactions with the dark sector component, the radiation satisfies the conservation equation:
\bea  \label{Eq:FLRW:Radiation EOM}
\Dot{\rho}_r + 4H\rho_r = 0 \, ,
\eea
Hence, we have introduced the pressure and energy density corresponding to the radiation component in the Universe in the above expressions for $p_{\rm total}$ and $\rho_{\rm total}$. The two Friedmann equations are:
\bea
\begin{aligned} \label{Eq:Friedmann equations}
-M_{\mathrm{Pl}}^2(2\Dot{H} + 3H^2) = p_{_{total}},~
3M_{\mathrm{Pl}}^2H^2 = \rho_{_{total}} \, ,
\end{aligned}
\eea
Substituting the total energy density and pressure in the above two Friedmann equations, we have:
\bea
& &
-\frac{(\alpha_0+1)}{(\alpha_0-1)}M_{\mathrm{Pl}}^2\left[ 2\Dot{H} + 3H^2 + 2\alpha_0\left( \frac{\dddot{\phi}}{\Dot{\phi}} \right) - \alpha_0(3\alpha_0+2)\left( \frac{\Ddot{\phi}}{\Dot{\phi}} \right)^2 \right]
\nonumber
\\
& & 
+ \frac{1}{2}\beta(\phi)\Dot{\phi}^2 + \frac{1}{4}\gamma(\phi)\Dot{\phi}^4 - V(\phi) + \ p_{_{DM}} + p_r = 0 \, ,
\label{Eq:Kinetic-coupling:FLRW:Friedmann eq.1}
\\ 
\nonumber 
\\
& & 
-\frac{3(\alpha_0+1)}{(\alpha_0-1)}M_{\mathrm{Pl}}^2\left[ 2\alpha_0\Dot{H} + (6\alpha_0-1)H^2 + 2\alpha_0^2\left( \frac{\dddot{\phi}}{\Dot{\phi}} \right) - 3\alpha_0^2\left( \frac{\Ddot{\phi}}{\Dot{\phi}} \right)^2 + 2\alpha_0(3\alpha_0-1)H\left( \frac{\Ddot{\phi}}{\Dot{\phi}} \right)  \right]
\nonumber
\\
& &
+ \ \frac{1}{2}\beta(\phi)\Dot{\phi}^2 + \frac{3}{4}\gamma(\phi)\Dot{\phi}^4 + V(\phi)
+ (1-\alpha_0)\rho_{_{DM}} + 3\alpha_0 \, p_{_{DM}} + \rho_r = 0 \, .
\label{Eq:Kinetic-coupling:FLRW:Friedmann eq.2}
\eea
We can remove the $\dddot{\phi}$ terms by combining the two equations leading to the following simplified expression:
\bea 
\begin{aligned} \label{Eq:Kinetic-coupling:Simplified expression}
&
-(\alpha_0-1)\left( \rho_{_{DM}} + \rho_r \right) + (3\alpha_0+1)V(\phi) + \frac{(1-3\alpha_0)}{2}\beta(\phi)\Dot{\phi}^2 - \frac{3(\alpha_0-1)}{4}\gamma(\phi)\Dot{\phi}^4
\\
&
-\frac{3(\alpha_0+1)(3\alpha_0-1)}{(\alpha_0-1)}M_{\mathrm{Pl}}^2\left[ H^2 + 2\alpha_0H\left( \frac{\Ddot{\phi}}{\Dot{\phi}} \right) + \alpha_0^2\left( \frac{\Ddot{\phi}}{\Dot{\phi}} \right)^2 \right] = 0 \, .
\end{aligned}
\eea 

It is possible to reduce the above set of equations to second-order differential equations in $\phi(t)$ and $a(t)$. The detailed expressions can be found in Appendix \ref{Appendix:Kinetic-coupling:Second-order EOMs}. Therefore, they do not suffer from the Ostrogradsky instability~\cite{Woodard:2015zca}. Hence, this theory belongs to the class of \emph{healthy} degenerate scalar-tensor theories.

\subsection{Autonomous system of equations for a specific kinetic-coupling model}
\label{sec:Autonomous02}

Like in Sec.~\eqref{sec:Autonomous01}, we now use the following dimensionless variables:
\begin{equation} \label{Eq:Kinetic-coupling:Dimensionless variables defined}
\begin{aligned}
&
x = \frac{\Dot{\phi}}{\sqrt{6}HM_{\mathrm{Pl}}} \, ,\ \ \ \ \ \ \ \ \ \ y = \beta(\phi) \, ,\ \ \ \ \ \ \ \ \ \ v = \gamma(\phi)H^2M_{\mathrm{Pl}}^2 \, ,\ \ \ \ \ \ \ \ \ \ w = \frac{V}{3H^{2}M_{\mathrm{Pl}}^{2}} \, ,
\\ 
&
b_{1} = M_{\mathrm{Pl}}\frac{V_{\phi}}{V} \, ,\ \ \ \ \ \ \ \ \ \ \ \ \ b_{2} = \frac{VV_{\phi\phi}}{V_{\phi}^{2}} \, ,\ \ \ \ \ \ \ \ \beta_{1} = M_{\mathrm{Pl}}\frac{\beta_{\phi}}{\beta} \, ,
\\
&
\beta_{2} = \frac{\beta\beta_{\phi\phi}}{\beta_{\phi}^{2}} \, ,\ \ \ \ \ \ \ \ \ \ \ \ \ \
\gamma_{1} = M_{\mathrm{Pl}}\frac{\gamma_{\phi}}{\gamma} \, , \ \ \ \ \ \ \ \gamma_{2} = \frac{\gamma\gamma_{\phi\phi}}{\gamma_{\phi}^{2}} \, .
\end{aligned}
\end{equation}
Using these variables, we can write the autonomous system of equations:
\begin{subequations} \label{Eq:Kinetic-coupling:Autonomous system}
\begin{align}
&
x' - \frac{3x\mathcal{G}}{2} + \frac{\mathcal{F}}{\sqrt{6}} = 0 \, , \\
&
y' - \sqrt{6}xy\beta_{1} = 0 \, , \\
&
v' + v( 3\mathcal{G} - \sqrt{6}\gamma_{1}x) = 0 \, , \\
&
w' - w(3\mathcal{G} + \sqrt{6}b_{1}x) = 0 \, , \\
&
\Omega_{_{DM}}' + 3\Omega_{_{DM}}(\omega_{_{DM}} + 1 - \mathcal{G}) + \overline{q} = 0 \, , \\
&
\Omega_{r}' + \Omega_{r}(4 - 3\mathcal{G}) = 0 \, , \\
&
b_{1}' + \sqrt{6}xb_{1}^{2}(1 - b_{2}) = 0\, , \\
&
\beta_{1}' + \sqrt{6}x\beta_{1}^{2}(1 - \beta_{2}) = 0\, , \\
&
\gamma_{1}' + \sqrt{6}x\gamma_{1}^{2}(1 - \gamma_{2}) = 0\, ,
\end{align}
\end{subequations}
where the variables $\mathcal{F}$ and $\mathcal{G}$ are defined in Appendix \ref{Appendix:Kinetic-coupling:Second-order EOMs}. Note that the slow-roll parameter $\epsilon = 3\mathcal{G}/2$. The scaled interaction strength ($\overline{q}$) takes the form:
\begin{eqnarray} \label{Eq:Kinetic-coupling:Scaled Interaction strength}
\overline{q} =
\frac{\overline{Q}}{3H^3M_{\mathrm{Pl}}^2} 
= \alpha_0(1 - 3\omega_{_{DM}})\Omega_{_{DM}}\frac{\mathcal{F}}{\sqrt{6}x}\, .
\end{eqnarray}
As mentioned earlier, we are considering a purely kinetic coupling, hence, $x = \Dot{\phi}/(\sqrt{6}HM_{\mathrm{Pl}}) \neq 0$.   
The simplified equation \eqref{Eq:Kinetic-coupling:Simplified expression}, in terms of the dimensionless variables is given by:
\begin{equation} \label{Eq:Kinetic-coupling:Dimensionless simplified expression}
\begin{aligned}
&
F^2 \alpha_0^2 (\alpha_0 + 1) (3\alpha_0 - 1) - 2 \sqrt{6} F \alpha_0 (\alpha_0 + 1) (3\alpha_0 - 1) x + 6 \left[ -1 + \Omega_{_{DM}}
\right. \\
+ & \left.
\ \Omega_r + \alpha_0^2 (3 + \Omega_{_{DM}} + \Omega_r - 3 w) - 2 \alpha_0 (-1 + \Omega_{_{DM}} + \Omega_r - w) + w \right] x^2
\\ 
+ & 
\ 54 (-1 + \alpha_0)^2 v x^6 + 6 (-1 + \alpha_0) (3\alpha_0 - 1) x^4 y = 0\, .
\end{aligned}
\end{equation}
Note that, in the limit $\alpha_0 = 0$, this expression reduces to the energy constraint equation in Eq.~\eqref{Eq:Field coupled:Dimensionless Energy constraint}, which remains the same for a non-interacting as well as field-coupled dark sector.
\begin{sidewaystable}
\setlength{\tabcolsep}{0.1em}
\hspace{-0.5cm}
\begin{tabular}{||>{\columncolor{gray!20}}c@{}|c|c|c|c|c|c|c|c|c|c||>{\columncolor{gray!20}}c@{}|@{}>{\columncolor{gray!20}}c@{}|@{}>{\columncolor{gray!20}}c@{}|@{}>{\columncolor{gray!20}}c@{}|@{}>{\columncolor{gray!20}}c@{}||}
\hline \hline
\textbf{F.P.} & $\mathbf{x^{*}}$\footnote{To ensure the purely kinetic coupling, $x^{*} \neq 0$.} & $\mathbf{y^{*}}$ & $\mathbf{v^{*}}$ & $\mathbf{w^{*}}$ & $\mathbf{b^{*}_{1}}$ & $\mathbf{b^{*}_{2}}$ & $\mathbf{\beta^{*}_{1}}$ & $\mathbf{\beta^{*}_{2}}$ & $\mathbf{\gamma^{*}_{1}}$ & $\mathbf{\gamma^{*}_{2}}$ & $\mathbf{\Omega^{*}_{_{DE}}}$ & $\mathbf{\Omega^{*}_{r}}$ & $\mathbf{\Omega^{*}_{_{DM}}}$ & $\mathbf{\epsilon^{*}}$ & \textbf{Behavior}$\mathbf{^{*}}$
\\
\hline\hline
& & & & & & & & & & & & & & & Attractor if
\\
2a & $\frac{\lambda}{2\sqrt{6}}$ & $\frac{8\alpha_0(9 - \alpha_0(5 + 18\alpha_0))}{\lambda^2(\alpha_0 - 1)}$ & 0 & $\frac{- \alpha_0(-3 + \alpha_0(1 + 6\alpha_0))}{3(\alpha_0 - 1)}$
& $\frac{-8}{\lambda}$ & 1 & 0 & - & 0 & -
& 0 & 1 & 0 & 2 & $\frac{2}{3} < \alpha_0 < 0.8507$,
\\ 
& & & & & & & & & & & & & & & else Saddle point
\\ 
\hline 
2b & $\frac{\lambda}{2\sqrt{6}}$ & $\frac{-24\alpha_0(\alpha_0 + 1)(2\alpha_0 - 1)}{\lambda^2(\alpha_0 - 1)}$ & $\frac{-64\alpha_0(-3 + \alpha_0(1 + 6\alpha_0))}{\lambda^4(\alpha_0 - 1)}$ & 0
& 0 & - & 0 & - & $\frac{8}{\lambda}$ & 1 & 0 & 1 & 0 & 2 & Saddle point
\\ 
\hline 
2c & $\frac{\lambda}{2\sqrt{6}}$ & 0 & $\frac{32\alpha_0(9 - \alpha_0(5 + 18\alpha_0))}{\lambda^4(\alpha_0- 1)}$ & $\frac{\alpha_0(\alpha_0 + 1)(2\alpha_0 - 1)}{2(\alpha_0 - 1)}$
& $\frac{-8}{\lambda}$ & 1 & 0 & - & $\frac{8}{\lambda}$ & 1 & 0 & 1 & 0 & 2 & Saddle point
\\ 
\hline \hline 
2d & $\frac{\lambda(2 - \alpha_0)}{3\sqrt{6}}$ & $\frac{27\alpha_0(16 - 15\alpha_0 - 42\alpha_0^2 + 25\alpha_0^3)}{\lambda^2 (\alpha_0 - 2)^4(\alpha_0 - 1)}$ & 0 & $\frac{\alpha_0( -8 + \alpha_0(-1 + 25\alpha_0))}{2(\alpha_0 - 2)^2}$
& $\frac{-18}{\lambda(\alpha_0 - 2)^2}$ & 1 & 0 & - & 0 & -& 0 & 0 & 1 & $\frac{-3}{\alpha_0 - 2}$ & Saddle point
\\ 
\hline 
2e & $\frac{\lambda(2 - \alpha_0)}{3\sqrt{6}}$ & $\frac{-27\alpha_0^2(5\alpha_0 - 1)^2}{\lambda^2(\alpha_0 - 2)^4(\alpha_0 - 1)}$ & $\frac{486\alpha_0( -8 + \alpha_0(- 1 + 25\alpha_0))}{\lambda^4(\alpha_0 - 2)^6}$ & 0
& 0 & - & 0 & - & $\frac{18}{\lambda(\alpha_0 - 2)^2}$ & 1 & 0 & 0 & 1 & $\frac{-3}{\alpha_0 - 2}$ & Saddle point
\\ 
\hline 
2f & $\frac{\lambda(2 - \alpha_0)}{3\sqrt{6}}$ & 0 & $\frac{243\alpha_0(16 - 15\alpha_0 - 42\alpha_0^2 + 25\alpha_0^3)}{\lambda^4(\alpha_0 - 2)^6(\alpha_0 - 1)}$ & $\frac{\alpha_0^2(5\alpha_0 - 1)^2}{4(\alpha_0 - 2)^2(\alpha_0 - 1)}$
& $\frac{-18}{\lambda(\alpha_0 - 2)^2}$ & 1 & 0 & - & $\frac{18}{\lambda(\alpha_0 - 2)^2}$  & 1 & 0 & 0 & 1 & $\frac{-3}{\alpha_0 - 2}$ & Saddle point
\\
\hline \hline 
2g & $\frac{\lambda}{3\sqrt{6}}$ & $\frac{-54(3\alpha_0 - 1)^2(\alpha_0 + 1)}{\lambda^2(\alpha_0 - 1)}$ & 0 & 0 
& 0 & - & 0 & - & 0 & - & 1 & 0 & 0 & 3 & Saddle point
\\ 
\hline 
2h & $\frac{\lambda}{9\sqrt{6}\alpha_0}$ & 0 & 0 & $\frac{(3\alpha_0 - 1)^3(3\alpha_0 + 1)(\alpha_0 + 1)}{(\alpha_0 - 1)}$ 
& $\frac{-162\alpha_0^2}{\lambda}$ & 1 & 0 & - & 0 & - & 1 & 0 & 0 & $9\alpha_0$ & Saddle point
\\
\hline 
2i & $\frac{\lambda(2\alpha_0 - 1)}{\sqrt{6}(3\alpha_0 -2 )}$ & 0 & $\frac{-4(3\alpha_0 - 2)^4(\alpha_0 + 1)(3\alpha_0 - 1)^3}{\lambda^4(2\alpha_0 - 1)^6}$ & 0 
& 0 & - & 0 & - & $\frac{2(3\alpha_0 - 2)^2}{\lambda(2\alpha_0 - 1)^2}$ & 1 & 1 & 0 & 0 & $\frac{3\alpha_0 - 2}{2\alpha_0 - 1}$ & Saddle point
\\
\hline 
2j & $\lambda$ & $\frac{3\alpha_0(\alpha_0 + 1)}{\lambda^2(\alpha_0 - 1)}$ & 0 & $\frac{(3\alpha_0 - 1)(\alpha_0 + 1)}{(\alpha_0 - 1)}$
& 0 & - & 0 & - & 0 & - & 1 & 0 & 0 & 0 & Attractor
\\
\hline
2k & $\lambda$ & $\frac{-(\alpha_0 + 1)(3\alpha_0 - 2)}{\lambda^2(\alpha_0 - 1)}$ & $\frac{(3\alpha_0 - 1)(\alpha_0 + 1)}{3\lambda^4(\alpha_0 - 1)}$ & 0 
& 0 & - & 0 & - & 0 & - & 1 & 0 & 0 & 0 & Attractor
\\ 
\hline 
& & & & & & & & & & & & & & & Attractor if
\\ 
2l & $\frac{\lambda \alpha_0}{\sqrt{6}}$ & $-\lambda^2\alpha_0^2\lambda_1$ & $\lambda_1$ & $-\frac{\lambda^4\alpha_0^4\lambda_1}{12}$ & $\frac{-2}{\lambda \alpha_0^2}$ & 1 & 0 & - & $\frac{2}{\lambda \alpha_0^2}$ & 1 & 1 & 0 & 0 & $\frac{1}{\alpha_0}$ & 
$\lambda_1 \neq 0,$
\\
& & & & & & & & & & & & & & & $\alpha_0 < 0, \alpha_0 \neq -1$
\\
\hline 
\hline
\end{tabular}
\caption{Fixed Points (F.P.) and stability analysis for the autonomous system describing kinetically-coupled DE and DM.}
\label{Table:Kinetic-coupling:Fixed points}
\end{sidewaystable}

We will now look at the fixed points of the autonomous system and discuss the stability of the system at these points. Note that the autonomous system is established for a general DM fluid with no assumptions on its energy density and pressure. For the stability analysis, we assume a pressureless DM fluid corresponding to $\omega_{_{DM}} = 0, \omega_{_{DM}}' = 0$. Refer to Appendix \ref{Appendix:Pressureless fluid constraints} for a discussion on how this assumption constrains the evolution of DM in the field description. Table~\eqref{Table:Kinetic-coupling:Fixed points} summarizes the fixed points and the system stability for the kinetic-coupling model. As mentioned earlier, the fixed points are categorized as the following:
\begin{itemize}
\item \emph{Radiation-dominated era:} Points $2a-2c$ represent the radiation-dominated fixed points. Points $2b$ and $2c$ are saddle points, while point $2a$ is a conditional saddle point. 
\item \emph{Matter-dominated era:} Points $2d-2f$ represent the matter-dominated fixed points and are all saddle points. For a non-accelerating matter-dominated era, this corresponds to $\epsilon > 1 \implies 2 > \alpha_0 > -1$ for these fixed points.
\item \emph{DE-dominated era:} Points $2g-2l$ represent the DE-dominated fixed points. Points $2j$ and $2k$ correspond to attractor fixed points with an accelerated expansion of the Universe. Point $2l$ is a conditional attractor fixed point. 
\end{itemize}
Interestingly, the fixed point $2h$, corresponding to a scalar field potential-dominated DE phase, is not an attractor for the kinetically interacting dark sector. This contrasts with the previous models, where the potential-dominated fixed points lead to an attractor de Sitter universe. The accelerated phase of the late Universe leading to an attractor de Sitter universe is demonstrated by fixed points $2j$ and $2k$, indicated by
$\epsilon^{*} = 0$.

Lastly, we can have consistent transitions from radiation to matter to DE-dominated epochs. For instance, we have transitions $2a \rightarrow 2d \rightarrow 2j$, $2b \rightarrow 2e \rightarrow 2k$ and $2c \rightarrow 2f \rightarrow 2l$.

\section{Comparing the background evolution of the two classes of models}
\label{Section:Numerical Evolution}

In the previous section, we showed that the two classes of coupling models can be expressed as an autonomous system of equations. We looked at the fixed point and stability analysis of the two classes of models. We also found that a transition from a matter-dominated to a DE-dominated epoch can arise in these classes of models. 

This leads us to the following questions: How robust is the result? In other words, can the transition occur for a generic form of $\beta(\phi), \gamma(\phi), V(\phi)$ and $\alpha(\phi, X)$? To go about answering this, we consider two classes of models ---- quintessence DE where we set $\beta(\phi) = 1\, ,~\gamma(\phi) = 0\, ,~V(\phi) = {V_0}/{\phi}$~\cite{2021-Johnson.Shankaranarayanan-Phys.Rev.D} and purely kinetic DE model where $\beta(\phi) = 1\, ,\gamma(\phi) = \gamma_0\, ,~V(\phi) = V_0$~\cite{Pedro:2007}. Any other model can be considered a combination of these two cases, as a generic model has both potential and kinetic terms. For these two scenarios, we perform a numerical analysis of the two models and compare the differences in their evolution. Here, we assume a linear field coupling $\alpha(\phi)$ for the field-coupling model:
\begin{eqnarray} \label{Eq:Numerical evolution:Field coupled:alpha form}
    \alpha(\phi) = \frac{c\,\phi}{M_{\rm Pl}}\, ,
\end{eqnarray}
where $c$ is a dimensionless constant representing the strength of the linear field coupling. In the case of kinetic-coupling models, we assume $\alpha(X)$ to be of the form Eq.~\eqref{Eq:Kinetic-coupling:Coupling form assumption}.
For both the models, we shall numerically solve the autonomous system of equations for the redshift range $1500 \leq z \leq 0$. Note that we shall plot the evolution of the cosmological parameters with respect to the number of e-foldings ($N$). 

From the analysis in the previous section, for both the field and kinetic couplings, it is evident that not all sets of initial conditions will result in a consistent transition from radiation-dominated to matter-dominated to DE-dominated epochs and an accelerated expansion of the late Universe. The initial conditions for the two coupling models must be set in a systematic manner such that they satisfy the following criteria:
\begin{itemize}
    \item They must ensure the conservation of the total energy density of the universe at the beginning of the evolution.
    \item The cosmological evolution of the system must be consistent with the present observations, resulting in a consistent transition between different epochs and an accelerated expansion of the late Universe. 
    \item In the limit that the coupling vanishes in the two models (which corresponds to $c = 0$ and $\alpha_0 = 0$, respectively), the resulting evolution in the non-interacting scenario must match the two models.
\end{itemize}
It is important to note that we don't need to fine-tune the initial conditions to obtain cosmological evolution consistent with the observations.  Although the evolution equations have been established for a general DM fluid, we shall assume it to be pressureless for numerical analysis in the rest of this section.

\subsection{Field coupling vs kinetic coupling for a quintessence DE field $\phi$}
\label{Section:Numerics01}

First, we will evolve the two models for a quintessence DE field $\phi$ described by an inverse potential. We assume the following form 
\begin{eqnarray} \label{Eq:Numerical evolution:Quintessence assumption}
\beta(\phi) = 1\, ,~\gamma(\phi) = 0\, ,~V(\phi) = \frac{V_0}{\phi}\, ,
\end{eqnarray}
in Eq.~\eqref{Eq:k-essence assumption}. 
For the field-coupling model, we choose the following initial conditions and parameter values:
\begin{eqnarray}
\label{Eq:Numerical evolution:Quintessence:Field coupled:Initial conditions}
\begin{aligned}
&
x_{i} = 1.5\times10^{-5}\, ,~~
y_{i} = 1\, ,~~
v_{i} = 0\, ,~~
w_{i} = 6.25\times10^{-10}\, ,
\\
&
\Omega_{r_i} = 0.4\, ,~~
\Omega_{_{DM_i}} = 1 - \Omega_{r_i} - x_{i}^2( y_{i} + 9x_{i}^2v_{i} ) - w_{i}\, ,
\\
&
b_{1_i} = -0.6\, ,~~
b_{2} = 2\, ,~~
\beta_{1} = 0\, ,~~
\beta_{2} = 0\, ,~~
\gamma_{1} = 0\, ,~~
\gamma_{2} = 0\, ,
\\
&
\alpha_{1_i} = \frac{c}{0.6}\, ,~~
\alpha_{2_i} = 0.6\, ,~~
\alpha_3 = 0\, .
\end{aligned}
\end{eqnarray}
For the kinetic-coupling model, we assume the following initial conditions and parameter values:
\begin{eqnarray} \label{Eq:Numerical evolution:Quintessence:Kinetic-coupling:Initial conditions}
\begin{aligned}
&
x_{i} = 1.5\times10^{-5}\, ,~~
y_{i} = 1\, ,~~
v_{i} = 0\, ,~~
\Omega_{r_i} = 0.4\, ,~~
\Omega_{_{DM_i}} = 0.6 - 8.5\times10^{-10}\, ,
\\
&
b_{1_i} = -0.6\, ,~~
b_{2} = 2\, ,~~
\beta_{1} = 0\, ,~~
\beta_{2} = 0\, ,~~
\gamma_{1} = 0\, ,~~
\gamma_{2} = 0\, .
\end{aligned}
\end{eqnarray}

\begin{figure}[!htb]
\minipage{0.49\textwidth}
  \includegraphics[width=1\linewidth]{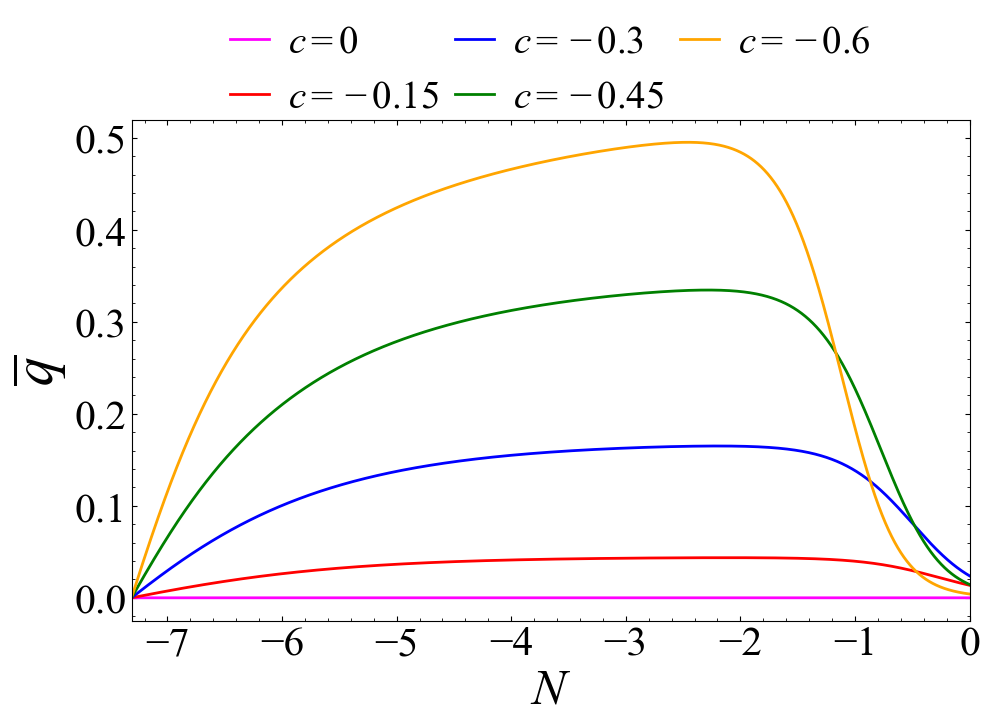}
\endminipage\hfill
\minipage{0.51\textwidth}
  \vspace{-2em}  
  \includegraphics[width=1\linewidth]{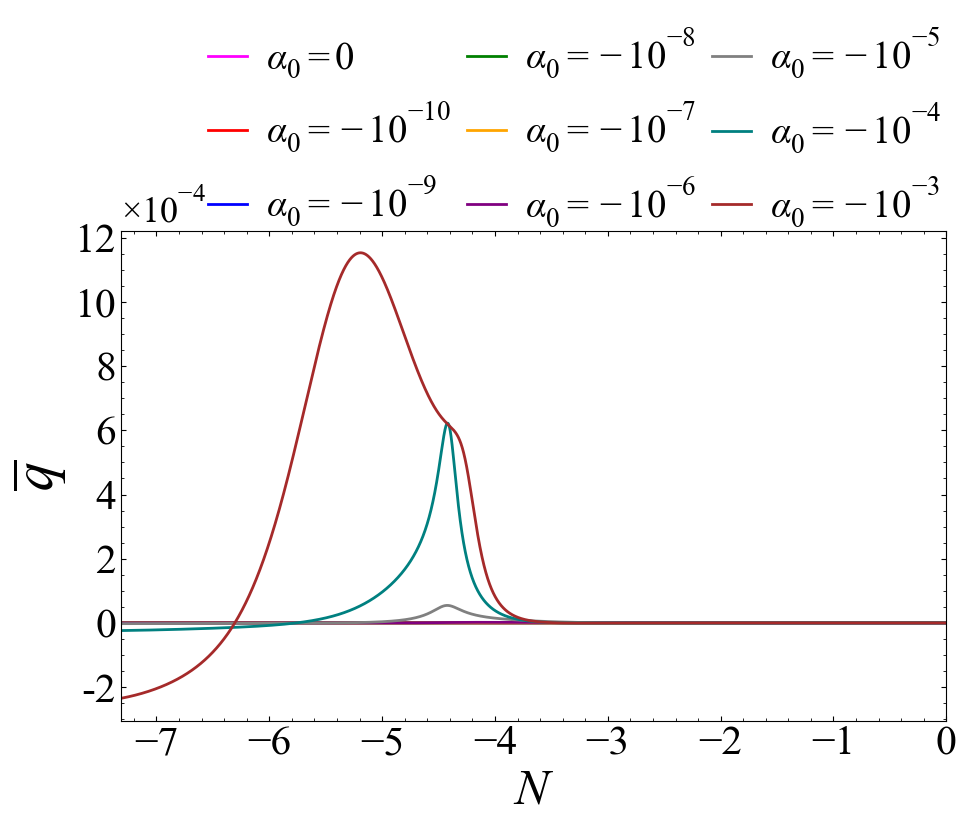}
\endminipage
\caption{Evolution of the scaled interaction term $\overline{q}$ for the quintessence DE field $\phi$ \eqref{Eq:Numerical evolution:Quintessence assumption} as a function of $N$. \emph{Left panel}: Field coupling, \emph{Right panel}: Kinetic coupling.}
\label{Fig:Quintessence:Interaction strength}
\end{figure}

\begin{figure}[!htb]
\minipage{0.49\textwidth}
  \includegraphics[width=1\linewidth]{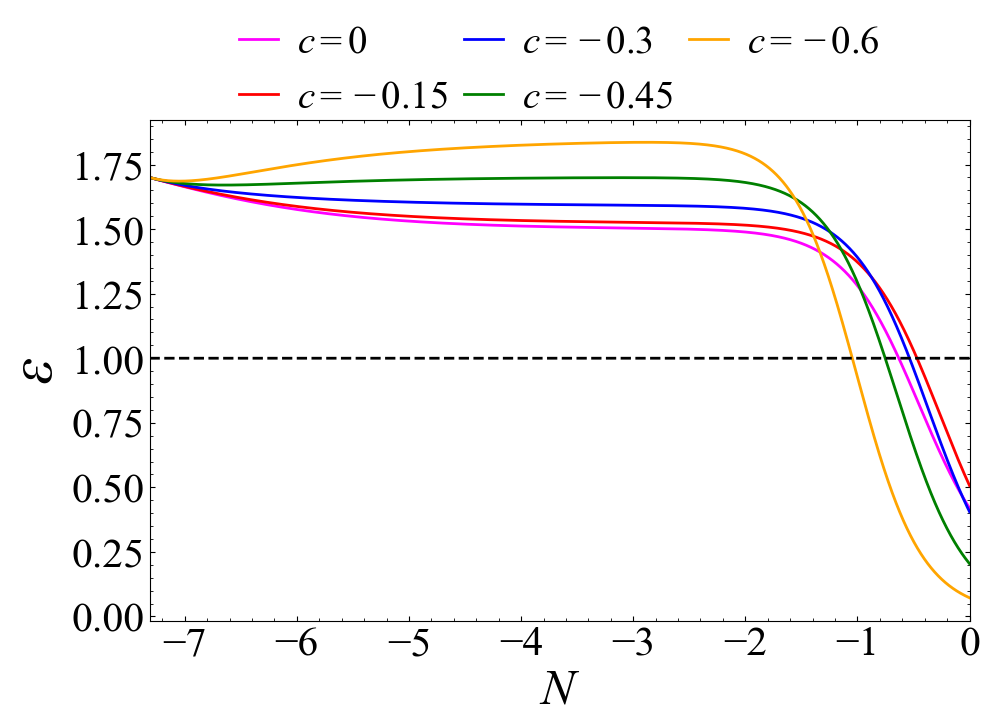}
\endminipage\hfill
\minipage{0.51\textwidth}
  \vspace{-2em}  
  \includegraphics[width=1\linewidth]{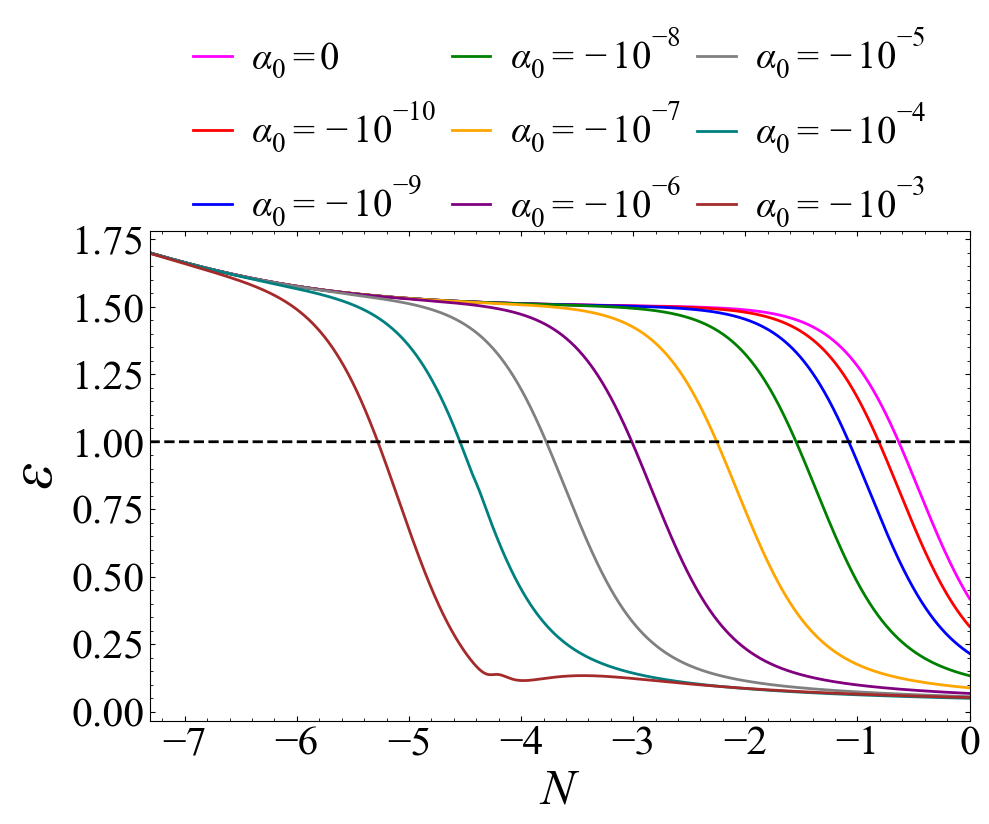}
\endminipage
\caption{Evolution of the slow-roll parameter $\epsilon$ for the quintessence DE field $\phi$ \eqref{Eq:Numerical evolution:Quintessence assumption} as a function of $N$. \emph{Left panel}: Field coupling, \emph{Right panel}: Kinetic coupling.}
\label{Fig:Quintessence:epsilon}
\end{figure}
$w_i$ (initial condition on the scaled potential $w$) is obtained by solving Eq.~\eqref{Eq:Kinetic-coupling:Dimensionless simplified expression}. Since this expression depends on the coupling strength ($\alpha_0$), $w_i$ is different for different kinetic coupling strengths. For the kinetic coupling model, we have set $\Omega_{_{DM_i}}$ (initial condition on the DM energy density parameter) such that it matches with the field-coupling model. Note that this is one set of initial conditions, and a range of initial conditions exist that can lead to an accelerated expansion of the late Universe and remain consistent with the observations. Therefore, these specific initial configurations of the cosmological parameters for the two coupling models are just representative values in performing the numerical evolution of the background.
\begin{figure}[!htb]
\minipage{0.49\textwidth}
  \includegraphics[width=1\linewidth]{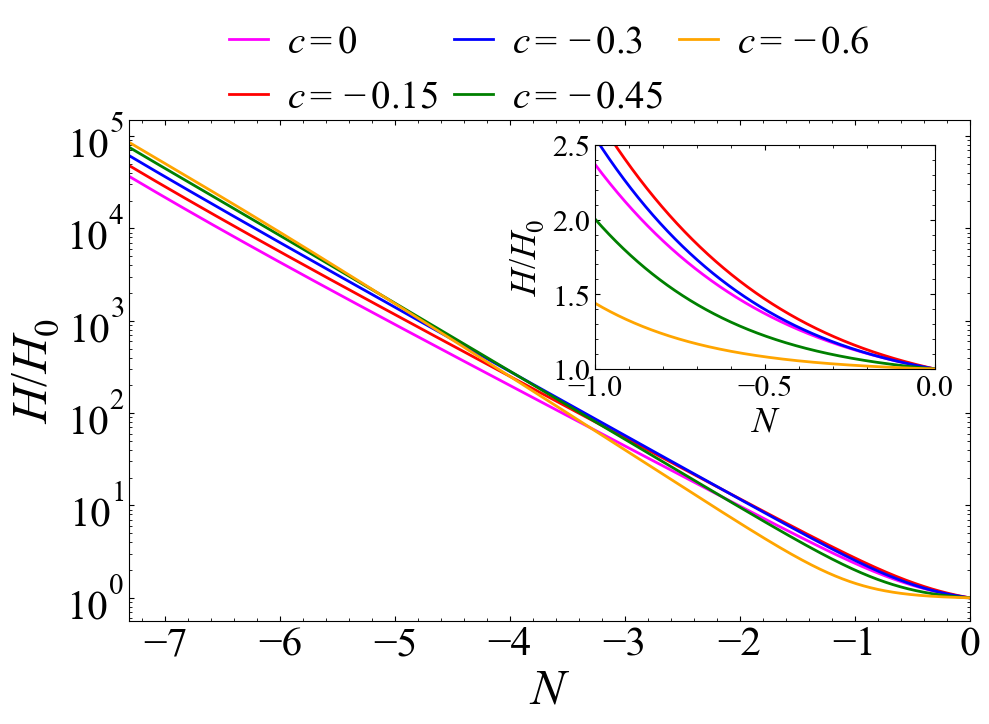}
\endminipage\hfill
\minipage{0.51\textwidth}
  \vspace{-2em}  
  \includegraphics[width=1\linewidth]{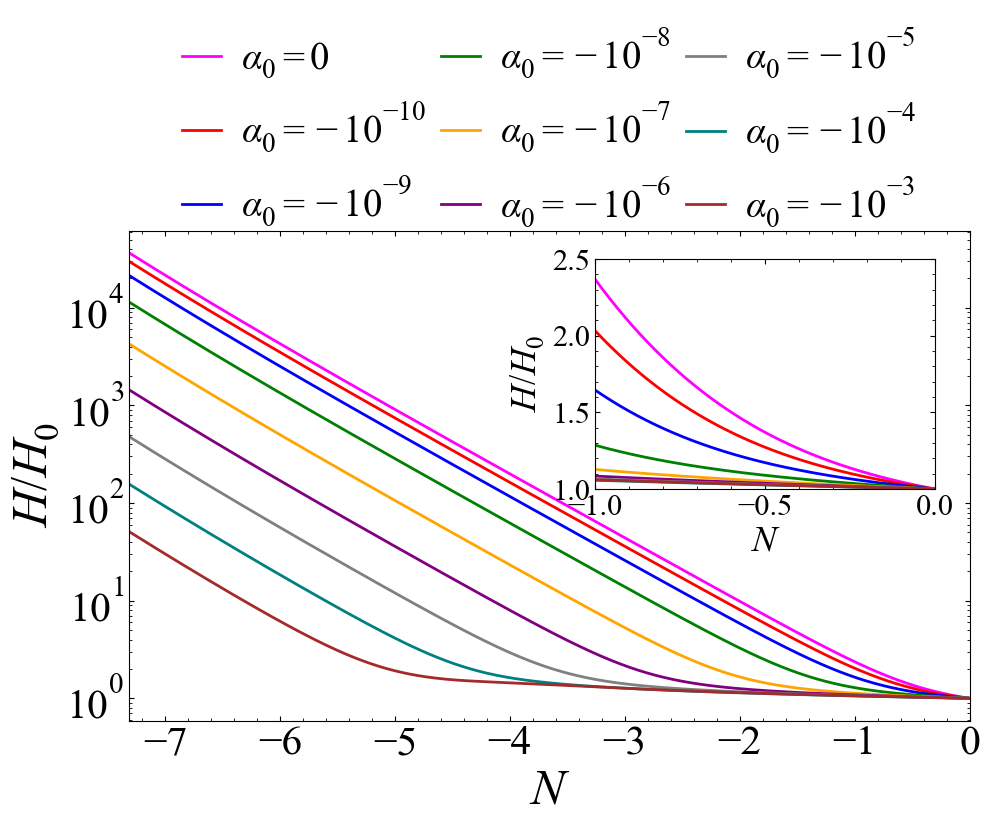}
\endminipage
\caption{Evolution of the scaled Hubble parameter $H/H_0$ for the quintessence DE field $\phi$ \eqref{Eq:Numerical evolution:Quintessence assumption} as a function of $N$. \emph{Left panel}: Field coupling, \emph{Right panel}: Kinetic coupling.}
\label{Fig:Quintessence:H}
\end{figure}
\begin{figure}[!htb]
\minipage{0.49\textwidth}
  \includegraphics[width=1\linewidth]{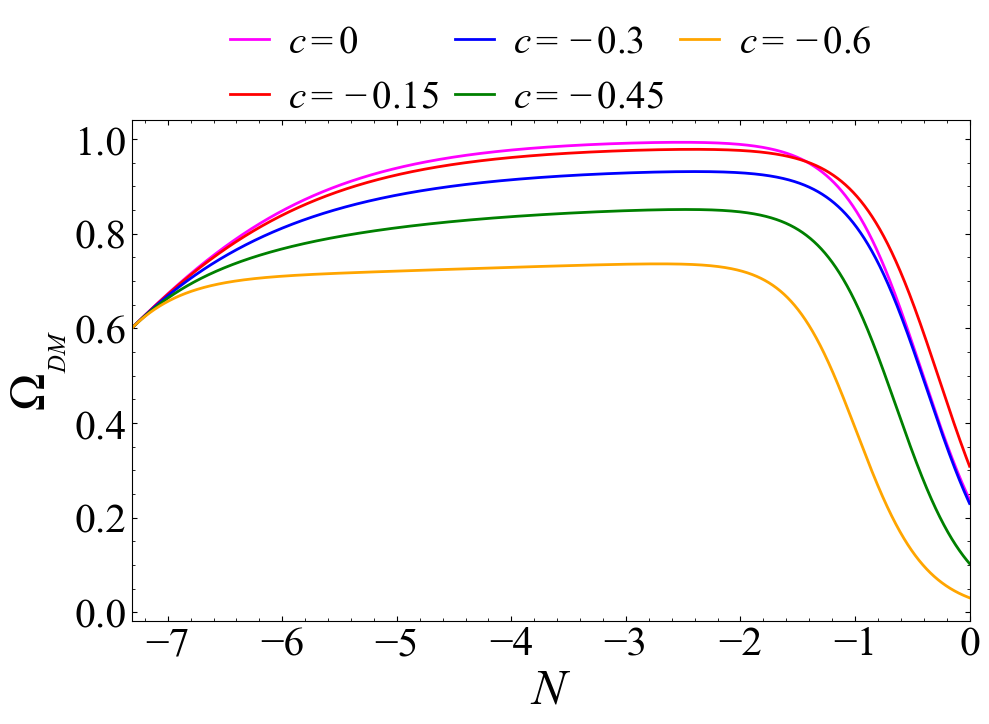}
\endminipage\hfill
\minipage{0.51\textwidth}
  \vspace{-2em}  
  \includegraphics[width=1\linewidth]{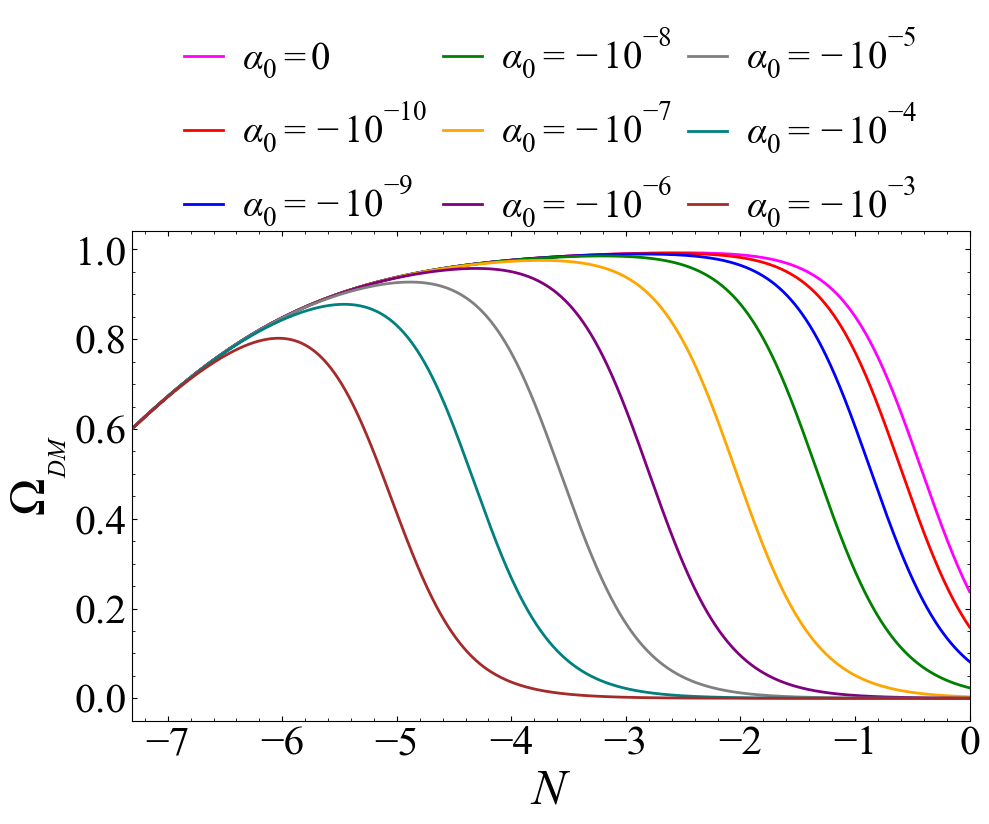}
\endminipage
\caption{Evolution of the matter energy density parameter $\Omega_{_{DM}}$ for the quintessence DE field $\phi$ \eqref{Eq:Numerical evolution:Quintessence assumption} as a function of $N$. \emph{Left panel}: Field coupling, \emph{Right panel}: Kinetic coupling.}
\label{Fig:Quintessence:Omega_DM}
\end{figure}

\ref{Fig:Quintessence:Interaction strength} contains the evolution of the scaled interaction term $\overline{q}$ as a function of $N$ for the two classes of models.  Comparing the two plots, we note that the interactions decay faster in the case of kinetic coupling than in the case of field coupling. Additionally, both, the field and kinetic-coupling interactions, are positive throughout the evolution. \ref{Fig:Quintessence:epsilon} shows the evolution of the slow-roll parameter $\epsilon$ for the two classes of models. We infer that the interactions cause an earlier transition from the matter-dominated to DE-dominated era, hence, an earlier accelerated expansion phase in the late Universe in both models. Notably, while the deviations of the field coupling from the non-interacting scenario are relatively small, the kinetic coupling leads to a significantly earlier onset of the accelerated expansion phase in the late Universe compared to the non-interacting scenario.

\begin{figure}[!htb]
\minipage{0.49\textwidth}
  \includegraphics[width=1\linewidth]{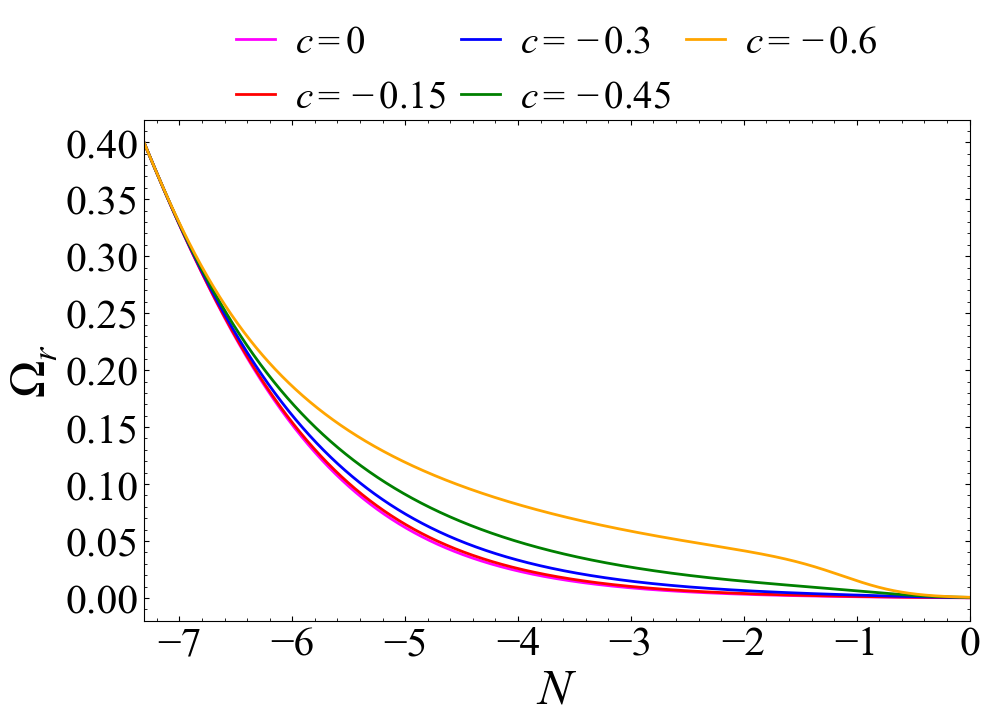}
\endminipage\hfill
\minipage{0.51\textwidth}
  \vspace{-2em}  
  \includegraphics[width=1\linewidth]{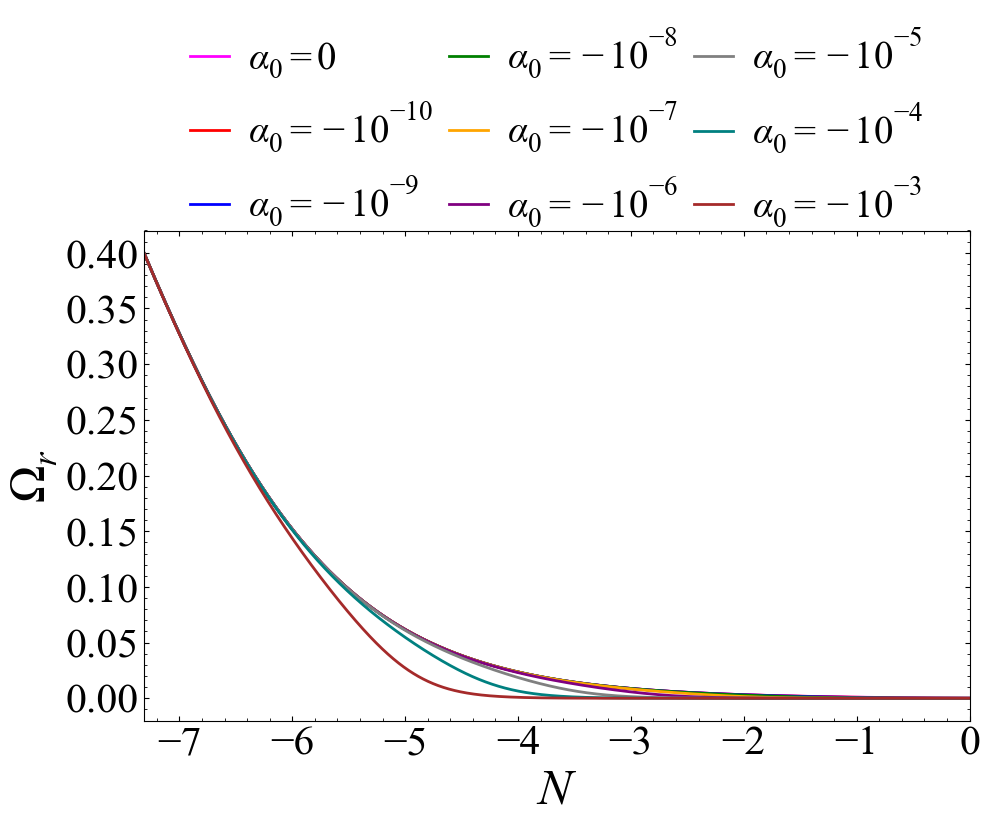}
\endminipage
\caption{Evolution of the radiation energy density parameter $\Omega_r$ for the quintessence DE field $\phi$ \eqref{Eq:Numerical evolution:Quintessence assumption} as a function of $N$. \emph{Left panel}: Field coupling, \emph{Right panel}: Kinetic coupling.}
\label{Fig:Quintessence:Omega_r}
\end{figure}

To investigate this further, we have plotted the evolution of the scaled Hubble parameter $H/H_0$, and the energy densities $\Omega_{_{DM}}$ and $\Omega_r$ in \ref{Fig:Quintessence:H}, \ref{Fig:Quintessence:Omega_DM}, \ref{Fig:Quintessence:Omega_r}, respectively. As the kinetic coupling strength $\alpha_0$ increases, the transition from matter-dominated to the DE-dominated era occurs significantly earlier. Furthermore, the scaled Hubble parameter's value is markedly lower at early times in the kinetic-coupling scenario than in the field-coupling and non-interacting scenarios. Our numerical analysis shows that despite the kinetic-coupling interactions lasting for a shorter duration than the field coupling, they significantly influence the background evolution.
This high sensitivity of cosmological evolution to the kinetic coupling strength suggests a promising avenue for establishing stringent constraints on this parameter from cosmological observations.

\subsection{Field coupling vs kinetic coupling for a k-essence DE field $\phi$}
\label{Section:Numerics02}

For our choice of the coupling function \eqref{Eq:Kinetic-coupling:Coupling form assumption}, it is evident that the DHOST terms in Eq.~\eqref{Eq:Field-kinetic coupled:DHOST terms} are purely a function of the kinetic term $X$. This gives us an opportunity to study purely kinetically driven scalar fields and investigate the impact of direct interactions on the dynamics of such a scalar field. Therefore, we shall evolve the two coupling models for a purely kinetically driven k-essence scalar field $\phi$ with a constant potential. For this, we assume the following form \cite{Pedro:2007}:
\begin{eqnarray} \label{Eq:Numerical evolution:k-essence assumption}
    \beta(\phi) = 1\, ,~~ \gamma(\phi) = \gamma_0\, ,~~ V(\phi) = V_0\, .
\end{eqnarray}
in Eq.~\eqref{Eq:k-essence assumption}. Once again, we note that the initial configuration of the cosmological parameters must satisfy the three criteria mentioned earlier. Taking them into account, we choose the following initial conditions and parameter values for the field-coupling model:

\begin{eqnarray}
\label{Eq:Numerical evolution:k-essence:Field coupled:Initial conditions}
\begin{aligned}
&
x_{i} = 1.5\times10^{-5}\, ,~~
y_{i} = 1\, ,~~
v_{i} = 1.296\times10^9\, ,~~
w_{i} = 6.25\times10^{-10}\, ,
\\
&
\Omega_{r_i} = 0.4\, ,~~
\Omega_{_{DM_i}} = 1 - \Omega_{r_i} - x_{i}^2( y_{i} + 9x_{i}^2v_{i} ) - w_{i}\, ,
\\
&
b_{1} = 0\, ,~~
b_{2} = 0\, ,~~
\beta_{1} = 0\, ,~~
\beta_{2} = 0\, ,~~
\gamma_{1} = 0\, ,~~
\gamma_{2} = 0\, ,
\\
&
\alpha_{1_i} = \frac{c}{0.6}\, ,~~
\alpha_{2_i} = 0.6\, ,~~
\alpha_3 = 0\, .
\end{aligned}
\end{eqnarray}
For the kinetic-coupling model, we assume the following initial conditions and parameter values:
\begin{eqnarray} \label{Eq:Numerical evolution:k-essence:Kinetic-coupling:Initial conditions}
& &
x_{i} = 1.5\times10^{-5}\, ,~~
y_{i} = 1\, ,~~
w_{i} = 6.25\times10^{-10}\, ,~~
\Omega_{r_i} = 0.4\, ,~~
\Omega_{_{DM_i}} = 0.6 - 1.44\times10^{-9}\, ,
\nonumber \\
& &
b_{1} = 0\, ,~~
b_{2} = 0\, ,~~
\beta_{1} = 0\, ,~~
\beta_{2} = 0\, ,~~
\gamma_{1} = 0\, ,~~
\gamma_{2} = 0\, .
\end{eqnarray}
\begin{figure}[!htb]
\minipage{0.49\textwidth}
  \includegraphics[width=1\linewidth]{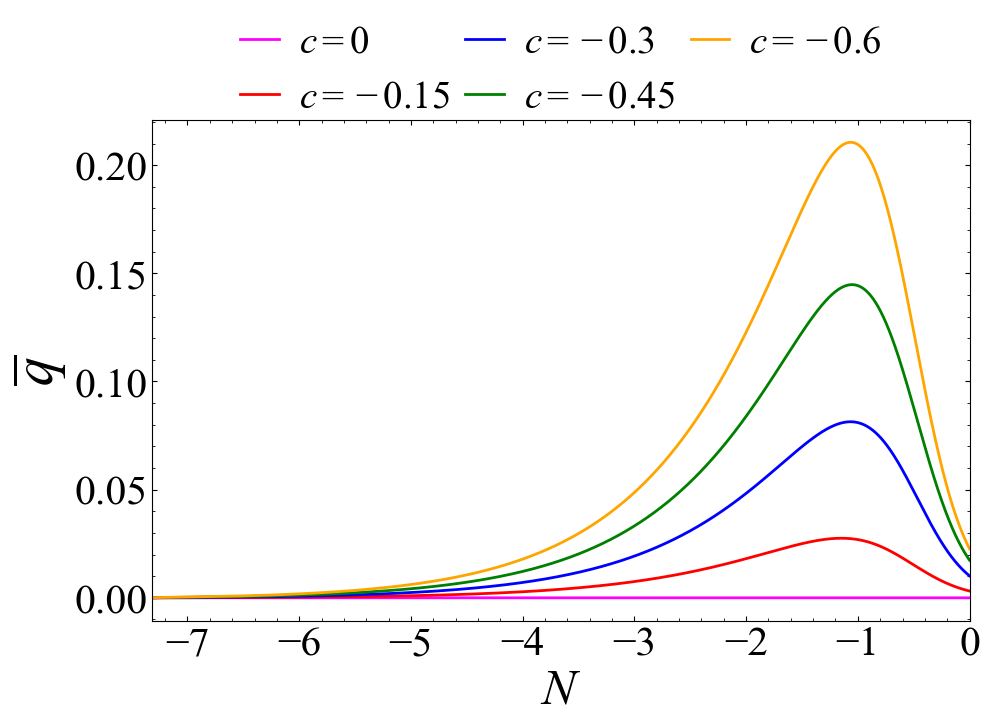}
\endminipage\hfill
\minipage{0.51\textwidth}
  \includegraphics[width=1\linewidth]{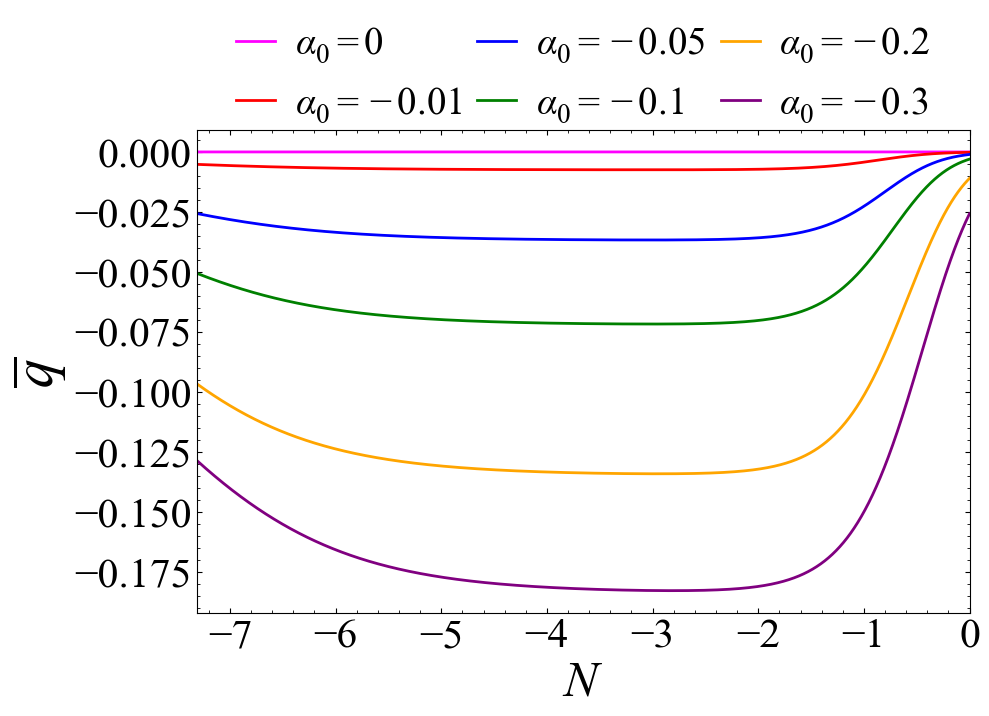}
\endminipage
\caption{Evolution of the scaled interaction term $\overline{q}$ for the k-essence DE field $\phi$ \eqref{Eq:Numerical evolution:k-essence assumption} as a function of $N$. \emph{Left panel}: Field coupling, \emph{Right panel}: Kinetic coupling.}
\label{Fig:k-essence:Interaction strength}
\end{figure}
\begin{figure}[!htb]
\minipage{0.49\textwidth}
  \includegraphics[width=1\linewidth]{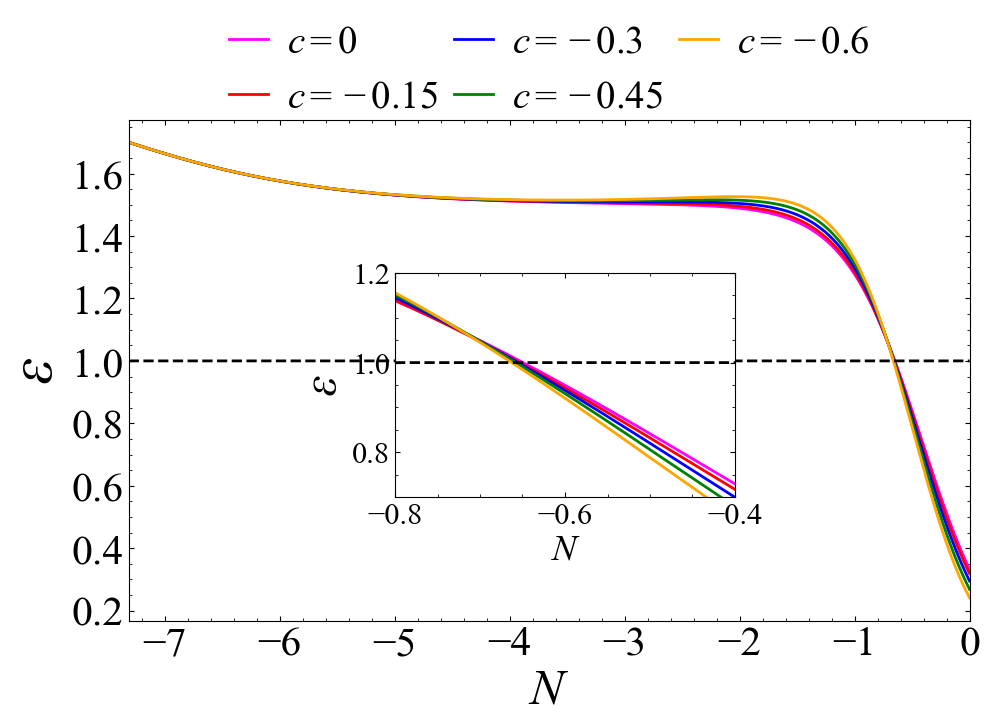}
\endminipage\hfill
\minipage{0.51\textwidth}
  \includegraphics[width=1\linewidth]{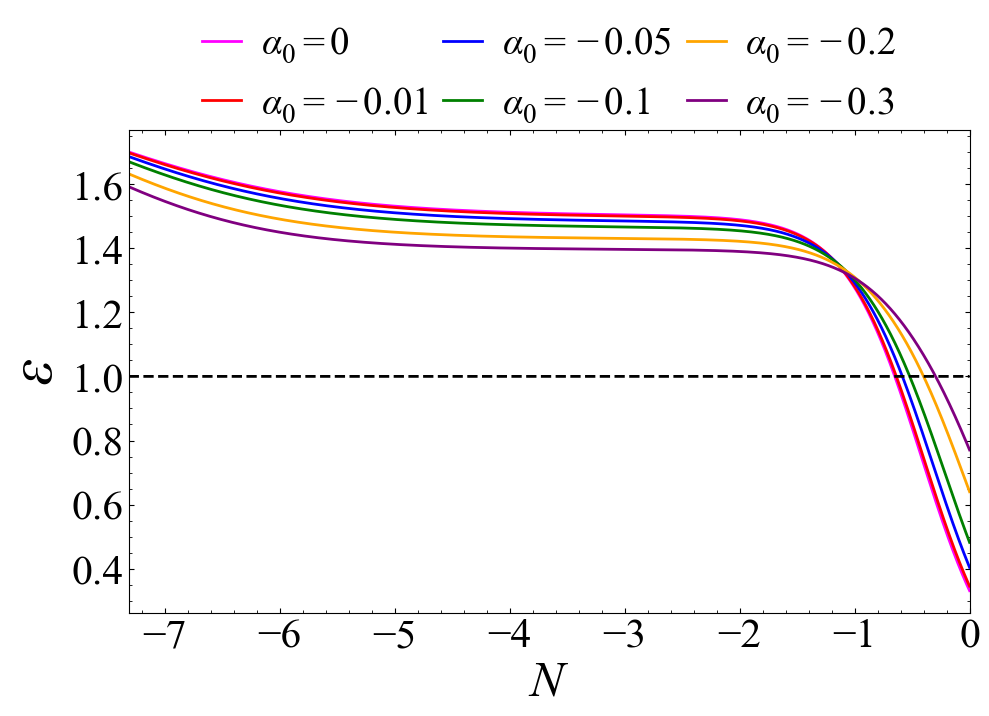}
\endminipage
\caption{Evolution of the slow-roll parameter $\epsilon$ for the k-essence DE field $\phi$ \eqref{Eq:Numerical evolution:k-essence assumption} as a function of $N$. \emph{Left panel}: Field coupling, \emph{Right panel}: Kinetic coupling.}
\label{Fig:k-essence:epsilon}
\end{figure}
We set $v_i$ by numerically solving Eq.~\eqref{Eq:Kinetic-coupling:Dimensionless simplified expression} for different kinetic coupling strengths $\alpha_0$. We note that since the variable $v$ is non-zero for the k-essence model, we have more choices to set different initial conditions for the numerical evolution.

\begin{figure}[!htb]
\minipage{0.49\textwidth}
  \includegraphics[width=1\linewidth]{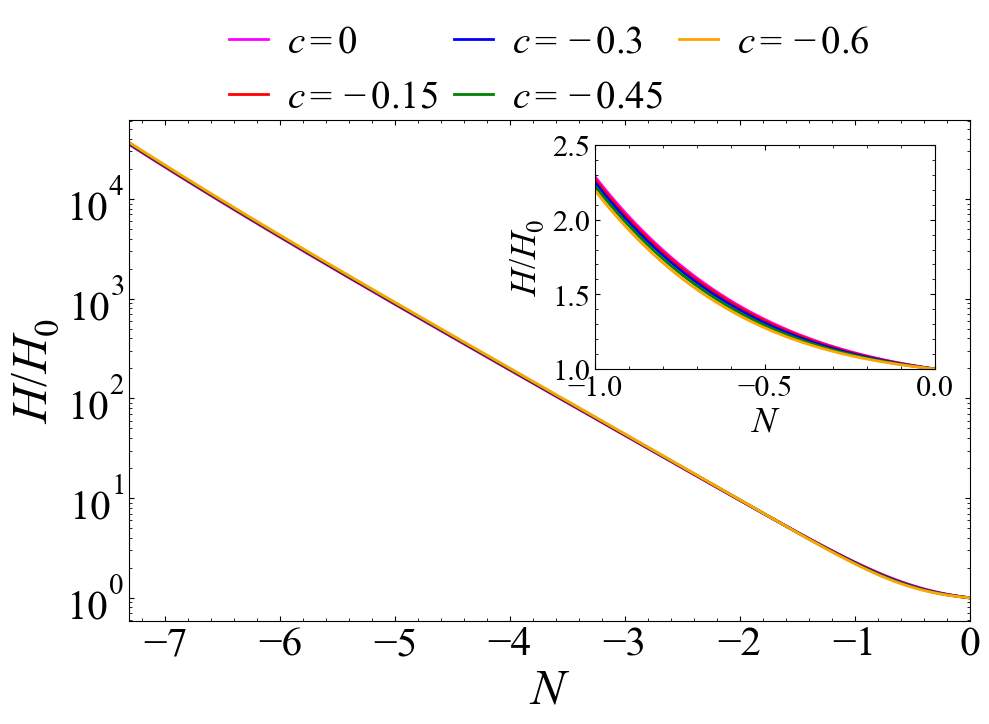}
\endminipage\hfill
\minipage{0.51\textwidth}
  \includegraphics[width=1\linewidth]{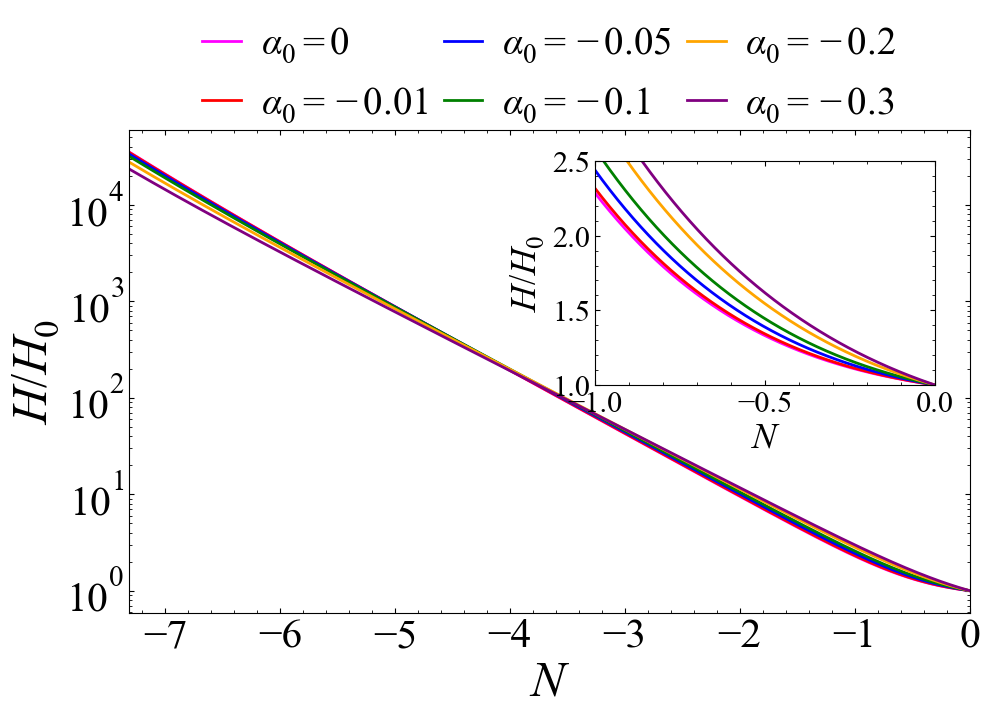}
\endminipage
\caption{Evolution of the scaled Hubble parameter $H/H_0$ for the k-essence DE field $\phi$ \eqref{Eq:Numerical evolution:k-essence assumption} as a function of $N$. \emph{Left panel}: Field coupling, \emph{Right panel}: Kinetic coupling.}
\label{Fig:k-essence:Hubble parameter}
\end{figure}

\begin{figure}[!htb]
\minipage{0.49\textwidth}
  \includegraphics[width=1\linewidth]{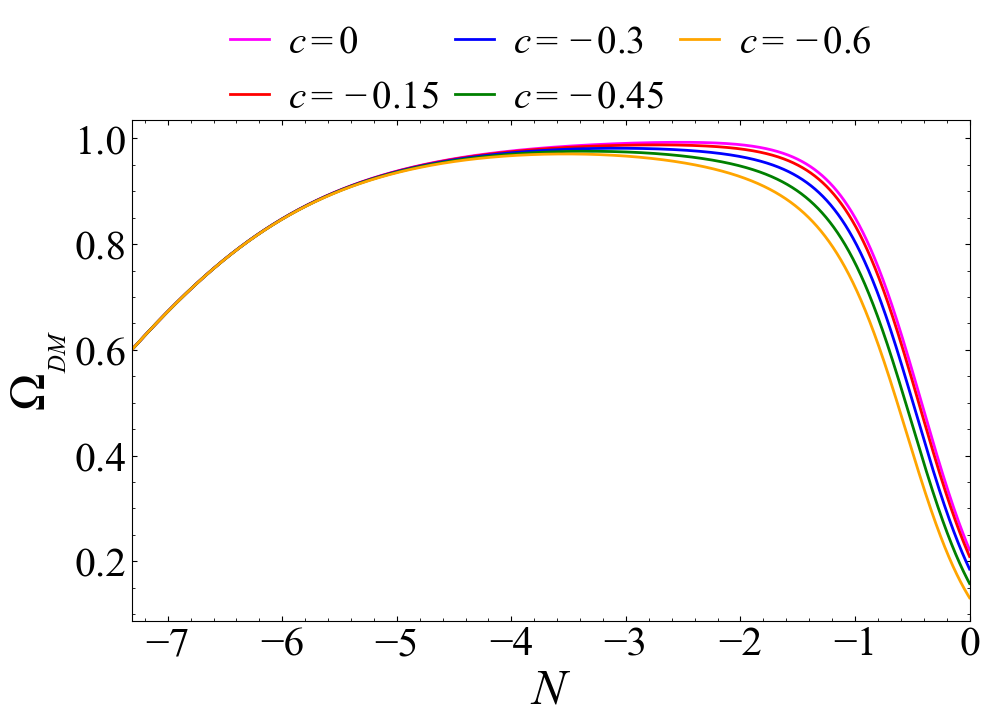}
\endminipage\hfill
\minipage{0.51\textwidth}
  \includegraphics[width=1\linewidth]{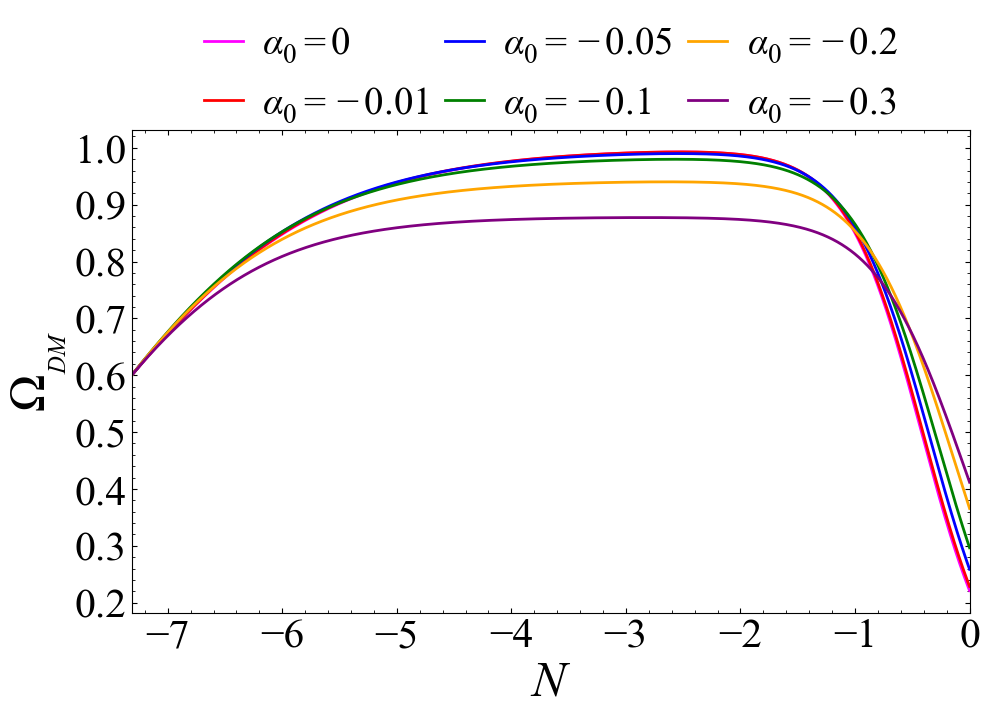}
\endminipage
\caption{Evolution of the matter energy density parameter $\Omega_{_{DM}}$ for the k-essence DE field $\phi$ \eqref{Eq:Numerical evolution:k-essence assumption} as a function of $N$. \emph{Left panel}: Field coupling, \emph{Right panel}: Kinetic coupling.}
\label{Fig:k-essence:Omega_DM}
\end{figure}

\begin{figure}[!htb]
\minipage{0.49\textwidth}
  \includegraphics[width=1\linewidth]{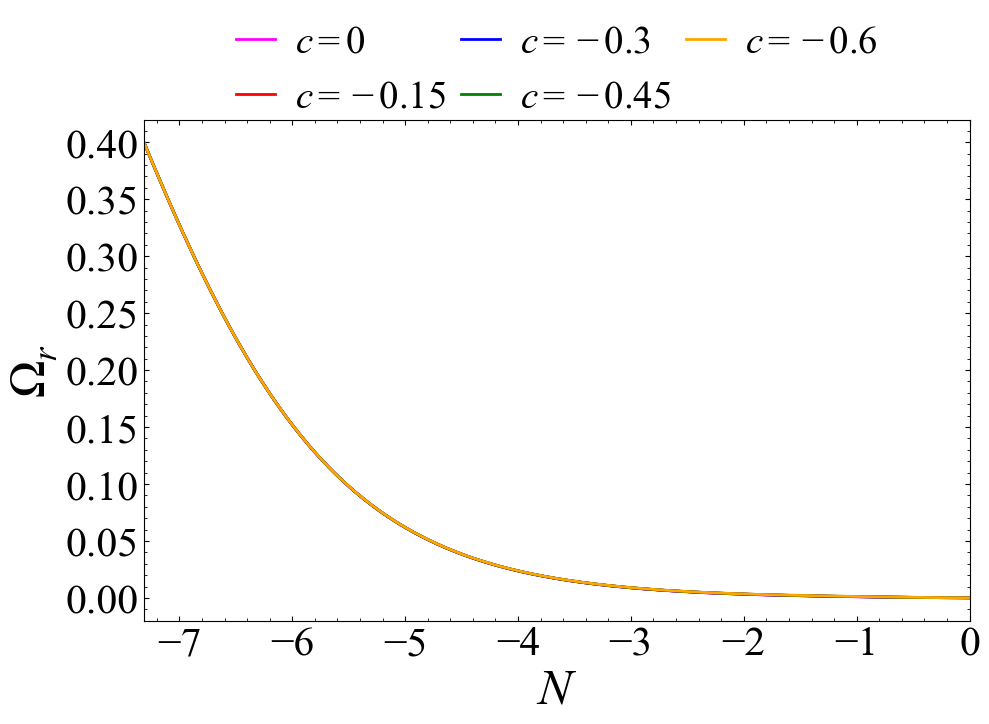}
\endminipage\hfill
\minipage{0.51\textwidth}
  \includegraphics[width=1\linewidth]{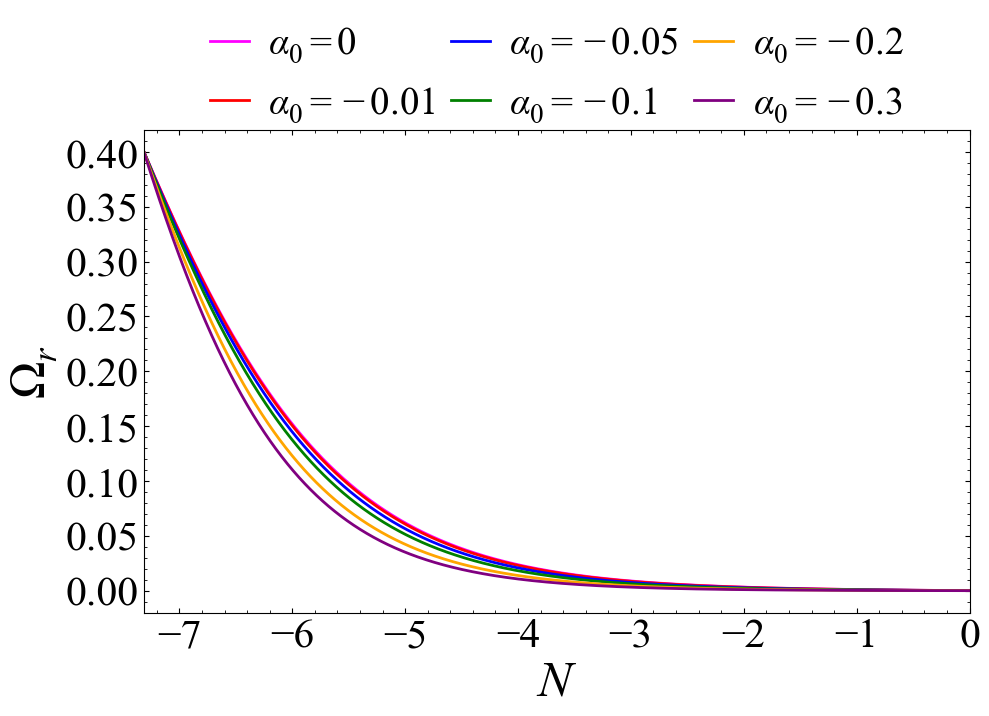}
\endminipage
\caption{Evolution of the radiation energy density parameter $\Omega_r$ for the k-essence DE field $\phi$ \eqref{Eq:Numerical evolution:k-essence assumption} as a function of $N$. \emph{Left panel}: Field coupling, \emph{Right panel}: Kinetic coupling.}
\label{Fig:k-essence:Omega_r}
\end{figure}

First, we shall compare the background evolution of the field-coupling quintessence and k-essence models. \ref{Fig:k-essence:Interaction strength}: left panel plot contains the evolution of the scaled interaction strength $\overline{q}$ (defined in Eq.~\eqref{Eq:Field coupled:Scaled Interaction term}) for the field-coupling k-essence model. We find that for the field-coupling model, $\overline{q}$ grows at a later time compared to the quintessence model. However, it remains positive throughout the evolution for both the DE field descriptions. \ref{Fig:k-essence:epsilon}, \ref{Fig:k-essence:Hubble parameter}, \ref{Fig:k-essence:Omega_DM}, and \ref{Fig:k-essence:Omega_r}: left panel plots contain the evolution of the cosmological parameters $\epsilon$, $H/H_0$, $\Omega_{_{DM}}$, and $\Omega_r$, respectively.
We find that the evolution of these parameters in the k-essence model has relatively smaller deviations from the non-interacting case than the quintessence DE model. However, the evolution of these parameters in the field-coupling models is similar for the two scalar field descriptions.
 
We now take a look at the kinetic-coupling k-essence model. The scaled interaction strength $\overline{q}$ (defined in Eq.~\eqref{Eq:Kinetic-coupling:Scaled Interaction strength}) plotted in \ref{Fig:k-essence:Interaction strength}: right panel plot, shows that the interactions remain negative throughout the evolution. This leads to certain deviations from the other models. First, the negative interactions, leading to an energy transfer from DE to DM, suggest that the DE-dominated phase starts later than the non-interacting and other scenarios.
Second, it also suggests a relatively later onset of the accelerated expansion phase in the late Universe. This is evident from the evolution of the slow-roll parameter $\epsilon$, and the matter energy density parameter $\Omega_{_{DM}}$ plotted in \ref{Fig:k-essence:epsilon} and \ref{Fig:k-essence:Omega_DM}, respectively. These results are in contrast with the kinetic-coupling quintessence field model, which predicts a significantly early onset of the accelerated expansion phase in the late Universe. The evolution of the scaled Hubble parameter $H/H_0$ is plotted in \ref{Fig:k-essence:Hubble parameter}. The right panel plot clearly shows that the scaled Hubble parameter, at early times, is larger than the non-interacting and coupled quintessence models, respectively.
 
It is important to note that, for a set of initial conditions different from Eq.~\eqref{Eq:Numerical evolution:k-essence:Kinetic-coupling:Initial conditions}, the kinetic-coupling k-essence model also predicts a significantly early transition from matter to DE-dominated era and an early accelerated expansion phase in the late Universe. The cosmological evolution of this model for the two sets of initial conditions has been compared in Appendix \ref{Appendix:Numerical evolution:Kinetic-coupling:Different ICs}.

\section{Linearized field equations in scalar sector}
\label{Section:Scalar perturbations}

Following our discussion of the background cosmology of the interacting dark sector models in the previous sections, we now study the cosmological evolution of these models in linear-order scalar perturbations. For this analysis, we keep the form of the coupling function $\alpha(\phi, X)$ arbitrary. To keep things transparent, we consider the spatially flat FLRW metric with first-order scalar perturbations in Newtonian gauge~\cite{1992-Mukhanov.etal-PRep}:
\begin{equation} \label{Eq:Scalar perturbations:Perturbed metric}
    \mathrm{d}s^{2} = -(1 + 2\Phi)\mathrm{d}t^{2} + a^{2}(t)(1 - 2\Psi)\mathrm{d}\textbf{x}^{2}\, .
\end{equation} 
where $\Phi \equiv \Phi(t,x,y,z)$ and 
$\Psi \equiv \Psi(t,x,y,z)$ are the Bardeen potentials. The scalar, vector, and tensor perturbations decouple in the linear order. Since the scalar perturbations couple to the matter-energy density ($\delta\rho$) and pressure ($\delta p$) leading to the growing inhomogeneities, we only consider scalar perturbations~\cite{1992-Mukhanov.etal-PRep}. The perturbed equation of motion for the DM fluid is given by:
\begin{equation} \label{Eq:Scalar perturbations:Perturbed fluid EOM}
    \Dot{\delta\rho}_{_{DM}} + 3\overline{H}(\delta\rho_{_{DM}} + \delta p_{_{DM}}) + (\overline{\rho}_{_{DM}} + \overline{p}_{_{DM}})\left( \frac{\nabla^{2}\delta u^{s}}{a^{2}} - 3\Dot{\Psi} \right) = -\delta \overline{Q}\, ,
\end{equation}
where $\delta \overline{Q}$ denotes the zeroth component of the perturbed interaction strength $\delta Q_{\nu}$ and $\delta u^{s} = -\delta\chi/\Dot{\overline{\chi}}$\,. In an FLRW universe, the perturbed interaction term in the fluid picture takes the following form:
\begin{equation} \label{Eq:Scalar perturbations:Perturbed interaction strength}
\begin{aligned}
\delta \overline{Q} = \
&
\Dot{\overline{\phi}}[\overline{\alpha}_{\phi} + \overline{\alpha}_{X}\Ddot{\overline{\phi}}][3\delta p_{_{DM}} - \delta \rho_{_{DM}}] + \left[\overline{\alpha}_{\phi}\Dot{\delta \phi} + \overline{\alpha}_{X}(\Dot{\overline{\phi}}\Ddot{\delta \phi} + \Ddot{\overline{\phi}}\Dot{\delta \phi} - 2\Phi\Ddot{\overline{\phi}}\Dot{\overline{\phi}} - \Dot{\Phi}\Dot{\overline{\phi}}^{2}) + \overline{\alpha}_{\phi\phi}\Dot{\overline{\phi}}\delta\phi
\right. \\
& \left.
+ \overline{\alpha}_{\phi X}(\Dot{\overline{\phi}}\Ddot{\overline{\phi}}\delta\phi + \Dot{\overline{\phi}}^{2}\Dot{\delta\phi} - \Phi\Dot{\overline{\phi}}^{3}) + \overline{\alpha}_{XX}\Ddot{\overline{\phi}}\Dot{\overline{\phi}}(\Dot{\overline{\phi}}\Dot{\delta\phi} - \Phi\Dot{\overline{\phi}}^{2})
\right][3\overline{p}_{_{DM}} - \overline{\rho}_{_{DM}}] \, .
\end{aligned}
\end{equation}

The first-order perturbed Einstein equations in the Newtonian gauge are given by:
\begin{subequations} \label{Eq:Scalar perturbations:Perturbed Einstein equations}
\begin{align}
&\Phi - \Psi = \frac{2}{[ 1 - \overline{\alpha}_{X}\Dot{\overline{\phi}}^2 ]^2}
\left[ \left( \overline{\alpha}_{X}\Ddot{\overline{\phi}}(1 + \overline{\alpha}_{X}\Dot{\overline{\phi}}^2) + \Dot{\overline{\phi}}^2(\overline{\alpha}_{\phi X} + \overline{\alpha}_{XX}\Ddot{\overline{\phi}}) \right)\delta\phi \right. \nonumber
\label{SubEq:Perturbations:ij} \\
&  \qquad \left. + \overline{\alpha}_{X}\Dot{\overline{\phi}}(1 - \overline{\alpha}_{X}\Dot{\overline{\phi}}^2)(\Dot{\delta\phi} + H\delta\phi) \right]\, , \\
& \Dot{\Psi} + H\Phi = \frac{1}{2M_{\mathrm{Pl}}^2}\left[ -\delta\sigma^{(\phi)} + (3\overline{p}_{_{DM}} - \overline{\rho}_{_{DM}})\overline{\alpha}_{X}\Dot{\overline{\phi}}\delta\phi - (\overline{p}_{_{DM}} + \overline{\rho}_{_{DM}})\delta u^{s} \right]\, , \\[8pt]
&
3H\Dot{\Psi} - \frac{\nabla^{2}\Psi}{a^{2}} + 3H^{2}\Phi = -\frac{1}{2M_{\mathrm{Pl}}^2} \left[\delta\rho_{_{DM}} - \delta\tensor{T}{^{(II)}_{0}^{0}} 
\right. \nonumber \\
& \left.
+ \overline{\alpha}_{X}\Dot{\overline{\phi}}^{2}(3\delta p_{_{DM}} - \delta\rho_{_{DM}}) + (3\overline{p}_{_{DM}} - \overline{\rho}_{_{DM}})(\overline{\alpha}_{\phi X}\Dot{\overline{\phi}}^{2}\delta\phi + [2\overline{\alpha}_{X} + \overline{\alpha}_{XX}\Dot{\overline{\phi}}^{2}][\Dot{\overline{\phi}}\Dot{\delta\phi} - \Phi\Dot{\overline{\phi}}^{2}]) \right]\, ,
\\
& 3\Ddot{\Psi} + \frac{\nabla^{2}\Phi}{a^{2}} + 6\Phi(\Dot{H} + H^{2}) + 3H(2\Dot{\Psi} + \Dot{\Phi}) =
\frac{1}{2M_{\mathrm{Pl}}^2}\left[\delta\rho_{_{DM}} + 3\delta p_{_{DM}} + \delta \tensor{T}{^{(II)}_{i}^{i}} - \delta\tensor{T}{^{(II)}_{0}^{0}} \right.  \\
& \left. 
+ \overline{\alpha}_{X}\Dot{\overline{\phi}}^{2}(3\delta p_{_{DM}} - \delta\rho_{_{DM}}) + (3\overline{p}_{_{DM}} - \overline{\rho}_{_{DM}})(\overline{\alpha}_{\phi X}\Dot{\overline{\phi}}^{2}\delta\phi + [2\overline{\alpha}_{X} + \overline{\alpha}_{XX}\Dot{\overline{\phi}}^{2}][\Dot{\overline{\phi}}\Dot{\delta\phi} - \Phi\Dot{\overline{\phi}}^{2}]) \right]\, \nonumber ,
\end{align}
\end{subequations}
where $\nabla_{i}\delta\sigma^{(\phi)} = \delta\tensor{T}{^{(II)}_{i}^{0}}$. Note that $\tensor{T}{^{(II)}_{\nu}^{\mu}} \equiv \tensor{T}{^{(II,\phi)}_{\nu}^{\mu}}$ comprises the Horndeski terms $G_2, G_3$ and the DHOST terms $A_1-A_5$\,.
Eq.~\eqref{SubEq:Perturbations:ij} implies that the metric perturbations $\Phi$ and $\Psi$ are not identical for the kinetic-coupling dark sector model. $\Phi$ and $\Psi$ are related to the pressure and density
perturbations of a generic perfect fluid {\it via} the other perturbed
Einstein's equations \eqref{Eq:Scalar perturbations:Perturbed Einstein equations}. The pressure perturbations, in general, can be
split into adiabatic and entropic (non-adiabatic) parts~\cite{1992-Mukhanov.etal-PRep}:
\begin{equation}
\delta p = c_{{\rm s}}^2 \delta\rho + \dot{\ov{p}} \Gamma \, ,
\label{eq:dPdR}
\end{equation}
where $c_s^2 \equiv \dot{\ov p}/\dot{\ov \rho}$ is the adiabatic
sound speed~\cite{1992-Mukhanov.etal-PRep}. The
non-adiabatic part is $\delta p_{\rm nad}\equiv {\dot{\ov
p}}\Gamma$, and
\begin{equation}
\label{defGamma}
\Gamma \equiv \frac{\delta p}{\dot{\ov p}} - 
\frac{\delta\rho}{\dot{\ov \rho}} \,.
\end{equation}
The entropic perturbation $\Gamma$, defined in this way, is
gauge-invariant, and represents the displacement between
hyper-surfaces of uniform pressure and uniform density. 

\subsection{Energy Transfer in DE-DM interactions}
\label{Subsection:Energy transfer}

To go about understanding the effect of interaction, we take a dot product of Eq.~\eqref{Eq:Interaction strength definition} with the four-velocity $u^{\nu}$ which gives us the energy-conservation equation. Depending on the sign of the interaction strength, the cosmic history and observables such as the Hubble parameter and individual energy densities in the past will be distinct~\cite{2021-Johnson.Shankaranarayanan-Phys.Rev.D,2022-Johnson.etal-JCAP}. 
For instance, as we saw in Sec.~\eqref{Section:Numerics01}, if we have a positive interaction corresponding to $\overline{Q} > 0$, this leads to an energy transfer from DM to DE. This is true for both field-coupling and kinetic-coupling models.  As can be seen from \ref{Fig:Quintessence:Omega_DM}, this transfer of energy leads to a quicker transition from matter-dominated to DE-dominated phase. Therefore we reach the following conclusion: Going back in time, starting with the same values of the cosmological parameters today, the relatively faster transition suggests that the matter-radiation equality epoch occurs at the smaller redshift as compared to the non-interacting case. 

Similarly, a negative interaction form ($\overline{Q} < 0$) leads to an energy transfer from DE to DM, resulting in a relatively slower transition from a matter-dominated universe to a DE-dominated one. This is evident from the numerical analysis of the kinetic-coupling model in Sec.~\eqref {Section:Numerics02}. From \ref{Fig:k-essence:Omega_DM} and \ref{Fig:k-essence:epsilon} (right panel), we see that the negative interactions lead to a relatively slower transition from matter-dominated to DE-dominated era and a later onset of the accelerated expansion phase in the late Universe. For a pressureless DM fluid, from Eq.~\eqref{Eq:FLRW:Interaction strength:Fluid representation}, we see that a positive interaction corresponds to $\Dot{\alpha}(\phi, X) < 0$ and vice versa.

\subsection{Momentum Transfer in DE-DM interactions}
\label{Subsection:Momentum Transfer}

To see the effect of the field-kinetic coupling model on the interactions in the dark sector, we consider the spatial evolution of the dark sector components. Let us now introduce the projection operator $q_{\mu\nu}$ defined as~\cite{2008-Tsagas.etal-PRep,2013-Pourtsidou.etal-Phys.Rev.D}:
\begin{equation} \label{Eq:M-transfer:Projection Operator}
    q_{\mu\nu} = u_{\mu}u_{\nu} + g_{\mu\nu}\, ,
\end{equation}
which projects the vector perpendicular to the fluid four-velocity $u^{\mu}$. With this, we can obtain the equations of motion governing the spatial evolution of the fields. Specifically, they satisfy the following condition:
\begin{equation} \label{Eq:M-transfer:Covariant EOM}
    q_{\mu}^{\nu}\nabla_{\nu}p_{_{DM}} + (\rho_{_{DM}} + p_{_{DM}})u^{\nu}\nabla_{\nu}u_{\mu} = q_{\mu}^{\nu}Q_{\nu}\, .
\end{equation}
Due to the homogeneity and isotropy of the FLRW background, this equation is identically zero for both models discussed in Sec.~\eqref{Section:Field-Coupling DE-DM} and \eqref{Section:Field-Kinetic Coupling DE-DM}. Therefore, we find a zero momentum exchange in the FLRW background. 
This is consistent with the numerical results in the previous sections, that the homogeneity and isotropy of the FLRW background on the direct interaction between DE and DM lead to energy transfer and no momentum transfer.

At the leading order in scalar perturbations, the perturbed interaction strength $\delta Q_i$ in Eq.~\eqref{Eq:Field-kinetic coupled:Interaction strength:Fluid representation} leads to:
\bea \label{Eq:Perturbed Spatial Interaction strength}
\delta Q_i = \partial_i\left[ (3\overline{p}_{_{DM}} - \overline{\rho}_{_{DM}})\delta\alpha(\phi, X)
 \right] = \partial_i\left[ \left( 3\overline{p}_{_{DM}} - \overline{\rho}_{_{DM}} \right)\left( 
\overline{\alpha}_{\phi}\delta\phi + \overline{\alpha}_X\Dot{\overline{\phi}}\Dot{\delta\phi} - \overline{\alpha}_X\Dot{\overline{\phi}}^2\Phi \right) \right]\, .
\eea
Let us now look at the first-order scalar perturbations in the Newtonian gauge. First, we define the momentum divergence for the DM fluid and the DE field, respectively, as:
\begin{equation} \label{Eq:M-transfer:M-divergence definition}
\theta_{_{DM}} = -\delta u^{s}\, ,~\theta_{\phi} = \frac{\delta\phi}{\Dot{\overline{\phi}}}\, .
\end{equation}
Perturbing Eq.~\eqref{Eq:M-transfer:Covariant EOM} about the perturbed FLRW metric gives us the following:
\begin{equation} \label{Eq:M-transfer:FLRW:Pertubed EOM}
    \partial_{i}[\delta p_{_{DM}} - \Dot{\overline{p}}_{_{DM}}\theta_{_{DM}} + (\overline{\rho}_{_{DM}} + \overline{p}_{_{DM}})(\Phi - \Dot{\theta}_{_{DM}})] = \delta Q_{i} - \partial_{i}(\theta_{_{DM}} \overline{Q})\, .
\end{equation}

For the field-coupling model, this equation reduces to the following form:
\begin{equation} \label{Eq:M-transfer:Field coupled:Perturbed FLRW EOM}
    (\overline{\rho}_{_{DM}} + \overline{p}_{_{DM}})(\Dot{\theta}_{_{DM}} - \Phi) + \Dot{\overline{p}}_{_{DM}}\theta_{_{DM}} - \delta p_{_{DM}} = (3\overline{p}_{_{DM}} - \overline{\rho}_{_{DM}})\overline{\alpha}_{\phi}\Dot{\overline{\phi}}(\theta_{_{DM}} - \theta_{\phi})\, ,
\end{equation}
For the pressureless DM description, it further reduces to:
\begin{equation} \label{Eq:M-transfer:Field coupled:Perturbed EOM:Pressureless DM}
    \Dot{\theta}_{_{DM}} = \Phi - \overline{\alpha}_{\phi}\Dot{\overline{\phi}}(\theta_{_{DM}} - \theta_{\phi})\, .
\end{equation}
It is interesting to note that this equation takes a similar form as the result for Type-1 theory established in Ref.~\cite{2013-Pourtsidou.etal-Phys.Rev.D}.

We now move on to the field-kinetic coupling model discussed in Sec.~\eqref{Section:Field-Kinetic Coupling DE-DM}. From Eq.~\eqref{Eq:M-transfer:FLRW:Pertubed EOM}, we can now write the following:
\begin{equation} \label{Eq:M-transfer:Field-kinetic coupled:Perturbed FLRW EOM}
    (\overline{\rho}_{_{DM}} + \overline{p}_{_{DM}})(\Dot{\theta}_{_{DM}} - \Phi) + \Dot{\overline{p}}_{_{DM}}\theta_{_{DM}} - \delta p_{_{DM}} = [3\overline{p}_{_{DM}} - \overline{\rho}_{_{DM}}][\Dot{\overline{\phi}}(\overline{\alpha}_{\phi} + \overline{\alpha}_{X}\Ddot{\overline{\phi}})(\theta_{_{DM}} - \theta_{\phi}) + \overline{\alpha}_{X}\Dot{\overline{\phi}}^{2}(\Phi - \Dot{\theta}_{\phi})]\, .
\end{equation}
For a pressureless DM description, the above expression reduces to the following:
\begin{equation} \label{Eq:M-transfer:Field-kinetic coupled:Perturbed EOM:Pressureless DM}
    \Dot{\theta}_{_{DM}} = (1 - \overline{\alpha}_{X}\Dot{\overline{\phi}}^{2})\Phi - \Dot{\overline{\phi}}(\overline{\alpha}_{\phi} + \overline{\alpha}_{X}\Ddot{\overline{\phi}})(\theta_{_{DM}} - \theta_{\phi}) + \overline{\alpha}_{X}\Dot{\overline{\phi}}^{2}\Dot{\theta}_{\phi}\, .
\end{equation}

Therefore, we find a non-zero momentum exchange between the DM and DE. The resulting expressions are particularly non-trivial when we take the kinetic dependence of the coupling into account. We plan to investigate this in detail in a later study.

\section{Conclusions and Discussions}
\label{Section:Conclusions}

When dissipative effects are negligible, the energy content of the Universe can be well-approximated as a sum of perfect fluids. Cosmological fluids are typically barotropic, with pressure depending solely on the energy density $p = p(\rho)$, and a single scalar function represents each fluid. Hence, scalar fields play a crucial role in describing the dark sector, which includes dark energy (DE) and dark matter (DM). However, it is essential to note that if the dark sector has non-gravitational interaction, the fluid description, while useful, will not fully capture the physics. This limitation underscores the need for a more fundamental framework that incorporates interactions directly at the level of the action.

In this work, we extended the analysis of interacting dark sector models with field-fluid mapping in Ref.~\cite{2021-Johnson.Shankaranarayanan-Phys.Rev.D} to include non-canonical scalar fields such as k-essence by considering the Horndeski models. We explore two types of coupling between dark energy and dark matter: field coupling and field-kinetic coupling. In the case of the field coupling, we see that the interaction has a similar form as in the case of the canonical scalar field presented in Ref.\cite{2021-Johnson.Shankaranarayanan-Phys.Rev.D}. This indicates that the field-fluid mapping and the resulting form of interaction are not unique to a specific model. We also showed that the interaction term is proportional to the trace of the energy-momentum tensor of the dark matter regardless of the nature of the coupling. Consequently, the interaction term vanishes for a dark radiation component with $p = \rho/3$, limiting its interaction to be purely gravitational. This result can provide a strong theoretical constraint on the dark radiation models \cite{Archidiacono:2022ich, Sobotka:2023bzr}. 

We looked at the background evolution with the field coupling and the field-kinetic coupling by constructing a dimensionless system of equations. We analyzed the background evolution for both cases with a quintessence dark energy model and a k-essence dark energy model. As can be seen from tables \eqref{Table:Field-coupling:Fixed points} and \eqref{Table:Kinetic-coupling:Fixed points}, these models lead to cosmological evolution consistent with observations. They all lead to a radiation-dominated saddle point followed by a matter-dominated saddle point that leads to an accelerated attractor. Regarding the evolution of the Hubble parameter, we see that for a given value at the redshift $z=0$, the field-kinetic coupling scenario leads to a lower value of the Hubble parameter in the past. Unlike the quintessence model, we also see that the k-essence model results in a purely kinetically driven accelerated expansion with both types of couplings. Furthermore, we showed that the evolution of the kinetic coupling is highly sensitive to the coupling strength as compared to the field-coupling model. This will help us obtain stringent constraints on the kinetic-coupling model. 

Looking at the evolution of the interaction function, we see that a higher rate of energy transfer from dark matter to dark energy results in a faster transition to the accelerated epoch. As mentioned above, the effect of the interaction strength is again more prominent in field-kinetic coupling. Since the background space-time is isotropic and homogeneous, there is no momentum transfer at the background level.

Even though one can see the different effects of the field-coupling and field-kinetic coupling on the background evolution, it is plausible that one model can mimic the evolution in the other model, which may make it difficult to choose one model over the other using observational data. However, we showed that these models result in a vastly different evolution of the first-order scalar perturbations. We see that both of these models lead to momentum transfer at the perturbed level. Looking at the perturbed field equations, we also see that, unlike the field coupling case, the field-kinetic coupling leads to a non-zero slip parameter $\Phi - \Psi$, which will significantly affect the cosmological observables.

Whether this analysis can be extended to models derived from a general disformal transformation applied to a Horndeski action remains to be answered. It would also be valuable to explore whether field-fluid mapping holds in the case of a Horndeski action with cubic terms, as this would encompass a broad range of possible interacting dark sector models. Further analysis is required for the models considered in this work, particularly for the scalar perturbations. This includes examining the relationship between energy and momentum transfer with adiabatic and entropic perturbations. Initial investigations of the perturbed equations suggest that the field-kinetic coupling might offer a more effective way to independently control background and perturbed interactions. This warrants further exploration, as it could simultaneously resolve cosmological tensions related to background and perturbative evolution, specifically the $ H_0 $ and $ \sigma_8 $ tensions.

Our numerical analysis shows that both these models can describe late Universe evolution consistent with observations while retaining their signatures. One needs to confront these models with observational data to constrain the model and cosmological parameters and determine if we can detect the signatures of dark sector interactions and if it resolves the cosmological tensions~\cite{Giare:2024smz}. Detailed data analysis of the background and perturbed evolution in these models will be carried out in future work.

\begin{acknowledgments}
The work is supported by the SERB-Core Research Grant (Project SERB/CRG/2022/002348) and the SPARC MoE grant SPARC/2019-2020/P2926/SL.
\end{acknowledgments}

\appendix
\section{Conformal transformation of the Horndeski action leading to a field-coupling in the dark sector} \label{Appendix:Field-Coupling Derivation}

Setting $\Tilde{G}_{4\Tilde{X}} = \Tilde{G}_{5} = 0$, we start with a subclass of the Horndeski action in Eq.~\eqref{Eq:Horndeski action} describing gravitation and the scalar field $\phi$, coupled minimally to the scalar field $\chi$. This reduces the action to the following:
\begin{equation} \label{Appendix:Eq:Field coupled:Horndeski action}
    S = \int \mathrm{d}^{4}x\sqrt{-\Tilde{g}}\left[ \frac{M^{2}_{\mathrm{Pl}}}{2}\Tilde{G}_{4}(\phi)\Tilde{R} + \Tilde{G}_{2}(\phi, \Tilde{X}) - \Tilde{G}_{3}(\phi, \Tilde{X})\Tilde{\Box}\phi + P_{1}(\chi, \Tilde{Y}) \right]\, ,
\end{equation}
where $\Tilde{X} = -\Tilde{g}^{\mu\nu}\Tilde{\nabla}_{\mu}\phi\Tilde{\nabla}_{\nu}\phi\,/2$, $\Tilde{\Box}\phi = \Tilde{g}^{\mu\nu}\Tilde{\nabla}_{\mu}\Tilde{\nabla}_{\nu}\phi$ and $\Tilde{Y} = -\Tilde{g}^{\mu\nu}\Tilde{\nabla}_{\mu}\chi\Tilde{\nabla}_{\nu}\chi\,/2$. Note that $\Tilde{G}_{2}, \Tilde{G}_{3}, \Tilde{G}_{4}$ are arbitrary functions of the scalar field $\phi$ and the corresponding kinetic term $\Tilde{X}$. $P_{1}$ is an arbitrary function of the scalar field $\chi$ and the corresponding kinetic term $\Tilde{Y}$.
To get rid of the non-minimal coupling, we perform a conformal transformation of the form:
\begin{equation} \label{Appendix:Eq:Conformal transformation}
    g_{\mu\nu} = \Omega^{2}(\phi)\Tilde{g}_{\mu\nu} = \Tilde{G}_{4}(\phi)\Tilde{g}_{\mu\nu}\, .
\end{equation}
Under this conformal transformation, the action in Eq.~\eqref{Appendix:Eq:Field coupled:Horndeski action} can be rewritten in the Einstein frame, in the following form:
\begin{equation} \label{Appendix:Eq:Field coupled:Transformed action}
\begin{aligned}
S = \int \mathrm{d}^{4}x\sqrt{-g}
&
\left[ \frac{M_{\mathrm{Pl}}^{2}}{2}R + \frac{1}{(\Tilde{G}_{4}(\phi))^{2}}\Tilde{G}_{2}\left( \phi, \Tilde{G}_{4}(\phi)X \right) +  \frac{3M^{2}_{\mathrm{Pl}}}{2}\left(\frac{\Tilde{G}_{4\phi}}{\Tilde{G}_{4}}\right)^{2}X \right. 
\\
& 
\left. - \frac{\Tilde{G}_{3}(\phi, \Tilde{G}_{4}(\phi)X)}{\Tilde{G}_{4}(\phi)}\left( \Box\phi + 2\frac{\Tilde{G}_{4\phi}}{\Tilde{G}_{4}}X \right) + \frac{1}{(\Tilde{G}_{4}(\phi))^{2}}P_{1}(\chi, \Tilde{G}_{4}Y) \right]\, ,
\end{aligned}
\end{equation}
where we have defined $X = -g^{\mu\nu}\nabla_{\mu}\phi\nabla_{\nu}\phi\,/2$, $\Box\phi = g^{\mu\nu}\nabla_{\mu}\nabla_{\nu}\phi$ and $Y = -g^{\mu\nu}\nabla_{\mu}\chi\nabla_{\nu}\chi\,/2$, respectively. To simplify the form of the above action, we define the functions $G_2, G_3$ and $\alpha$ as: 
\begin{subequations} \label{Appendix:Eq:Field coupled:Redefinitions}
\begin{align}
&  
G_{2}(\phi, X) = \frac{\Tilde{G}_{2}\left( \phi, \Tilde{G}_{4}(\phi)X \right)}{(\Tilde{G}_{4}(\phi))^{2}} +  \frac{3M^{2}_{\mathrm{Pl}}}{2}\left(\frac{\Tilde{G}_{4\phi}}{\Tilde{G}_{4}}\right)^{2}X - 2\left( \frac{\Tilde{G}_{3}(\phi, \Tilde{G}_{4}X)\Tilde{G}_{4\phi}}{(\Tilde{G}_{4}(\phi))^{2}} \right)X\, , \\
&
G_{3}(\phi, X) = \frac{\Tilde{G}_{3}(\phi, \Tilde{G}_{4}(\phi)X)}{\Tilde{G}_{4}(\phi)}\, , \\
&
\mathrm{e}^{2 \alpha(\phi)} = \frac{1}{\Tilde{G}_{4}(\phi)}\, .
\end{align}
\end{subequations}
Finally, using these definitions, we can rewrite the action in Eq.~\eqref{Appendix:Eq:Field coupled:Transformed action} to a simpler form:
\begin{equation} \label{Appendix:Eq:Final action}
    S = \int \mathrm{d}^{4}x\sqrt{-g} \left[ \frac{M^{2}_{\mathrm{Pl}}}{2}R + G_{2}(\phi, X) -G_{3}(\phi, X)\Box\phi  + \mathrm{e}^{4\alpha(\phi)}P_{1}(\chi, Z(\phi, Y))  \right]\, ,
\end{equation}
where we have defined $Z(\phi, Y) = \mathrm{e}^{-2 \alpha(\phi)}Y$. This action describes a theory of two interacting scalar fields ($\phi$, $\chi$) with a Horndeski description of the scalar field $\phi$, a k-essence description of the scalar field $\chi$ and an arbitrary field coupling described by the function $\alpha(\phi)$ in the Einstein frame. 

\section{Extended conformal transformation of the quadratic Horndeski action leading to a field-kinetic coupling in the dark sector} \label{Appendix:Field-Kinetic Coupling Derivation}

We start by setting $\Tilde{G}_5 = 0$ in the Horndeski action \eqref{Eq:Horndeski action} describing gravitation and the scalar field $\phi$, coupled minimally to the scalar field $\chi$. The action reduces to the following:
\begin{equation} \label{Appendix:Eq:Horndeski quadratic action}
\begin{aligned}
S = \int \mathrm{d}^{4}x\sqrt{-\Tilde{g}}
&
\left[\Tilde{G}_{2}(\phi, \Tilde{X}) - \Tilde{G}_{3}(\phi, \Tilde{X})\Tilde{\Box}\phi + \frac{M^{2}_{\mathrm{Pl}}}{2}\left[\Tilde{G}_{4}(\phi, \Tilde{X})\Tilde{R} + \Tilde{G}_{4\Tilde{X}}[(\Tilde{\Box}\phi)^{2} - \Tilde{\nabla}_{\mu}\Tilde{\nabla}_{\nu}\phi\Tilde{\nabla}^{\mu}\Tilde{\nabla}^{\nu}\phi]\right]\right. \\
& \left. 
+ P_{1}(\chi, \Tilde{Y})\right]\, .
\end{aligned}
\end{equation}
Using an extended conformal transformation of the form:
\begin{equation} \label{Appendix:Eq:Extended conformal transformation}
\Tilde{g}_{\mu\nu} = \Tilde{G}^{-1}_{4}\left( \phi, \tilde{X}\right)g_{\mu\nu} = \mathrm{e}^{2 \alpha(\phi,X)}g_{\mu\nu} \, ,
\end{equation}
we can rewrite the action \eqref{Appendix:Eq:Horndeski quadratic action} in the Einstein frame. Under the extended conformal transformation, the quantity $\Tilde{G}_{4\Tilde{X}}$ can be rewritten as:
\begin{equation} \label{Appendix:Eq:G4X transformation}
    \Tilde{G}_{4\Tilde{X}} = -\left( \frac{2\alpha_{X}}{1 - 2X\alpha_{X}} \right)\, ,
\end{equation}
and the metric related quantities transform as follows:
\begin{subequations} \label{Appendix:Eq:Metric quantities transformations}
\begin{align}
& \Tilde{\Gamma}^{\sigma}_{\mu\nu} = \Gamma^{\sigma}_{\mu\nu} + \delta^{\sigma}_{\mu}\nabla_{\nu}\alpha + \delta^{\sigma}_{\nu}\nabla_{\mu}\alpha - g_{\mu\nu}g^{\sigma\beta}\nabla_{\beta}\alpha\, ,
\\
&
\Tilde{R} = \mathrm{e}^{-2 \alpha(\phi,X)}R - 6\mathrm{e}^{- \alpha(\phi,X)}g^{\mu\nu}\nabla_{\mu}\nabla_{\nu}\left[ \mathrm{e}^{- \alpha(\phi,X)} \right]\, .
\end{align}
\end{subequations}
The scalar field dependent quantities transform as follows:
\begin{subequations}  \label{Appendix:Eq:Extended conformal:Scalar field terms}
\begin{align}
& \Tilde{X} = \mathrm{e}^{-2\alpha(\phi,X)}X\, ,
\\
&
\Tilde{\nabla}_{\mu}\Tilde{\nabla}_{\nu}\phi = \nabla_{\mu}\nabla_{\nu}\phi - \left( \nabla_{\mu}\phi\nabla_{\nu}\alpha + \nabla_{\nu}\phi\nabla_{\mu}\alpha - g_{\mu\nu}\nabla^{\sigma}\phi\nabla_{\sigma}\alpha \right)\, ,
\\
&
\Tilde{\Box}\phi = \mathrm{e}^{- 2\alpha(\phi,X)}\Box\phi + \mathrm{e}^{-4\alpha(\phi,X)}\nabla^{\sigma}\phi\nabla_{\sigma}\mathrm{e}^{2\alpha(\phi,X)}\, .
\end{align}
\end{subequations}
Using Eqs.~\eqref{Appendix:Eq:Extended conformal transformation}-\eqref{Appendix:Eq:Extended conformal:Scalar field terms}, the action in Eq.~\eqref{Appendix:Eq:Horndeski quadratic action} transforms to the following form:
\bea \label{Appendix:Eq:Field-kinetic coupled:Tranformed action}
    S = \int \mathrm{d}^{4}x\sqrt{-g}
& &
    \left[ \frac{M_{\mathrm{Pl}}^{2}}{2}R + \left[ \mathrm{e}^{4\alpha(\phi, X)}\Tilde{G}_{2}\left(\phi, X\mathrm{e}^{-2\alpha(\phi, X)} \right) + 4\mathrm{e}^{2\alpha(\phi, X)}X\alpha_{\phi}\Tilde{G}_{3}\left(\phi, X\mathrm{e}^{-2\alpha(\phi, X)} \right)  \right.\right.
    \nonumber \\
& & \left.\left.
    + 6M^{2}_{\mathrm{Pl}}\alpha_{\phi}^{2}X \right] \right.
    \nonumber \\
& & \left.
    - \Box\phi\left[ \mathrm{e}^{2\alpha(\phi, X)}\Tilde{G}_{3}\left(\phi, X\mathrm{e}^{-2\alpha(\phi, X)} \right) - 4M^{2}_{\mathrm{Pl}}X\alpha_{\phi}\left(\frac{\alpha_{X}}{1 - 2X\alpha_{X}}\right) \right] \right. 
    \nonumber \\
& & \left.
    + \phi_{\mu\nu}\phi^{\mu}\phi^{\nu}\left[ - 2\mathrm{e}^{2\alpha(\phi, X)}\alpha_{X}\Tilde{G}_{3} + 6M_{\mathrm{Pl}}^{2}\alpha_{\phi}\alpha_X - 4M^{2}_{\mathrm{Pl}}\alpha_{\phi}\left( 
    \frac{\alpha_{X}}{1 - 2X\alpha_{X}} \right) \right] \right. 
    \nonumber \\
& & \left.
    + \left[ \phi_{\mu\nu}\phi^{\mu\nu} - (\Box\phi)^{2} \right]\left[ M_{\mathrm{Pl}}^{2}\left( \frac{\alpha_{X}}{1 - 2X\alpha_{X}} \right) \right] \right. 
    \nonumber \\
& & \left.
    + [\phi_{\mu\nu}\phi^{\mu}\phi^{\nu}]^{2}\left[ -2M^{2}_{\mathrm{Pl}}\alpha_X^2\left( \frac{\alpha_{X}}{1 - 2X\alpha_{X}} \right) \right] \right. 
    \nonumber \\
& & \left.
    + [\Box\phi\phi_{\mu\nu}\phi^{\mu}\phi^{\nu}]\left[ 2M^{2}_{\mathrm{Pl}}\alpha_X\left( \frac{\alpha_{X}}{1 - 2X\alpha_{X}} \right) 
 \right] \right. 
 \nonumber \\
& & \left.
    + [\phi_{\mu\nu}\phi^{\nu}\phi^{\mu\sigma}\phi_{\sigma}]\left[ M^{2}_{\mathrm{Pl}}\alpha_X^2 + 4M^{2}_{\mathrm{Pl}}\alpha_X^2\left( \frac{X\alpha_{X}}{1 - 2X\alpha_{X}} \right) \right] \right. 
    \nonumber \\
& & \left.
 + \mathrm{e}^{4\alpha(\phi, X)}P_{1}\left(\chi, Y\mathrm{e}^{-2\alpha(\phi, X)} \right)\right]\, ,
\eea
where $ \phi_{\mu} \equiv \nabla_{\mu}\phi, \phi_{\mu\nu} \equiv \nabla_{\nu}\nabla_{\mu}\phi$ and $Y = -g^{\mu\nu}\nabla_{\mu}\chi\nabla_{\nu}\chi\,/2 \equiv -g^{\mu\nu}\chi_{\mu}\chi_{\nu}\,/2$.
We can rewrite the integral for $\phi_{\mu\nu}\phi^{\mu}\phi^{\nu}$ as:
\begin{equation}
    \int \mathrm{d}^{4}x\sqrt{-g}[D(\phi, X) + XD_{X}(\phi, X)]\phi_{\mu\nu}\phi^{\mu}\phi^{\nu} = \int \mathrm{d}^{4}x\sqrt{-g}[XD(\phi, X)\Box\phi - 2X^{2}D_{\phi}(\phi, X)]\, .
\end{equation}
Therefore, we define the function $D(\phi, X)$ in the following manner:
\bea 
D(\phi, X) + XD_{X}(\phi, X) = 
& & 
- 2\alpha_{X}\mathrm{e}^{2\alpha(\phi, X)}\Tilde{G}_{3}\left(\phi, X\mathrm{e}^{-2\alpha(\phi, X)} \right) + 6M_{\mathrm{Pl}}^{2}\left( \alpha_{\phi}\alpha_{X} \right)
\nonumber \\ 
& & 
- 4M^{2}_{\mathrm{Pl}}\alpha_{\phi}\left( 
\frac{\alpha_{X}}{1 - 2X\alpha_{X}} \right)\, .
\eea 
For a simpler form of Eq.~\eqref{Appendix:Eq:Field-kinetic coupled:Tranformed action}, we define the following functions: 
\bea \label{Appendix:Eq:Field-kinetic coupling:Redefinitions}
\begin{aligned}
G_{2}(\phi, X) =
&
\mathrm{e}^{4\alpha(\phi, X)}\Tilde{G}_{2}\left(\phi, \mathrm{e}^{-2\alpha(\phi, X)}X\right) + 4X\mathrm{e}^{2\alpha(\phi, X)}\alpha_{\phi}\Tilde{G}_{3}\left( \phi,X\mathrm{e}^{-2\alpha(\phi, X)} \right)
\\
& + 6M^{2}_{\mathrm{Pl}}\alpha_{\phi}^{2}X - 2X^{2}D_{\phi}(\phi, X)\, ,
\\
G_{3}(\phi, X) =
&
\mathrm{e}^{2\alpha(\phi, X)}\Tilde{G}_{3}\left(\phi, X\mathrm{e}^{-2\alpha(\phi, X)} \right) - 4M^{2}_{\mathrm{Pl}}X\alpha_{\phi}\left(\frac{\alpha_{X}}{1 - 2X\alpha_{X}}\right) - XD(\phi, X)\, .
\end{aligned}
\eea
With these definitions, the action \eqref{Appendix:Eq:Horndeski quadratic action} takes the final form:
\bea \label{Appendix:Eq:Field-kinetic coupled:Final action}
S = \int \mathrm{d}^{4}x\sqrt{-g}
& &
\left[ \frac{M^{2}_{\mathrm{Pl}}}{2}R + G_{2}(\phi, X) - G_{3}(\phi, X)\Box\phi + A_{1}(\phi, X)\phi_{\mu\nu}\phi^{\mu\nu} + A_{2}(\phi, X)(\Box\phi)^{2} 
\right.
\nonumber \\
& & \left.
+ A_{3}(\phi, X)(\phi_{\mu\nu}\phi^{\mu}\phi^{\nu})^{2} + A_{4}(\phi, X)\Box\phi\phi_{\mu\nu}\phi^{\mu}\phi^{\nu} + A_{5}(\phi, X)\phi_{\mu\nu}\phi^{\nu}\phi^{\mu\sigma}\phi_{\sigma}
\right.
\nonumber \\
& & \left. 
 + \mathrm{e}^{4\alpha(\phi, X)}P_{1}\left(\chi, \mathrm{e}^{-2\alpha(\phi, X)}Y \right)\right]\, ,
\eea
with the functions $A_1-A_5$ defined in Eq.~\eqref{Eq:Field-kinetic coupled:DHOST terms}. The above action is identical to the action \eqref{Eq:Field-kinetic coupled:Final action} in Sec.~\eqref{Section:Field-Kinetic Coupling DE-DM}.
Note that the terms $A_1-A_5$ in the above action correspond to a quadratic DHOST type Ia description of the scalar field $\phi$. It must be noted that for invertibility of the extended conformal transformation \eqref{Appendix:Eq:Extended conformal transformation}, $\alpha(\phi, X)$ must satisfy relation \eqref{Eq:Extended Conformal:Invertibility condition}.

\section{Constraints on the evolution of the corresponding DM scalar field for a pressureless DM fluid}
\label{Appendix:Pressureless fluid constraints}

Since the extended conformal transformation of the quadratic Horndeski theories discussed in Sec.~\eqref{Section:Field-Kinetic Coupling DE-DM} generalizes the results of Sec.~\eqref{Section:Field-Coupling DE-DM}, we shall look at the constraints on the evolution of the DM scalar field in the field-kinetic coupling scenario. As discussed in Sec.~\eqref{Section:Field-Kinetic Coupling DE-DM}, we start with a non-canonical description for the DM scalar field $\chi$ and perform an extended conformal transformation \eqref{Eq:Extended conformal transformation}\,. This leads to the following final form of the DM scalar field action in the Einstein frame:
\bea 
S_{_{DM}} = \int \mathrm{d}^{4}x\sqrt{-g}\,\mathrm{e}^{4\alpha(\phi, X)}P_{1}\left(\chi, Z \right)\,,
\eea 
where we have defined $Z = Y\mathrm{e}^{-2\alpha(\phi, X)}$\, and $Y = -\chi_{\mu}\chi^{\mu}/2$ denotes the kinetic term for the DM scalar field $\chi$. Performing the field-to-fluid mapping gives us the following relation between the DM scalar field and the energy density and pressure, respectively, for the corresponding DM fluid \eqref{Eq:Field-kinetic coupled:rho & p}:
\bea
& &
\rho_{_{DM}} = \mathrm{e}^{2\alpha(\phi,X)}\left[ 2Y \, P_{1Z} - \mathrm{e}^{2\alpha(\phi,X)} \, P_{1} \right] \, ,
\nonumber \\
& &
p_{_{DM}} = \mathrm{e}^{4\alpha(\phi,X)} \, P_{1} \, .
\eea
To better understand the constraints on the DM scalar field for a corresponding pressureless DM fluid, we assume a non-canonical description for the scalar field $\chi$ given by the following:
\bea 
P_1(\chi, Z) = \xi(\chi)\,Z - U(\chi)\,,
\eea 
where $U(\chi)$ denotes the DM scalar field potential and $\xi(\chi) > 0$ to prevent any ghost instabilities. From the above equations, it is evident that for $p_{_{DM}} = 0$, we have:
\bea 
\xi(\chi)\,Z = U(\chi)\,.
\eea 
We note that under this constraint, the energy density of the DM fluid reduces to:
\bea
\rho_{_{DM}} = 2\mathrm{e}^{2\alpha(\phi,X)}\,\xi(\chi)\,Y = \mathrm{e}^{2\alpha(\phi,X)}\,\xi(\chi)\,\Dot{\chi}^2\,.
\eea 
From the above expression for the FLRW metric, we see that assuming a pressureless dark matter (DM) fluid, the energy density $\rho_{_{DM}}$ of the fluid remains non-zero. Furthermore, the equation clearly shows that $\rho_{_{DM}}$ is always positive, even in the case of a canonical DM scalar field $\chi$. It is important to note that non-linear constraints may arise for more general k-essence models.

\section{Second-order equations of motion for the kinetic-coupling model} 
\label{Appendix:Kinetic-coupling:Second-order EOMs}

Using the set of equations \eqref{Eq:Kinetic-coupling:FLRW:DM EOM}-\eqref{Eq:Kinetic-coupling:Simplified expression} established in Sec.~\eqref{Subsection:Kinetic model}, the equations of motion for the scalar field $\phi(t)$ and the scale factor $a(t)$ can be reduced to the following forms:
\bea  \label{Appendix:Eq:Kinetic-coupling:phi EOM:2nd order}
\Ddot{\phi} = 
& &
\frac{\Ddot{\phi}_{\rm Numerator}}{\Ddot{\phi}_{\rm Denominator}}\, ,
\\
\Ddot{\phi}_{\rm Numerator} = 
& &
\Dot{\phi}\left[
\ 4(3\alpha_0+1)V_{\phi}\Dot{\phi} - 2(3\alpha_0-1)\beta_{\phi}\Dot{\phi}^3 - 3(\alpha_0-1)\gamma_{\phi}\Dot{\phi}^5 - 2H\left( 6(3\alpha_0 - 1)\beta(\phi)\Dot{\phi}^2
\right. \right.
\nonumber \\ 
& &
\left. \left. 
{} \ \ \
+ 3(3\alpha_0 - 2)\gamma(\phi)\Dot{\phi}^4 + 4\alpha_0\left( 3p_{_{DM}} + \rho_{r} - 9V(\phi) \right) \right) \right]\, ,
\nonumber \\
\Ddot{\phi}_{\rm Denominator} = 
& &
8\alpha_0\left[ 3\alpha_0p_{_{DM}} - (\alpha_0 - 1)\rho_{_{DM}} + \rho_r + 3V(\phi) \right] + 4(3\alpha_0 - 1)\beta(\phi)\Dot{\phi}^2 + 6(3\alpha_0 - 2)\gamma(\phi)\Dot{\phi}^4\, .
\nonumber
\eea
\bea  \label{Appendix:Eq:Kinetic-coupling:Hdot equation}
\Dot{H} = 
& &  \
\frac{\Dot{H}_{\rm Numerator}}{\Dot{H}_{\rm Denominator}}\, ,
 \nonumber \\
\Dot{H}_{\rm Numerator} = 
& - &
1296 \alpha_0 (\alpha_0 + 1)^2(3\alpha_0 - 1)M_{\mathrm{Pl}}^4H^4\Dot{\phi}^4 + 288\alpha_0^2(\alpha_0 - 1)^2p_{_{DM}}^2 \Dot{\phi}^4 \\ 
& + &
16(\alpha_0 - 1)^3\alpha_0\rho_{_{DM}}\rho_{r}\Dot{\phi}^4+16 (\alpha_0 - 1)^2 \alpha_0 (3\alpha_0 - 1) \rho_{r}^2 \Dot{\phi}^4
\nonumber \\ 
& - & 
48 (\alpha_0 - 1)^3 \alpha_0 \rho_{_{DM}} V(\phi) \Dot{\phi}^4-96 (\alpha_0 - 1)^2 \alpha_0 (3\alpha_0 - 1) \rho_{r} V(\phi) \Dot{\phi}^4
\nonumber \\ 
& + &
144 (\alpha_0 - 1)^2 \alpha_0 (3\alpha_0 - 1) V(\phi)^2 \Dot{\phi}^4+24 (\alpha_0 - 1)^3 \alpha_0 \beta(\phi) \rho_{_{DM}} \Dot{\phi}^6
\nonumber \\ 
& + & 
16 (\alpha_0 - 1)^2 (-1+9 \alpha_0^2) \beta(\phi) \rho_{r} \Dot{\phi}^6-48 (\alpha_0 - 1)^2 (-1+9 \alpha_0^2) \beta(\phi) V(\phi) \Dot{\phi}^6
\nonumber \\ 
& + &
12 (\alpha_0 - 1)^2 (-2+3 \alpha_0+9 \alpha_0^2) \beta(\phi)^2 \Dot{\phi}^8+12 (\alpha_0 - 1)^3 \alpha_0 \gamma(\phi) \rho_{_{DM}} \Dot{\phi}^8
\nonumber \\ 
& + & 
24 (\alpha_0 - 1)^2 (-2+\alpha_0+3 \alpha_0^2) \gamma(\phi) \rho_{r} \Dot{\phi}^8-72 (\alpha_0 - 1)^2 (-2+\alpha_0+3 \alpha_0^2) \gamma(\phi) V(\phi) \Dot{\phi}^8
\nonumber \\ 
& + &
12 (\alpha_0 - 1)^2 (-7+6 \alpha_0+9 \alpha_0^2) \beta(\phi) \gamma(\phi) \Dot{\phi}^{10}+9 (\alpha_0 - 1)^2 (-4+3 \alpha_0 (1+\alpha_0)) \gamma(\phi)^2 \Dot{\phi}^{12}
\nonumber \\ 
& - & 
48 (-1+\alpha_0^2) M_{\mathrm{Pl}}^2 H^2 \Dot{\phi}^4 \left( 3 (\alpha_0 - 1) \alpha_0 \rho_{_{DM}}+16 \alpha_0^2 \rho_{r}+3 (3\alpha_0 - 1) \beta(\phi) \Dot{\phi}^2
\right. \nonumber \\ 
& + &  \left.
9 (\alpha_0 - 1) \gamma(\phi) \Dot{\phi}^4 \right)+48 (\alpha_0 - 1) \alpha_0 (1+\alpha_0) (3\alpha_0 - 1) M_{\mathrm{Pl}}^2 \Dot{\phi}^8 \beta_{\phi\phi}
 \nonumber \\ 
& + & 
72 (\alpha_0 - 1)^2 \alpha_0 (1+\alpha_0) M_{\mathrm{Pl}}^2 \Dot{\phi}^{10} \gamma_{\phi\phi}-96 (\alpha_0 - 1) \alpha_0 (1+\alpha_0) (1+3 \alpha_0) M_{\mathrm{Pl}}^2 \Dot{\phi}^6 V_{\phi\phi}
\nonumber \\ 
& - & 
5184 \alpha_0^2 (1+\alpha_0)^2 (3\alpha_0 - 1) M_{\mathrm{Pl}}^4 H^3 \Dot{\phi}^3 \Ddot{\phi}-48 \alpha_0 (-1+\alpha_0^2) M_{\mathrm{Pl}}^2 H \Dot{\phi}^3 \left[ 6 (\alpha_0 - 1) \alpha_0 \rho_{_{DM}}
\right. \nonumber \\ 
& - &  \left.
4 \alpha_0 (3\alpha_0 - 1) (\rho_{r}-9 V(\phi))-6 (-2+3 \alpha_0+9 \alpha_0^2) \beta(\phi) \Dot{\phi}^2
\right. \nonumber \\ 
& - & \left.
3 (-8+9 \alpha_0 (1+\alpha_0)) \gamma(\phi) \Dot{\phi}^4 \right] \Ddot{\phi}
\nonumber \\ 
& - & 
48 \alpha_0 (1+\alpha_0) M_{\mathrm{Pl}}^2 \Dot{\phi}^2 \left( 54 \alpha_0^2 (1+\alpha_0) (3\alpha_0 - 1) M_{\mathrm{Pl}}^2 H^2+(\alpha_0 - 1) (\alpha_0^2 \left( (21-93 \alpha_0) p_{_{DM}}
\right. \right. \nonumber \\ 
& - &  \left. \left.
5 (\alpha_0 - 1) \rho_{_{DM}}-12 (3\alpha_0 - 1) (\rho_{r}-3 V(\phi)) \right)-(3\alpha_0 - 1) (4+9 \alpha_0 (1+2 \alpha_0)) \beta(\phi) \Dot{\phi}^2
\right. \nonumber \\ 
& - &  \left.
3 (-8+3 \alpha_0 (1+\alpha_0 (2+3 \alpha_0))) \gamma(\phi) \Dot{\phi}^4) \right) \Ddot{\phi}^2+5184 \alpha_0^4 (1+\alpha_0)^2 (3\alpha_0 - 1) M_{\mathrm{Pl}}^4 H \Dot{\phi} \Ddot{\phi}^3
\nonumber \\ 
& + &
3888 \alpha_0^5 (1+\alpha_0)^2 (3\alpha_0 - 1) M_{\mathrm{Pl}}^4 \Ddot{\phi}^4+576 \alpha_0^2 (-1+\alpha_0^2) M_{\mathrm{Pl}}^2 \Dot{p}_{_{DM}} \Dot{\phi}^3 (H \Dot{\phi}+\alpha_0 \Ddot{\phi})
\nonumber \\ 
& + & 
48 (\alpha_0 - 1) \alpha_0 (1+\alpha_0) (3\alpha_0 - 1) M_{\mathrm{Pl}}^2 \beta_{\phi} \Dot{\phi}^6 \left( 6 H \Dot{\phi}+(5+3 \alpha_0) \Ddot{\phi} \right)
\nonumber \\ 
& - &  
96 \alpha_0 (-1+\alpha_0^2) M_{\mathrm{Pl}}^2 V_{\phi} \Dot{\phi}^4 (18 \alpha_0 H \Dot{\phi}+(1+9 \alpha_0^2) \Ddot{\phi})
\nonumber \\ 
& + &
72 \alpha_0 (-1+\alpha_0^2) M_{\mathrm{Pl}}^2 \gamma_{\phi} \Dot{\phi}^8 \left( 2 (-2+3 \alpha_0) H \Dot{\phi} + (-9+\alpha_0 (8+3 \alpha_0)) \Ddot{\phi} \right)
 \nonumber \\ 
& + & 
12 (\alpha_0 - 1) p_{_{DM}} \Dot{\phi}^3 \left( (\alpha_0 - 1) \Dot{\phi} \left( 4 \alpha_0 \left( 9 (1+\alpha_0) M_{\mathrm{Pl}}^2 H^2+(\alpha_0 - 1) \rho_{_{DM}}
\right. \right. \right. \nonumber \\ 
& + &  \left.  \left. \left.
(-1+5 \alpha_0) (\rho_{r}-3 V(\phi)) \right)+2 (-2+3 \alpha_0 (1+5 \alpha_0)) \beta(\phi) \Dot{\phi}^2
\right. \right. \nonumber \\
& + &  \left. \left.
3 (-4+\alpha_0 (3+5 \alpha_0)) \gamma(\phi) \Dot{\phi}^4 \right) + 24 \alpha_0^2 (1+\alpha_0) (-5+17 \alpha_0) M_{\mathrm{Pl}}^2 H \Ddot{\phi} \right)
\, , 
\nonumber \\
\Dot{H}_{\rm Denominator} = 
& & 
24 (1+\alpha_0) M_{\mathrm{Pl}}^2 \Dot{\phi}^2 \left[ 36 \alpha_0 (\alpha_0 + 1) (3\alpha_0 - 1) M_{\mathrm{Pl}}^2 (H\Dot{\phi} + \alpha_0\Ddot{\phi} )^2
\right. \nonumber \\
& + &  \left.
(\alpha_0 - 1)\Dot{\phi}^2 \left[ 4 \alpha_0(\alpha_0 - 1)(\rho_{_{DM}}+\rho_{r}) + 12\alpha_0(3\alpha_0 + 1)V(\phi)
\right. \right. \nonumber \\
& - &  \left. \left.
2(3\alpha_0 - 1)(3\alpha_0 - 2)\beta(\phi)\Dot{\phi}^2 - 3(\alpha_0 - 1) (3\alpha_0 - 4)\gamma(\phi)\Dot{\phi}^4
\right]
\right]
\nonumber
\eea 
Note that, the above pair of equations are second-order differential equations. Hence, they do not suffer from the Ostrogradsky instability \cite{Woodard:2015zca}.

The dimensionless variables $\mathcal{G}$ and $\mathcal{F}$ used to write the autonomous system of equations in Sec.~\eqref{sec:Autonomous02} are given by:
\bea   \label{Appendix:Eq:Kinetic-coupling:Dimensionless G}
\mathcal{G} = 
& &
\frac{\mathcal{G}_{\rm Numerator}}{\mathcal{G}_{\rm Denominator}}\, , 
\\ 
\mathcal{G}_{\rm Numerator} = 
& - &
9 F^4 \alpha_0^5 (\alpha_0 + 1)^2 (3\alpha_0 - 1) + 12 \sqrt{6} F^3 \alpha_0^4 (\alpha_0 + 1)^2 (3\alpha_0 - 1) x
\nonumber \\
& - &
2 F^2 \alpha_0 (\alpha_0 + 1) x^2 \left[\alpha_0 (\alpha_0 (5 \Omega_{_{DM}} + \alpha_0 \left(5 \alpha_0 \Omega_{_{DM}} - 2 (18 + 5 \Omega_{_{DM}} + 24 \Omega_r - 72 w)
\right. \right.
\nonumber  \\
& + & \left. \left.
18 \alpha_0 (-3 + 2 \Omega_r - 6 w) \right) + 6 (3 + 2 \Omega_r - 6 w)) 
\right.
\nonumber \\
& + & \left. 
36 (-11 + 3 \alpha_0 (-1 + \alpha_0 (3\alpha_0 - 1))) v x^4)
\right.
\nonumber \\
& + & \left. 
288 v x^4 + 2 (-1 + \alpha_0) (3\alpha_0 - 1) (4 + 9 \alpha_0 (1 + 2 \alpha_0)) x^2 y  \right]
\nonumber \\
& + &
4 F \alpha_0 (\alpha_0 + 1) x^3 \left[ \sqrt{6}\alpha_0 \left[ -3 \Omega_{_{DM}} + 2 (9 + \Omega_r - 9 w)
\right. \right.
\nonumber \\
& + & \left. \left.
\alpha_0 \left(-36 + 6 \Omega_{_{DM}} - 6 \omega_{_{DM}}' \Omega_{_{DM}} + (-3 + 6 \omega_{_{DM}}') \alpha_0 \Omega_{_{DM}} - 8 \Omega_r + 72 w 
\right. \right. \right. 
\nonumber \\
& + & \left. \left. \left.
6 \alpha_0 (\Omega_r - 9 (1 + w)) \right) \right] - 6 b_1 (-1 + \alpha_0) (1 + 9 \alpha_0^2) w x 
\right.
\nonumber \\
& + & \left.
18 \sqrt{6} (8 - 17 \alpha_0 + 9 \alpha_0^3) v x^4 + 54 \gamma_1 (-1 + \alpha_0) (-9 + \alpha_0 (8 + 3 \alpha_0)) v x^5
\right. 
\nonumber \\
& + & \left.
6 \sqrt{6} (-1 + \alpha_0) (3\alpha_0 - 1) (2 + 3 \alpha_0) x^2 y + 
      6 \beta_1 (-1 + \alpha_0) (3\alpha_0 - 1) (5 + 3 \alpha_0) x^3 y \right]
\nonumber \\
& + &
12 x^4 \left( 9\alpha_0(\alpha_0+1)^2(3\alpha_0 - 1) + 3\alpha_0(\alpha_0 - 1)(\alpha_0 + 1)(\alpha_0 - 1 - 4\alpha_0\omega'_{_{DM}})\Omega_{_{DM}}
\right.
\nonumber \\
& + & \left.
16\alpha_0^2(\alpha_0 - 1)(\alpha_0 + 1)\Omega_r - \alpha_0(\alpha_0 - 1)^2(3\alpha_0 - 1)(\Omega_r - 3w)^2
\right. 
\nonumber \\
& - & \left.
\alpha_0(\alpha_0 - 1)^3\Omega_{_{DM}}(\Omega_r - 3w)
\right. 
\nonumber \\
& + & \left.
36 \sqrt{6} b_1 \alpha_0^2 (-1 + \alpha_0^2) w x - 36 \sqrt{6} \gamma_1 (-1 + \alpha_0) \alpha_0 (\alpha_0 + 1) (-2 + 3 \alpha_0) v x^5
\right.
\nonumber \\
& - & \left.
81 (-1 + \alpha_0)^2 (-4 + 3 \alpha_0 (\alpha_0 + 1)) v^2 x^8 - 12 \sqrt{6} \beta_1 (-1 + \alpha_0) \alpha_0 (\alpha_0 + 1) 
(3\alpha_0 - 1) x^3 y
\right. 
\nonumber \\
& - & \left.  
18 (-1 + \alpha_0)^2 v x^6 (6 \gamma_1^2 \gamma_2 \alpha_0 (\alpha_0 + 1) + (-7 + 6 \alpha_0 + 9 \alpha_0^2) y)
\right.
\nonumber \\
& + & \left.
(-1 + \alpha_0) x^2 (12 b_1^2 b_2 \alpha_0 (\alpha_0 + 1) (1 + 3 \alpha_0) w + \alpha_0 (-3 \Omega_{_{DM}} - 3 \alpha_0^2(\Omega_{_{DM}} + 6 (\Omega_r - 3 w))
\right. 
\nonumber \\
& + & \left. 
6 \alpha_0 (3 + \Omega_{_{DM}} + 3 \Omega_r - 9 w) + 2 (6 + \Omega_r - 3 w)) y - 2 (3 + \Omega_r - 3 w) y)
\right.
\nonumber \\
& + & \left.
3 (-1 + \alpha_0) x^4 (-3 (-1 + \alpha_0) v (-4 (3 + \Omega_r - 3 w) + \alpha_0 ((-1 + \alpha_0) \Omega_{_{DM}} + 2 \Omega_r + 6 \alpha_0 \Omega_r 
\right.
\nonumber \\
& - & \left.
6 (2 + w + 3 \alpha_0 w))) - (3\alpha_0 - 1) y (4 \beta_1^2 \beta_2 \alpha_0 (\alpha_0 + 1) + (-1 + \alpha_0) (2 + 3 \alpha_0) y)) 
\right)
\nonumber \\
& + &
6 \Omega_{_{DM}} x^2 \omega_{_{DM}} \left[-54 (-1 + \alpha_0)^2 (-4 + \alpha_0 (3 + 5 \alpha_0)) v x^6
\nonumber \right. \\
& - & \left.
6 (-1 + \alpha_0)^2 (-2 + 3 \alpha_0 (1 + 5 \alpha_0)) x^4 y
\nonumber \right. \\
& + & \left.
F^2 \alpha_0^3 (-1 + \alpha_0^2) (7 - 35 \alpha_0 + 12 \alpha_0 \omega_{_{DM}}) - 2 \sqrt{6} F \alpha_0^2 (-1 + \alpha_0^2) x (5 - 13 \alpha_0 + 12 \alpha_0 \omega_{_{DM}})
\nonumber \right. \\
& - & \left.
6 (-1 + \alpha_0) \alpha_0 x^2 (-3 + \Omega_{_{DM}} + \Omega_r + \alpha_0^2 (-9 + \Omega_{_{DM}} + 5 \Omega_r - 15 w) 
\nonumber \right. \\
& - & \left.
2 \alpha_0 (6 + \Omega_{_{DM}} + 3 \Omega_r - 9 w) - 3 w + 6 \alpha_0 (-2 + \alpha_0 (-2 + \Omega_{_{DM}}) - \Omega_{_{DM}}) \omega_{_{DM}})\right]
\, ,
\nonumber \\
\mathcal{G}_{\rm Denominator} = 
& &
18 (\alpha_0 + 1) x^2 \left[ \alpha_0(\alpha_0 + 1)(3\alpha_0 - 1)(F\alpha_0 - \sqrt{6}x)^2 + 2\alpha_0(\alpha_0 - 1)^2x^2(\Omega_{_{DM}} + \Omega_r)
\right.
\nonumber \\ 
& - & \left.
2(\alpha_0 - 1)(3\alpha_0 - 2)(3\alpha_0 - 1)x^4y - 18(\alpha_0 - 1)^2(3\alpha_0 - 4)x^6v
\right.
\nonumber \\
& + & \left.
6\alpha_0(\alpha_0 - 1)(3\alpha_0 + 1)x^2w
\right]
\, ,
\nonumber 
\eea 
\bea 
\begin{aligned} \label{Appendix:Eq:Kinetic-coupling:Dimensionless F}
\mathcal{F} = 
& \
\frac{\mathcal{F}_{\rm Numerator}}{\mathcal{F}_{\rm Denominator}}\, ,
\\
\mathcal{F}_{\rm Numerator} = 
& \
x \left[- \ 3b_1(3\alpha_0 + 1)wx + 9\left[ \sqrt{6}(3\alpha_0 - 2) +3\gamma_1(\alpha_0 - 1)x \right]vx^4
\right. \\
& \ \left.
{} \ \ \
+ 3(3\alpha_0 - 1)(\sqrt{6} + \beta_1x)x^2y + \sqrt{6}\alpha_0(\Omega_r - 9w + 3\Omega_{_{DM}}\omega_{_{DM}}) \right]\, , 
\\
\mathcal{F}_{\rm Denominator} = 
& \ 
- x^2 (18 v x^2 + y) + \alpha_0 (\Omega_{_{DM}} + \Omega_r + 3 (w + 9 v x^4 + x^2 y)) + \alpha_0^2 \Omega_{_{DM}} (3\omega_{_{DM}} - 1)\, .
\end{aligned}
\eea 
%

\section{Comparing the numerical evolution of the kinetic-coupling k-essence model for different sets of initial conditions}
\label{Appendix:Numerical evolution:Kinetic-coupling:Different ICs}

For the form of the purely kinetic-coupling \eqref{Eq:Kinetic-coupling:Coupling form assumption} and the k-essence model \eqref{Eq:Numerical evolution:k-essence assumption}, we have a choice of setting different values, either for $v_i$ or $w_i$ by solving Eq.~\eqref{Eq:Kinetic-coupling:Dimensionless simplified expression} for different values of $\alpha_0$. This is an additional choice because of the k-essence description of the DE field $\phi$ compared to the quintessence models. Therefore, we compare the numerical evolution of the kinetic-coupling k-essence model for two sets of initial conditions. One set of conditions is given by Eq.~\eqref{Eq:Numerical evolution:k-essence:Kinetic-coupling:Initial conditions} where $v_i$ is different for different values of $\alpha_0$. The other set is given by the following:
\begin{eqnarray} \label{Appendix:Eq:Numerical evolution:k-essence:Kinetic-coupling:Initial conditions02}
& &
x_{i} = 1.5\times10^{-5}\, ,~~
y_{i} = 1\, ,~~
v_{i} = 1.296\times10^{9}\, ,~~
\Omega_{r_i} = 0.4\, ,~~
\Omega_{_{DM_i}} = 0.6 - 1.44\times10^{-9}\, ,
\nonumber \\
& &
b_{1} = 0\, ,~~
b_{2} = 0\, ,~~
\beta_{1} = 0\, ,~~
\beta_{2} = 0\, ,~~
\gamma_{1} = 0\, ,~~
\gamma_{2} = 0\, ,
\end{eqnarray}
and $w_i$ is computed by solving Eq.~\eqref{Eq:Kinetic-coupling:Dimensionless simplified expression} for different values of $\alpha_0$. We will investigate the impact of different initial conditions on the cosmological evolution for the same coupling model.

\begin{figure}[!htb]
\minipage{0.5\textwidth}
  \includegraphics[width=1\linewidth]{KC-Kv-q.png}
\endminipage\hfill
\minipage{0.5\textwidth}
  \vspace{-2em}  
  \includegraphics[width=1\linewidth]{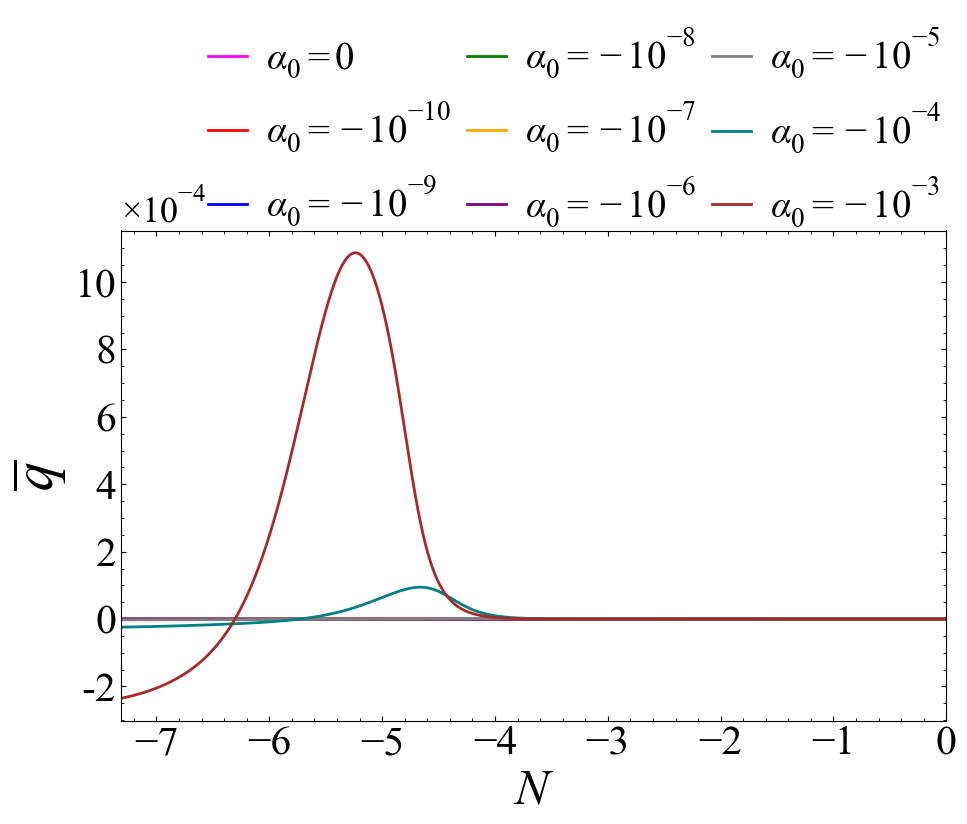}
\endminipage
\caption{Evolution of the scaled interaction term $\overline{q}$ for the kinetic-coupling k-essence model \eqref{Eq:Numerical evolution:k-essence assumption} for two different sets of initial conditions (I.C.). \emph{Left panel}: I.C. \eqref{Eq:Numerical evolution:k-essence:Kinetic-coupling:Initial conditions}, \emph{Right panel}: I.C. \eqref{Appendix:Eq:Numerical evolution:k-essence:Kinetic-coupling:Initial conditions02}.}
\label{App:Fig:Interaction strength}
\end{figure}

\begin{figure}[!htb]
\minipage{0.5\textwidth}
  \includegraphics[width=1\linewidth]{KC-Kv-epsilon.png}
\endminipage\hfill
\minipage{0.5\textwidth}  
\vspace{-2em}  
  \includegraphics[width=1\linewidth]{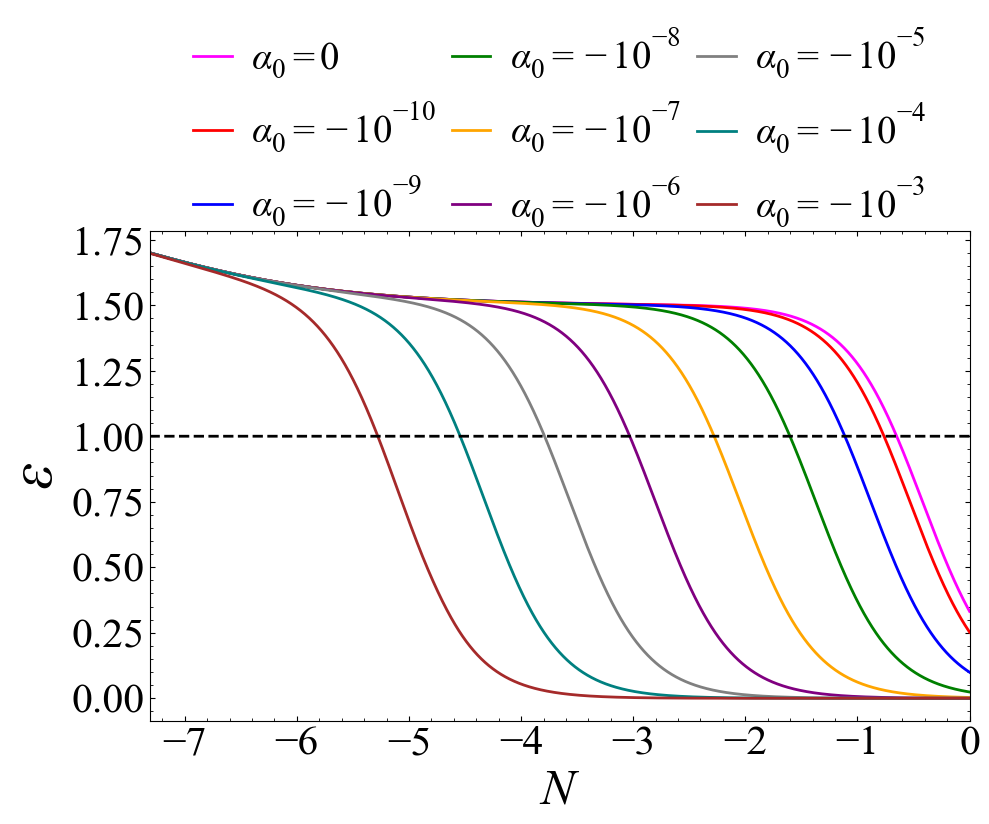}
\endminipage
\caption{Evolution of the slow-roll parameter $\epsilon$ for the kinetic-coupling k-essence model \eqref{Eq:Numerical evolution:k-essence assumption} for two different sets of initial conditions (I.C.). \emph{Left panel}: I.C. \eqref{Eq:Numerical evolution:k-essence:Kinetic-coupling:Initial conditions}, \emph{Right panel}: I.C. \eqref{Appendix:Eq:Numerical evolution:k-essence:Kinetic-coupling:Initial conditions02}.}
\label{App:Fig:epsilon}
\end{figure}

\begin{figure}[!htb]
\minipage{0.5\textwidth}
  \includegraphics[width=1\linewidth]{KC-Kv-Omega_DM.png}
\endminipage\hfill
\minipage{0.5\textwidth}
  \vspace{-2em}  
  \includegraphics[width=1\linewidth]{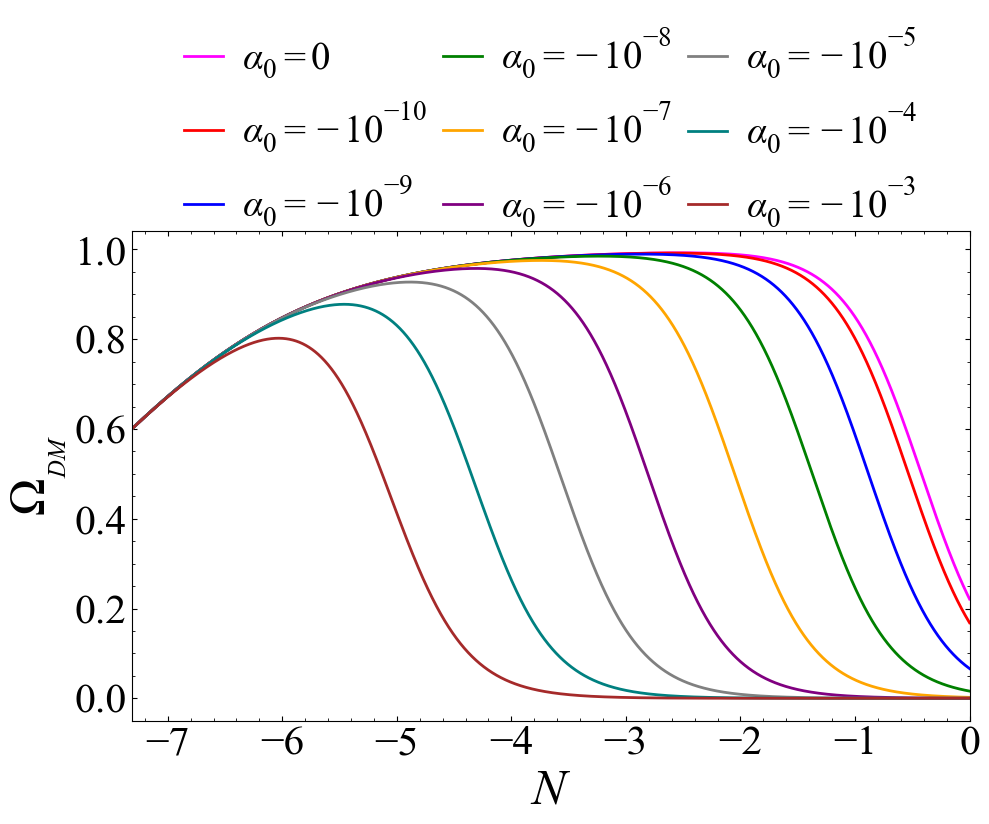}
\endminipage
\caption{Evolution of the matter energy density parameter $\Omega_{_{DM}}$ for the kinetic-coupling k-essence model \eqref{Eq:Numerical evolution:k-essence assumption} for two different sets of initial conditions (I.C.). \emph{Left panel}: I.C. \eqref{Eq:Numerical evolution:k-essence:Kinetic-coupling:Initial conditions}, \emph{Right panel}: I.C. \eqref{Appendix:Eq:Numerical evolution:k-essence:Kinetic-coupling:Initial conditions02}.}
\label{App:Fig:Omega_DM}
\end{figure}

\ref{App:Fig:Interaction strength} contains the plots for the interaction strength $\overline{q}$ for the two sets of initial conditions. We find that one set leads to negative interactions ($\overline{q} < 0$), while the other leads to positive interactions ($\overline{q} > 0$). The cosmological parameters $\epsilon$ and $\Omega_{_{DM}}$ are plotted in \ref{App:Fig:epsilon} and \ref{App:Fig:Omega_DM}, respectively. We see that the negative interactions lead to a slower transition and a later onset of the accelerated expansion phase in the late Universe, compared to the non-interacting case.
\begin{figure}[!htb]
\minipage{0.5\textwidth}
  \includegraphics[width=1\linewidth]{KC-Kv-H.png}
\endminipage\hfill
\minipage{0.5\textwidth}
  \vspace{-2em}  
  \includegraphics[width=1\linewidth]{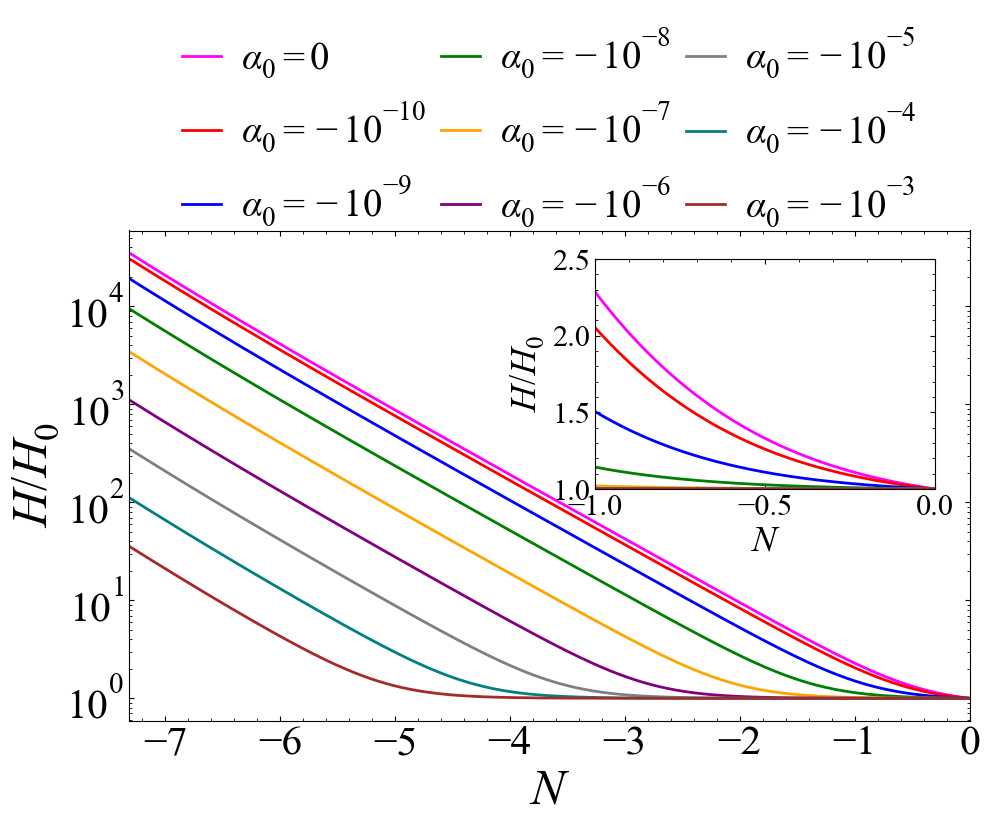}
\endminipage
\caption{Evolution of the scaled Hubble parameter $H/H_0$ for the kinetic-coupling k-essence model \eqref{Eq:Numerical evolution:k-essence assumption} for two different sets of initial conditions (I.C.). \emph{Left panel}: I.C. \eqref{Eq:Numerical evolution:k-essence:Kinetic-coupling:Initial conditions}, \emph{Right panel}: I.C. \eqref{Appendix:Eq:Numerical evolution:k-essence:Kinetic-coupling:Initial conditions02}.}
\label{App:Fig:Hubble}
\end{figure}
\begin{figure}[!htb]
\minipage{0.5\textwidth}
  \includegraphics[width=1\linewidth]{KC-Kv-Omega_r.png}
\endminipage\hfill
\minipage{0.5\textwidth}
  \vspace{-2em}  
  \includegraphics[width=1\linewidth]{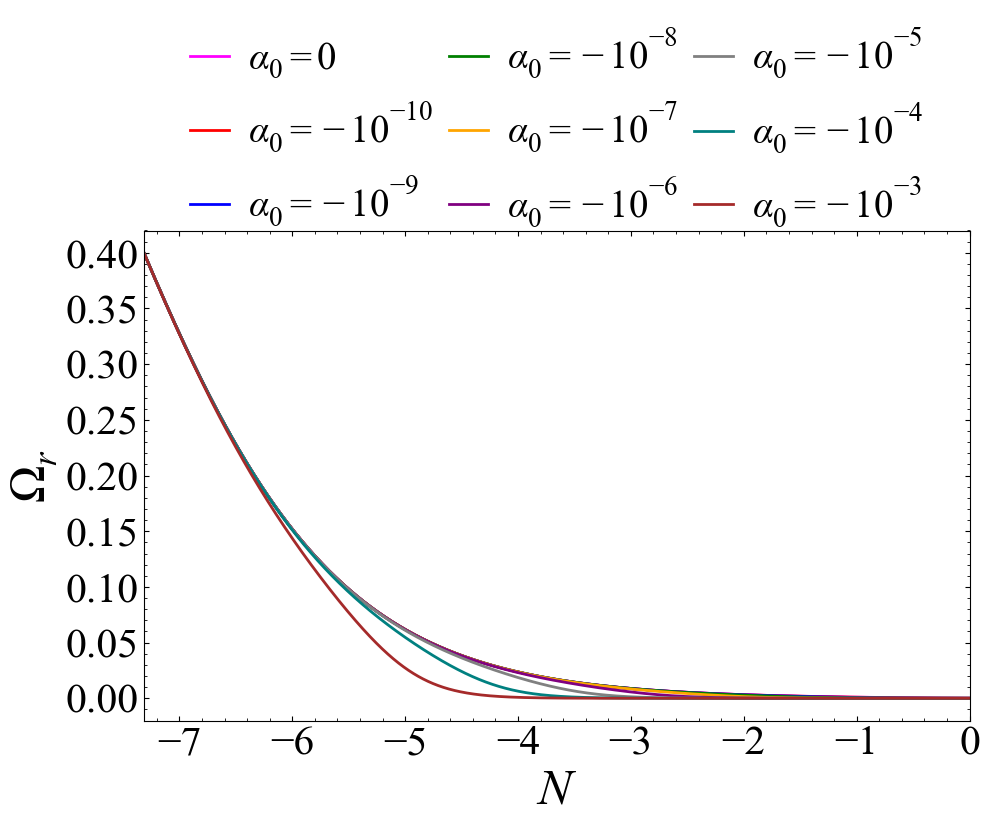}
\endminipage
\caption{Evolution of the radiation energy density parameter $\Omega_r$ for the kinetic-coupling k-essence model \eqref{Eq:Numerical evolution:k-essence assumption} for two different sets of initial conditions (I.C.). \emph{Left panel}: I.C. \eqref{Eq:Numerical evolution:k-essence:Kinetic-coupling:Initial conditions}, \emph{Right panel}: I.C. \eqref{Appendix:Eq:Numerical evolution:k-essence:Kinetic-coupling:Initial conditions02}.}
\label{App:Fig:Omega_r}
\end{figure}
On the other hand, the positive interactions lead to the opposite scenario. The scaled Hubble parameter $H/H_0$ plotted in \ref{App:Fig:Hubble}, has a larger value for negative interactions and a smaller value for positive interactions, in the past, compared to the non-interacting scenario. Thus, we find that the same model can predict a markedly fast transition as well as a slower transition from matter-dominated to DE-dominated era depending on the initial conditions.


\begin{thebibliography}{83}%
\makeatletter
\providecommand \@ifxundefined [1]{%
 \@ifx{#1\undefined}
}%
\providecommand \@ifnum [1]{%
 \ifnum #1\expandafter \@firstoftwo
 \else \expandafter \@secondoftwo
 \fi
}%
\providecommand \@ifx [1]{%
 \ifx #1\expandafter \@firstoftwo
 \else \expandafter \@secondoftwo
 \fi
}%
\providecommand \natexlab [1]{#1}%
\providecommand \enquote  [1]{``#1''}%
\providecommand \bibnamefont  [1]{#1}%
\providecommand \bibfnamefont [1]{#1}%
\providecommand \citenamefont [1]{#1}%
\providecommand \href@noop [0]{\@secondoftwo}%
\providecommand \href [0]{\begingroup \@sanitize@url \@href}%
\providecommand \@href[1]{\@@startlink{#1}\@@href}%
\providecommand \@@href[1]{\endgroup#1\@@endlink}%
\providecommand \@sanitize@url [0]{\catcode `\\12\catcode `\$12\catcode `\&12\catcode `\#12\catcode `\^12\catcode `\_12\catcode `\%12\relax}%
\providecommand \@@startlink[1]{}%
\providecommand \@@endlink[0]{}%
\providecommand \url  [0]{\begingroup\@sanitize@url \@url }%
\providecommand \@url [1]{\endgroup\@href {#1}{\urlprefix }}%
\providecommand \urlprefix  [0]{URL }%
\providecommand \Eprint [0]{\href }%
\providecommand \doibase [0]{http://dx.doi.org/}%
\providecommand \selectlanguage [0]{\@gobble}%
\providecommand \bibinfo  [0]{\@secondoftwo}%
\providecommand \bibfield  [0]{\@secondoftwo}%
\providecommand \translation [1]{[#1]}%
\providecommand \BibitemOpen [0]{}%
\providecommand \bibitemStop [0]{}%
\providecommand \bibitemNoStop [0]{.\EOS\space}%
\providecommand \EOS [0]{\spacefactor3000\relax}%
\providecommand \BibitemShut  [1]{\csname bibitem#1\endcsname}%
\let\auto@bib@innerbib\@empty
\bibitem [{\citenamefont {Johnson}\ and\ \citenamefont {Shankaranarayanan}(2021)}]{2021-Johnson.Shankaranarayanan-Phys.Rev.D}%
  \BibitemOpen
  \bibfield  {author} {\bibinfo {author} {\bibfnamefont {J.~P.}\ \bibnamefont {Johnson}}\ and\ \bibinfo {author} {\bibfnamefont {S.}~\bibnamefont {Shankaranarayanan}},\ }\href {\doibase 10.1103/PhysRevD.103.023510} {\bibfield  {journal} {\bibinfo  {journal} {Phys. Rev. D}\ }\textbf {\bibinfo {volume} {103}},\ \bibinfo {pages} {023510} (\bibinfo {year} {2021})},\ \Eprint {http://arxiv.org/abs/2006.04618} {arXiv:2006.04618 [gr-qc]} \BibitemShut {NoStop}%
\bibitem [{\citenamefont {Padmanabhan}(2000)}]{2000-Padmanabhan-TheoreticalAstrophysicsVolume}%
  \BibitemOpen
  \bibfield  {author} {\bibinfo {author} {\bibfnamefont {T.}~\bibnamefont {Padmanabhan}},\ }\href {https://books.google.co.in/books?id=-yvq5BEsFvcC} {\emph {\bibinfo {title} {Theoretical Astrophysics: Volume 3, Galaxies and Cosmology}}},\ Theoretical Astrophysics\ (\bibinfo  {publisher} {Cambridge University Press},\ \bibinfo {year} {2000})\BibitemShut {NoStop}%
\bibitem [{\citenamefont {Mukhanov}(2005)}]{2005-Mukhanov-PhysicalFoundationsCosmology}%
  \BibitemOpen
  \bibfield  {author} {\bibinfo {author} {\bibfnamefont {V.}~\bibnamefont {Mukhanov}},\ }\href {\doibase 10.1017/CBO9780511790553} {\emph {\bibinfo {title} {Physical Foundations of Cosmology}}}\ (\bibinfo  {publisher} {Cambridge University Press},\ \bibinfo {year} {2005})\BibitemShut {NoStop}%
\bibitem [{\citenamefont {Weinberg}(2008)}]{2008-Weinberg-Cosmology}%
  \BibitemOpen
  \bibfield  {author} {\bibinfo {author} {\bibfnamefont {S.}~\bibnamefont {Weinberg}},\ }\href {https://global.oup.com/academic/product/cosmology-9780198526827?cc=us&lang=en&} {\emph {\bibinfo {title} {Cosmology}}},\ Cosmology\ (\bibinfo  {publisher} {OUP Oxford},\ \bibinfo {year} {2008})\BibitemShut {NoStop}%
\bibitem [{\citenamefont {Peebles}\ and\ \citenamefont {Ratra}(2003)}]{Peebles:2002gy}%
  \BibitemOpen
  \bibfield  {author} {\bibinfo {author} {\bibfnamefont {P.~J.~E.}\ \bibnamefont {Peebles}}\ and\ \bibinfo {author} {\bibfnamefont {B.}~\bibnamefont {Ratra}},\ }\href {\doibase 10.1103/RevModPhys.75.559} {\bibfield  {journal} {\bibinfo  {journal} {Rev. Mod. Phys.}\ }\textbf {\bibinfo {volume} {75}},\ \bibinfo {pages} {559} (\bibinfo {year} {2003})},\ \Eprint {http://arxiv.org/abs/astro-ph/0207347} {arXiv:astro-ph/0207347} \BibitemShut {NoStop}%
\bibitem [{\citenamefont {Padmanabhan}(2003)}]{Padmanabhan:2002ji}%
  \BibitemOpen
  \bibfield  {author} {\bibinfo {author} {\bibfnamefont {T.}~\bibnamefont {Padmanabhan}},\ }\href {\doibase 10.1016/S0370-1573(03)00120-0} {\bibfield  {journal} {\bibinfo  {journal} {Phys. Rept.}\ }\textbf {\bibinfo {volume} {380}},\ \bibinfo {pages} {235} (\bibinfo {year} {2003})},\ \Eprint {http://arxiv.org/abs/hep-th/0212290} {arXiv:hep-th/0212290} \BibitemShut {NoStop}%
\bibitem [{\citenamefont {Tsujikawa}(2010)}]{Tsujikawa:2010zza}%
  \BibitemOpen
  \bibfield  {author} {\bibinfo {author} {\bibfnamefont {S.}~\bibnamefont {Tsujikawa}},\ }\href {\doibase 10.1007/978-3-642-10598-2_3} {\bibfield  {journal} {\bibinfo  {journal} {Lect. Notes Phys.}\ }\textbf {\bibinfo {volume} {800}},\ \bibinfo {pages} {99} (\bibinfo {year} {2010})},\ \Eprint {http://arxiv.org/abs/1101.0191} {arXiv:1101.0191 [gr-qc]} \BibitemShut {NoStop}%
\bibitem [{\citenamefont {Bertone}\ and\ \citenamefont {Hooper}(2018)}]{Bertone:2016nfn}%
  \BibitemOpen
  \bibfield  {author} {\bibinfo {author} {\bibfnamefont {G.}~\bibnamefont {Bertone}}\ and\ \bibinfo {author} {\bibfnamefont {D.}~\bibnamefont {Hooper}},\ }\href {\doibase 10.1103/RevModPhys.90.045002} {\bibfield  {journal} {\bibinfo  {journal} {Rev. Mod. Phys.}\ }\textbf {\bibinfo {volume} {90}},\ \bibinfo {pages} {045002} (\bibinfo {year} {2018})},\ \Eprint {http://arxiv.org/abs/1605.04909} {arXiv:1605.04909 [astro-ph.CO]} \BibitemShut {NoStop}%
\bibitem [{\citenamefont {Joyce}\ \emph {et~al.}(2016)\citenamefont {Joyce}, \citenamefont {Lombriser},\ and\ \citenamefont {Schmidt}}]{Joyce:2016vqv}%
  \BibitemOpen
  \bibfield  {author} {\bibinfo {author} {\bibfnamefont {A.}~\bibnamefont {Joyce}}, \bibinfo {author} {\bibfnamefont {L.}~\bibnamefont {Lombriser}}, \ and\ \bibinfo {author} {\bibfnamefont {F.}~\bibnamefont {Schmidt}},\ }\href {\doibase 10.1146/annurev-nucl-102115-044553} {\bibfield  {journal} {\bibinfo  {journal} {Ann. Rev. Nucl. Part. Sci.}\ }\textbf {\bibinfo {volume} {66}},\ \bibinfo {pages} {95} (\bibinfo {year} {2016})},\ \Eprint {http://arxiv.org/abs/1601.06133} {arXiv:1601.06133 [astro-ph.CO]} \BibitemShut {NoStop}%
\bibitem [{\citenamefont {Sofue}(2020)}]{Sofue:2020rnl}%
  \BibitemOpen
  \bibfield  {author} {\bibinfo {author} {\bibfnamefont {Y.}~\bibnamefont {Sofue}},\ }\href {\doibase 10.3390/galaxies8020037} {\bibfield  {journal} {\bibinfo  {journal} {Galaxies}\ }\textbf {\bibinfo {volume} {8}},\ \bibinfo {pages} {37} (\bibinfo {year} {2020})},\ \Eprint {http://arxiv.org/abs/2004.11688} {arXiv:2004.11688 [astro-ph.GA]} \BibitemShut {NoStop}%
\bibitem [{\citenamefont {Shankaranarayanan}\ and\ \citenamefont {Johnson}(2022)}]{2022-Shanki.Joseph-GRG}%
  \BibitemOpen
  \bibfield  {author} {\bibinfo {author} {\bibfnamefont {S.}~\bibnamefont {Shankaranarayanan}}\ and\ \bibinfo {author} {\bibfnamefont {J.~P.}\ \bibnamefont {Johnson}},\ }\href {\doibase 10.1007/s10714-022-02927-2} {\bibfield  {journal} {\bibinfo  {journal} {Gen. Rel. Grav.}\ }\textbf {\bibinfo {volume} {54}},\ \bibinfo {pages} {44} (\bibinfo {year} {2022})},\ \Eprint {http://arxiv.org/abs/2204.06533} {arXiv:2204.06533 [gr-qc]} \BibitemShut {NoStop}%
\bibitem [{\citenamefont {Weinberg}(1989)}]{Weinberg:1988cp}%
  \BibitemOpen
  \bibfield  {author} {\bibinfo {author} {\bibfnamefont {S.}~\bibnamefont {Weinberg}},\ }\href {\doibase 10.1103/RevModPhys.61.1} {\bibfield  {journal} {\bibinfo  {journal} {Rev. Mod. Phys.}\ }\textbf {\bibinfo {volume} {61}},\ \bibinfo {pages} {1} (\bibinfo {year} {1989})}\BibitemShut {NoStop}%
\bibitem [{\citenamefont {Copeland}\ \emph {et~al.}(2006)\citenamefont {Copeland}, \citenamefont {Sami},\ and\ \citenamefont {Tsujikawa}}]{2006-Copeland.etal-Int.J.Mod.Phys.}%
  \BibitemOpen
  \bibfield  {author} {\bibinfo {author} {\bibfnamefont {E.~J.}\ \bibnamefont {Copeland}}, \bibinfo {author} {\bibfnamefont {M.}~\bibnamefont {Sami}}, \ and\ \bibinfo {author} {\bibfnamefont {S.}~\bibnamefont {Tsujikawa}},\ }\href {\doibase 10.1142/S021827180600942X} {\bibfield  {journal} {\bibinfo  {journal} {Int. J. Mod. Phys.}\ }\textbf {\bibinfo {volume} {D15}},\ \bibinfo {pages} {1753} (\bibinfo {year} {2006})},\ \Eprint {http://arxiv.org/abs/hep-th/0603057} {arXiv:hep-th/0603057 [hep-th]} \BibitemShut {NoStop}%
\bibitem [{\citenamefont {Riess}\ and\ \citenamefont {Others}(1998)}]{1998-Riess.Others-Astron.J.}%
  \BibitemOpen
  \bibfield  {author} {\bibinfo {author} {\bibfnamefont {A.~G.}\ \bibnamefont {Riess}}\ and\ \bibinfo {author} {\bibnamefont {Others}},\ }\href {\doibase 10.1086/300499} {\bibfield  {journal} {\bibinfo  {journal} {Astron. J.}\ }\textbf {\bibinfo {volume} {116}},\ \bibinfo {pages} {1009} (\bibinfo {year} {1998})},\ \Eprint {http://arxiv.org/abs/astro-ph/9805201} {arXiv:astro-ph/9805201 [astro-ph]} \BibitemShut {NoStop}%
\bibitem [{\citenamefont {Perlmutter}\ and\ \citenamefont {Others}(1999)}]{1999-Perlmutter.Others-Astrophys.J.}%
  \BibitemOpen
  \bibfield  {author} {\bibinfo {author} {\bibfnamefont {S.}~\bibnamefont {Perlmutter}}\ and\ \bibinfo {author} {\bibnamefont {Others}},\ }\href {\doibase 10.1086/307221} {\bibfield  {journal} {\bibinfo  {journal} {Astrophys. J.}\ }\textbf {\bibinfo {volume} {517}},\ \bibinfo {pages} {565} (\bibinfo {year} {1999})},\ \Eprint {http://arxiv.org/abs/astro-ph/9812133} {arXiv:astro-ph/9812133 [astro-ph]} \BibitemShut {NoStop}%
\bibitem [{\citenamefont {Spergel}\ and\ \citenamefont {Others}(2007)}]{2007-Spergel.Others-Astrophys.J.Suppl.}%
  \BibitemOpen
  \bibfield  {author} {\bibinfo {author} {\bibfnamefont {D.~N.}\ \bibnamefont {Spergel}}\ and\ \bibinfo {author} {\bibnamefont {Others}},\ }\href {\doibase 10.1086/513700} {\bibfield  {journal} {\bibinfo  {journal} {Astrophys. J. Suppl.}\ }\textbf {\bibinfo {volume} {170}},\ \bibinfo {pages} {377} (\bibinfo {year} {2007})},\ \Eprint {http://arxiv.org/abs/astro-ph/0603449} {arXiv:astro-ph/0603449 [astro-ph]} \BibitemShut {NoStop}%
\bibitem [{\citenamefont {Akrami}\ and\ \citenamefont {Others}(2018)}]{2018-Akrami.Others-Astron.Astrophys.}%
  \BibitemOpen
  \bibfield  {author} {\bibinfo {author} {\bibfnamefont {Y.}~\bibnamefont {Akrami}}\ and\ \bibinfo {author} {\bibnamefont {Others}} (\bibinfo {collaboration} {Planck}),\ }\href {\doibase 10.1051/0004-6361/201833880} {\bibfield  {journal} {\bibinfo  {journal} {Astron. Astrophys.}\ }\textbf {\bibinfo {volume} {641}},\ \bibinfo {pages} {A1} (\bibinfo {year} {2018})},\ \Eprint {http://arxiv.org/abs/1807.06205} {arXiv:1807.06205 [astro-ph.CO]} \BibitemShut {NoStop}%
\bibitem [{\citenamefont {Aghanim}\ \emph {et~al.}(2020)\citenamefont {Aghanim} \emph {et~al.}}]{2020-Aghanim.others-Astron.Astrophys.}%
  \BibitemOpen
  \bibfield  {author} {\bibinfo {author} {\bibfnamefont {N.}~\bibnamefont {Aghanim}} \emph {et~al.} (\bibinfo {collaboration} {Planck}),\ }\href {\doibase 10.1051/0004-6361/201833910} {\bibfield  {journal} {\bibinfo  {journal} {Astron. Astrophys.}\ }\textbf {\bibinfo {volume} {641}},\ \bibinfo {pages} {A6} (\bibinfo {year} {2020})},\ \Eprint {http://arxiv.org/abs/1807.06209} {arXiv:1807.06209 [astro-ph.CO]} \BibitemShut {NoStop}%
\bibitem [{\citenamefont {Brout}\ \emph {et~al.}(2022)\citenamefont {Brout} \emph {et~al.}}]{2022-Brout.others-Astrophys.J.}%
  \BibitemOpen
  \bibfield  {author} {\bibinfo {author} {\bibfnamefont {D.}~\bibnamefont {Brout}} \emph {et~al.},\ }\href {\doibase 10.3847/1538-4357/ac8e04} {\bibfield  {journal} {\bibinfo  {journal} {Astrophys. J.}\ }\textbf {\bibinfo {volume} {938}},\ \bibinfo {pages} {110} (\bibinfo {year} {2022})},\ \Eprint {http://arxiv.org/abs/2202.04077} {arXiv:2202.04077 [astro-ph.CO]} \BibitemShut {NoStop}%
\bibitem [{\citenamefont {Perivolaropoulos}\ and\ \citenamefont {Skara}(2022)}]{2022-Perivolaropoulos.Skara-NewAstron.Rev.}%
  \BibitemOpen
  \bibfield  {author} {\bibinfo {author} {\bibfnamefont {L.}~\bibnamefont {Perivolaropoulos}}\ and\ \bibinfo {author} {\bibfnamefont {F.}~\bibnamefont {Skara}},\ }\href {\doibase 10.1016/j.newar.2022.101659} {\bibfield  {journal} {\bibinfo  {journal} {New Astron. Rev.}\ }\textbf {\bibinfo {volume} {95}},\ \bibinfo {pages} {101659} (\bibinfo {year} {2022})},\ \Eprint {http://arxiv.org/abs/2105.05208} {arXiv:2105.05208 [astro-ph.CO]} \BibitemShut {NoStop}%
\bibitem [{\citenamefont {Marra}\ \emph {et~al.}(2013)\citenamefont {Marra}, \citenamefont {Amendola}, \citenamefont {Sawicki},\ and\ \citenamefont {Valkenburg}}]{2013-Marra.etal-Phys.Rev.Lett.}%
  \BibitemOpen
  \bibfield  {author} {\bibinfo {author} {\bibfnamefont {V.}~\bibnamefont {Marra}}, \bibinfo {author} {\bibfnamefont {L.}~\bibnamefont {Amendola}}, \bibinfo {author} {\bibfnamefont {I.}~\bibnamefont {Sawicki}}, \ and\ \bibinfo {author} {\bibfnamefont {W.}~\bibnamefont {Valkenburg}},\ }\href {\doibase 10.1103/PhysRevLett.110.241305} {\bibfield  {journal} {\bibinfo  {journal} {Phys. Rev. Lett.}\ }\textbf {\bibinfo {volume} {110}},\ \bibinfo {pages} {241305} (\bibinfo {year} {2013})},\ \Eprint {http://arxiv.org/abs/1303.3121} {arXiv:1303.3121 [astro-ph.CO]} \BibitemShut {NoStop}%
\bibitem [{\citenamefont {Bennett}\ \emph {et~al.}(2014)\citenamefont {Bennett}, \citenamefont {Larson}, \citenamefont {Weiland},\ and\ \citenamefont {Hinshaw}}]{2014-Bennett.etal-Astrophys.J.}%
  \BibitemOpen
  \bibfield  {author} {\bibinfo {author} {\bibfnamefont {C.~L.}\ \bibnamefont {Bennett}}, \bibinfo {author} {\bibfnamefont {D.}~\bibnamefont {Larson}}, \bibinfo {author} {\bibfnamefont {J.~L.}\ \bibnamefont {Weiland}}, \ and\ \bibinfo {author} {\bibfnamefont {G.}~\bibnamefont {Hinshaw}},\ }\href {\doibase 10.1088/0004-637X/794/2/135} {\bibfield  {journal} {\bibinfo  {journal} {Astrophys. J.}\ }\textbf {\bibinfo {volume} {794}},\ \bibinfo {pages} {135} (\bibinfo {year} {2014})},\ \Eprint {http://arxiv.org/abs/1406.1718} {arXiv:1406.1718 [astro-ph.CO]} \BibitemShut {NoStop}%
\bibitem [{\citenamefont {Riess}\ \emph {et~al.}(2016)\citenamefont {Riess} \emph {et~al.}}]{2016-Riess.others-Astrophys.J.}%
  \BibitemOpen
  \bibfield  {author} {\bibinfo {author} {\bibfnamefont {A.~G.}\ \bibnamefont {Riess}} \emph {et~al.},\ }\href {\doibase 10.3847/0004-637X/826/1/56} {\bibfield  {journal} {\bibinfo  {journal} {Astrophys. J.}\ }\textbf {\bibinfo {volume} {826}},\ \bibinfo {pages} {56} (\bibinfo {year} {2016})},\ \Eprint {http://arxiv.org/abs/1604.01424} {arXiv:1604.01424 [astro-ph.CO]} \BibitemShut {NoStop}%
\bibitem [{\citenamefont {Riess}\ \emph {et~al.}(2019)\citenamefont {Riess}, \citenamefont {Casertano}, \citenamefont {Yuan}, \citenamefont {Macri},\ and\ \citenamefont {Scolnic}}]{2019-Riess.etal-Astrophys.J.}%
  \BibitemOpen
  \bibfield  {author} {\bibinfo {author} {\bibfnamefont {A.~G.}\ \bibnamefont {Riess}}, \bibinfo {author} {\bibfnamefont {S.}~\bibnamefont {Casertano}}, \bibinfo {author} {\bibfnamefont {W.}~\bibnamefont {Yuan}}, \bibinfo {author} {\bibfnamefont {L.~M.}\ \bibnamefont {Macri}}, \ and\ \bibinfo {author} {\bibfnamefont {D.}~\bibnamefont {Scolnic}},\ }\href {\doibase 10.3847/1538-4357/ab1422} {\bibfield  {journal} {\bibinfo  {journal} {Astrophys. J.}\ }\textbf {\bibinfo {volume} {876}},\ \bibinfo {pages} {85} (\bibinfo {year} {2019})},\ \Eprint {http://arxiv.org/abs/1903.07603} {arXiv:1903.07603 [astro-ph.CO]} \BibitemShut {NoStop}%
\bibitem [{\citenamefont {Gupta}(2023)}]{2023-Gupta-Mon.Not.Roy.Astron.Soc.}%
  \BibitemOpen
  \bibfield  {author} {\bibinfo {author} {\bibfnamefont {R.~P.}\ \bibnamefont {Gupta}},\ }\href {\doibase 10.1093/mnras/stad2032} {\bibfield  {journal} {\bibinfo  {journal} {Mon. Not. Roy. Astron. Soc.}\ }\textbf {\bibinfo {volume} {524}},\ \bibinfo {pages} {3385} (\bibinfo {year} {2023})},\ \Eprint {http://arxiv.org/abs/2309.13100} {arXiv:2309.13100 [astro-ph.CO]} \BibitemShut {NoStop}%
\bibitem [{\citenamefont {Di~Valentino}\ \emph {et~al.}(2021{\natexlab{a}})\citenamefont {Di~Valentino}, \citenamefont {Mena}, \citenamefont {Pan}, \citenamefont {Visinelli}, \citenamefont {Yang}, \citenamefont {Melchiorri}, \citenamefont {Mota}, \citenamefont {Riess},\ and\ \citenamefont {Silk}}]{2021-DiValentino.etal-CQG}%
  \BibitemOpen
  \bibfield  {author} {\bibinfo {author} {\bibfnamefont {E.}~\bibnamefont {Di~Valentino}}, \bibinfo {author} {\bibfnamefont {O.}~\bibnamefont {Mena}}, \bibinfo {author} {\bibfnamefont {S.}~\bibnamefont {Pan}}, \bibinfo {author} {\bibfnamefont {L.}~\bibnamefont {Visinelli}}, \bibinfo {author} {\bibfnamefont {W.}~\bibnamefont {Yang}}, \bibinfo {author} {\bibfnamefont {A.}~\bibnamefont {Melchiorri}}, \bibinfo {author} {\bibfnamefont {D.~F.}\ \bibnamefont {Mota}}, \bibinfo {author} {\bibfnamefont {A.~G.}\ \bibnamefont {Riess}}, \ and\ \bibinfo {author} {\bibfnamefont {J.}~\bibnamefont {Silk}},\ }\href {\doibase 10.1088/1361-6382/ac086d} {\bibfield  {journal} {\bibinfo  {journal} {Class. Quant. Grav.}\ }\textbf {\bibinfo {volume} {38}},\ \bibinfo {pages} {153001} (\bibinfo {year} {2021}{\natexlab{a}})},\ \Eprint {http://arxiv.org/abs/2103.01183} {arXiv:2103.01183 [astro-ph.CO]} \BibitemShut {NoStop}%
\bibitem [{\citenamefont {Kamionkowski}\ and\ \citenamefont {Riess}(2023)}]{Kamionkowski:2022pkx}%
  \BibitemOpen
  \bibfield  {author} {\bibinfo {author} {\bibfnamefont {M.}~\bibnamefont {Kamionkowski}}\ and\ \bibinfo {author} {\bibfnamefont {A.~G.}\ \bibnamefont {Riess}},\ }\href {\doibase 10.1146/annurev-nucl-111422-024107} {\bibfield  {journal} {\bibinfo  {journal} {Ann. Rev. Nucl. Part. Sci.}\ }\textbf {\bibinfo {volume} {73}},\ \bibinfo {pages} {153} (\bibinfo {year} {2023})},\ \Eprint {http://arxiv.org/abs/2211.04492} {arXiv:2211.04492 [astro-ph.CO]} \BibitemShut {NoStop}%
\bibitem [{\citenamefont {Chevallier}\ and\ \citenamefont {Polarski}(2001)}]{2001-Chevallier.Polarski-Int.J.Mod.Phys.D}%
  \BibitemOpen
  \bibfield  {author} {\bibinfo {author} {\bibfnamefont {M.}~\bibnamefont {Chevallier}}\ and\ \bibinfo {author} {\bibfnamefont {D.}~\bibnamefont {Polarski}},\ }\href {\doibase 10.1142/S0218271801000822} {\bibfield  {journal} {\bibinfo  {journal} {Int. J. Mod. Phys. D}\ }\textbf {\bibinfo {volume} {10}},\ \bibinfo {pages} {213} (\bibinfo {year} {2001})},\ \Eprint {http://arxiv.org/abs/gr-qc/0009008} {arXiv:gr-qc/0009008} \BibitemShut {NoStop}%
\bibitem [{\citenamefont {Linder}(2003)}]{2003-Linder-Phys.Rev.Lett.}%
  \BibitemOpen
  \bibfield  {author} {\bibinfo {author} {\bibfnamefont {E.~V.}\ \bibnamefont {Linder}},\ }\href {\doibase 10.1103/PhysRevLett.90.091301} {\bibfield  {journal} {\bibinfo  {journal} {Phys. Rev. Lett.}\ }\textbf {\bibinfo {volume} {90}},\ \bibinfo {pages} {091301} (\bibinfo {year} {2003})},\ \Eprint {http://arxiv.org/abs/astro-ph/0208512} {arXiv:astro-ph/0208512} \BibitemShut {NoStop}%
\bibitem [{\citenamefont {Tripathi}\ \emph {et~al.}(2017)\citenamefont {Tripathi}, \citenamefont {Sangwan},\ and\ \citenamefont {Jassal}}]{2017-Tripathi.etal-JCAP}%
  \BibitemOpen
  \bibfield  {author} {\bibinfo {author} {\bibfnamefont {A.}~\bibnamefont {Tripathi}}, \bibinfo {author} {\bibfnamefont {A.}~\bibnamefont {Sangwan}}, \ and\ \bibinfo {author} {\bibfnamefont {H.~K.}\ \bibnamefont {Jassal}},\ }\href {\doibase 10.1088/1475-7516/2017/06/012} {\bibfield  {journal} {\bibinfo  {journal} {JCAP}\ }\textbf {\bibinfo {volume} {06}},\ \bibinfo {pages} {012} (\bibinfo {year} {2017})},\ \Eprint {http://arxiv.org/abs/1611.01899} {arXiv:1611.01899 [astro-ph.CO]} \BibitemShut {NoStop}%
\bibitem [{\citenamefont {Ratra}\ and\ \citenamefont {Peebles}(1988)}]{1988-Ratra.Peebles-Phys.Rev.D}%
  \BibitemOpen
  \bibfield  {author} {\bibinfo {author} {\bibfnamefont {B.}~\bibnamefont {Ratra}}\ and\ \bibinfo {author} {\bibfnamefont {P.~J.~E.}\ \bibnamefont {Peebles}},\ }\href {\doibase 10.1103/PhysRevD.37.3406} {\bibfield  {journal} {\bibinfo  {journal} {Phys. Rev. D}\ }\textbf {\bibinfo {volume} {37}},\ \bibinfo {pages} {3406} (\bibinfo {year} {1988})}\BibitemShut {NoStop}%
\bibitem [{\citenamefont {Armendariz-Picon}\ \emph {et~al.}(2000)\citenamefont {Armendariz-Picon}, \citenamefont {Mukhanov},\ and\ \citenamefont {Steinhardt}}]{2000-Armendariz-Picon.etal-Phys.Rev.Lett.}%
  \BibitemOpen
  \bibfield  {author} {\bibinfo {author} {\bibfnamefont {C.}~\bibnamefont {Armendariz-Picon}}, \bibinfo {author} {\bibfnamefont {V.~F.}\ \bibnamefont {Mukhanov}}, \ and\ \bibinfo {author} {\bibfnamefont {P.~J.}\ \bibnamefont {Steinhardt}},\ }\href {\doibase 10.1103/PhysRevLett.85.4438} {\bibfield  {journal} {\bibinfo  {journal} {Phys. Rev. Lett.}\ }\textbf {\bibinfo {volume} {85}},\ \bibinfo {pages} {4438} (\bibinfo {year} {2000})},\ \Eprint {http://arxiv.org/abs/astro-ph/0004134} {arXiv:astro-ph/0004134} \BibitemShut {NoStop}%
\bibitem [{\citenamefont {Armendariz-Picon}\ \emph {et~al.}(2001)\citenamefont {Armendariz-Picon}, \citenamefont {Mukhanov},\ and\ \citenamefont {Steinhardt}}]{2001-Armendariz-Picon.etal-Phys.Rev.D}%
  \BibitemOpen
  \bibfield  {author} {\bibinfo {author} {\bibfnamefont {C.}~\bibnamefont {Armendariz-Picon}}, \bibinfo {author} {\bibfnamefont {V.~F.}\ \bibnamefont {Mukhanov}}, \ and\ \bibinfo {author} {\bibfnamefont {P.~J.}\ \bibnamefont {Steinhardt}},\ }\href {\doibase 10.1103/PhysRevD.63.103510} {\bibfield  {journal} {\bibinfo  {journal} {Phys. Rev. D}\ }\textbf {\bibinfo {volume} {63}},\ \bibinfo {pages} {103510} (\bibinfo {year} {2001})},\ \Eprint {http://arxiv.org/abs/astro-ph/0006373} {arXiv:astro-ph/0006373} \BibitemShut {NoStop}%
\bibitem [{\citenamefont {Rajvanshi}\ \emph {et~al.}(2021)\citenamefont {Rajvanshi}, \citenamefont {Singh}, \citenamefont {Jassal},\ and\ \citenamefont {Bagla}}]{2021-Rajvanshi.etal-Class.Quant.Grav.}%
  \BibitemOpen
  \bibfield  {author} {\bibinfo {author} {\bibfnamefont {M.~P.}\ \bibnamefont {Rajvanshi}}, \bibinfo {author} {\bibfnamefont {A.}~\bibnamefont {Singh}}, \bibinfo {author} {\bibfnamefont {H.~K.}\ \bibnamefont {Jassal}}, \ and\ \bibinfo {author} {\bibfnamefont {J.~S.}\ \bibnamefont {Bagla}},\ }\href {\doibase 10.1088/1361-6382/ac1b49} {\bibfield  {journal} {\bibinfo  {journal} {Class. Quant. Grav.}\ }\textbf {\bibinfo {volume} {38}},\ \bibinfo {pages} {195001} (\bibinfo {year} {2021})},\ \Eprint {http://arxiv.org/abs/2104.00982} {arXiv:2104.00982 [astro-ph.CO]} \BibitemShut {NoStop}%
\bibitem [{\citenamefont {Bamba}\ \emph {et~al.}(2012)\citenamefont {Bamba}, \citenamefont {Capozziello}, \citenamefont {Nojiri},\ and\ \citenamefont {Odintsov}}]{Bamba:2012cp}%
  \BibitemOpen
  \bibfield  {author} {\bibinfo {author} {\bibfnamefont {K.}~\bibnamefont {Bamba}}, \bibinfo {author} {\bibfnamefont {S.}~\bibnamefont {Capozziello}}, \bibinfo {author} {\bibfnamefont {S.}~\bibnamefont {Nojiri}}, \ and\ \bibinfo {author} {\bibfnamefont {S.~D.}\ \bibnamefont {Odintsov}},\ }\href {\doibase 10.1007/s10509-012-1181-8} {\bibfield  {journal} {\bibinfo  {journal} {Astrophys. Space Sci.}\ }\textbf {\bibinfo {volume} {342}},\ \bibinfo {pages} {155} (\bibinfo {year} {2012})},\ \Eprint {http://arxiv.org/abs/1205.3421} {arXiv:1205.3421 [gr-qc]} \BibitemShut {NoStop}%
\bibitem [{\citenamefont {Sch\"oneberg}\ \emph {et~al.}(2022{\natexlab{a}})\citenamefont {Sch\"oneberg}, \citenamefont {Franco~Abell\'an}, \citenamefont {P\'erez~S\'anchez}, \citenamefont {Witte}, \citenamefont {Poulin},\ and\ \citenamefont {Lesgourgues}}]{Schoneberg:2021qvd}%
  \BibitemOpen
  \bibfield  {author} {\bibinfo {author} {\bibfnamefont {N.}~\bibnamefont {Sch\"oneberg}}, \bibinfo {author} {\bibfnamefont {G.}~\bibnamefont {Franco~Abell\'an}}, \bibinfo {author} {\bibfnamefont {A.}~\bibnamefont {P\'erez~S\'anchez}}, \bibinfo {author} {\bibfnamefont {S.~J.}\ \bibnamefont {Witte}}, \bibinfo {author} {\bibfnamefont {V.}~\bibnamefont {Poulin}}, \ and\ \bibinfo {author} {\bibfnamefont {J.}~\bibnamefont {Lesgourgues}},\ }\href {\doibase 10.1016/j.physrep.2022.07.001} {\bibfield  {journal} {\bibinfo  {journal} {Phys. Rept.}\ }\textbf {\bibinfo {volume} {984}},\ \bibinfo {pages} {1} (\bibinfo {year} {2022}{\natexlab{a}})},\ \Eprint {http://arxiv.org/abs/2107.10291} {arXiv:2107.10291 [astro-ph.CO]} \BibitemShut {NoStop}%
\bibitem [{\citenamefont {Abdalla}\ \emph {et~al.}(2022)\citenamefont {Abdalla} \emph {et~al.}}]{Abdalla:2022yfr}%
  \BibitemOpen
  \bibfield  {author} {\bibinfo {author} {\bibfnamefont {E.}~\bibnamefont {Abdalla}} \emph {et~al.},\ }\href {\doibase 10.1016/j.jheap.2022.04.002} {\bibfield  {journal} {\bibinfo  {journal} {JHEAp}\ }\textbf {\bibinfo {volume} {34}},\ \bibinfo {pages} {49} (\bibinfo {year} {2022})},\ \Eprint {http://arxiv.org/abs/2203.06142} {arXiv:2203.06142 [astro-ph.CO]} \BibitemShut {NoStop}%
\bibitem [{\citenamefont {Wang}\ \emph {et~al.}(2016)\citenamefont {Wang}, \citenamefont {Abdalla}, \citenamefont {Atrio-Barandela},\ and\ \citenamefont {Pavon}}]{Wang:2016lxa}%
  \BibitemOpen
  \bibfield  {author} {\bibinfo {author} {\bibfnamefont {B.}~\bibnamefont {Wang}}, \bibinfo {author} {\bibfnamefont {E.}~\bibnamefont {Abdalla}}, \bibinfo {author} {\bibfnamefont {F.}~\bibnamefont {Atrio-Barandela}}, \ and\ \bibinfo {author} {\bibfnamefont {D.}~\bibnamefont {Pavon}},\ }\href {\doibase 10.1088/0034-4885/79/9/096901} {\bibfield  {journal} {\bibinfo  {journal} {Rept. Prog. Phys.}\ }\textbf {\bibinfo {volume} {79}},\ \bibinfo {pages} {096901} (\bibinfo {year} {2016})},\ \Eprint {http://arxiv.org/abs/1603.08299} {arXiv:1603.08299 [astro-ph.CO]} \BibitemShut {NoStop}%
\bibitem [{\citenamefont {Wang}\ \emph {et~al.}(2024{\natexlab{a}})\citenamefont {Wang}, \citenamefont {Abdalla}, \citenamefont {Atrio-Barandela},\ and\ \citenamefont {Pav\'on}}]{Wang:2024vmw}%
  \BibitemOpen
  \bibfield  {author} {\bibinfo {author} {\bibfnamefont {B.}~\bibnamefont {Wang}}, \bibinfo {author} {\bibfnamefont {E.}~\bibnamefont {Abdalla}}, \bibinfo {author} {\bibfnamefont {F.}~\bibnamefont {Atrio-Barandela}}, \ and\ \bibinfo {author} {\bibfnamefont {D.}~\bibnamefont {Pav\'on}},\ }\href {\doibase 10.1088/1361-6633/ad2527} {\bibfield  {journal} {\bibinfo  {journal} {Rept. Prog. Phys.}\ }\textbf {\bibinfo {volume} {87}},\ \bibinfo {pages} {036901} (\bibinfo {year} {2024}{\natexlab{a}})},\ \Eprint {http://arxiv.org/abs/2402.00819} {arXiv:2402.00819 [astro-ph.CO]} \BibitemShut {NoStop}%
\bibitem [{\citenamefont {Sch\"oneberg}\ \emph {et~al.}(2022{\natexlab{b}})\citenamefont {Sch\"oneberg}, \citenamefont {Franco~Abell\'an}, \citenamefont {P\'erez~S\'anchez}, \citenamefont {Witte}, \citenamefont {Poulin},\ and\ \citenamefont {Lesgourgues}}]{2022-Schoeneberg.etal-Phys.Rept.}%
  \BibitemOpen
  \bibfield  {author} {\bibinfo {author} {\bibfnamefont {N.}~\bibnamefont {Sch\"oneberg}}, \bibinfo {author} {\bibfnamefont {G.}~\bibnamefont {Franco~Abell\'an}}, \bibinfo {author} {\bibfnamefont {A.}~\bibnamefont {P\'erez~S\'anchez}}, \bibinfo {author} {\bibfnamefont {S.~J.}\ \bibnamefont {Witte}}, \bibinfo {author} {\bibfnamefont {V.}~\bibnamefont {Poulin}}, \ and\ \bibinfo {author} {\bibfnamefont {J.}~\bibnamefont {Lesgourgues}},\ }\href {\doibase 10.1016/j.physrep.2022.07.001} {\bibfield  {journal} {\bibinfo  {journal} {Phys. Rept.}\ }\textbf {\bibinfo {volume} {984}},\ \bibinfo {pages} {1} (\bibinfo {year} {2022}{\natexlab{b}})},\ \Eprint {http://arxiv.org/abs/2107.10291} {arXiv:2107.10291 [astro-ph.CO]} \BibitemShut {NoStop}%
\bibitem [{\citenamefont {Di~Valentino}\ \emph {et~al.}(2021{\natexlab{b}})\citenamefont {Di~Valentino}, \citenamefont {Mena}, \citenamefont {Pan}, \citenamefont {Visinelli}, \citenamefont {Yang}, \citenamefont {Melchiorri}, \citenamefont {Mota}, \citenamefont {Riess},\ and\ \citenamefont {Silk}}]{2021-DiValentino.etal-Class.Quant.Grav.}%
  \BibitemOpen
  \bibfield  {author} {\bibinfo {author} {\bibfnamefont {E.}~\bibnamefont {Di~Valentino}}, \bibinfo {author} {\bibfnamefont {O.}~\bibnamefont {Mena}}, \bibinfo {author} {\bibfnamefont {S.}~\bibnamefont {Pan}}, \bibinfo {author} {\bibfnamefont {L.}~\bibnamefont {Visinelli}}, \bibinfo {author} {\bibfnamefont {W.}~\bibnamefont {Yang}}, \bibinfo {author} {\bibfnamefont {A.}~\bibnamefont {Melchiorri}}, \bibinfo {author} {\bibfnamefont {D.~F.}\ \bibnamefont {Mota}}, \bibinfo {author} {\bibfnamefont {A.~G.}\ \bibnamefont {Riess}}, \ and\ \bibinfo {author} {\bibfnamefont {J.}~\bibnamefont {Silk}},\ }\href {\doibase 10.1088/1361-6382/ac086d} {\bibfield  {journal} {\bibinfo  {journal} {Class. Quant. Grav.}\ }\textbf {\bibinfo {volume} {38}},\ \bibinfo {pages} {153001} (\bibinfo {year} {2021}{\natexlab{b}})},\ \Eprint {http://arxiv.org/abs/2103.01183} {arXiv:2103.01183 [astro-ph.CO]} \BibitemShut {NoStop}%
\bibitem [{\citenamefont {Wang}\ \emph {et~al.}(2024{\natexlab{b}})\citenamefont {Wang}, \citenamefont {Abdalla}, \citenamefont {Atrio-Barandela},\ and\ \citenamefont {Pav\'on}}]{2024-Wang-Rept.Prog.Phys.}%
  \BibitemOpen
  \bibfield  {author} {\bibinfo {author} {\bibfnamefont {B.}~\bibnamefont {Wang}}, \bibinfo {author} {\bibfnamefont {E.}~\bibnamefont {Abdalla}}, \bibinfo {author} {\bibfnamefont {F.}~\bibnamefont {Atrio-Barandela}}, \ and\ \bibinfo {author} {\bibfnamefont {D.}~\bibnamefont {Pav\'on}},\ }\href {\doibase 10.1088/1361-6633/ad2527} {\bibfield  {journal} {\bibinfo  {journal} {Rept. Prog. Phys.}\ }\textbf {\bibinfo {volume} {87}},\ \bibinfo {pages} {036901} (\bibinfo {year} {2024}{\natexlab{b}})},\ \Eprint {http://arxiv.org/abs/2402.00819} {arXiv:2402.00819 [astro-ph.CO]} \BibitemShut {NoStop}%
\bibitem [{\citenamefont {Giar\`e}\ \emph {et~al.}(2024{\natexlab{a}})\citenamefont {Giar\`e}, \citenamefont {Sabogal}, \citenamefont {Nunes},\ and\ \citenamefont {Di~Valentino}}]{2024-Giare-}%
  \BibitemOpen
  \bibfield  {author} {\bibinfo {author} {\bibfnamefont {W.}~\bibnamefont {Giar\`e}}, \bibinfo {author} {\bibfnamefont {M.~A.}\ \bibnamefont {Sabogal}}, \bibinfo {author} {\bibfnamefont {R.~C.}\ \bibnamefont {Nunes}}, \ and\ \bibinfo {author} {\bibfnamefont {E.}~\bibnamefont {Di~Valentino}},\ }\href@noop {} {\  (\bibinfo {year} {2024}{\natexlab{a}})},\ \Eprint {http://arxiv.org/abs/2404.15232} {arXiv:2404.15232 [astro-ph.CO]} \BibitemShut {NoStop}%
\bibitem [{\citenamefont {Johnson}\ \emph {et~al.}(2022)\citenamefont {Johnson}, \citenamefont {Sangwan},\ and\ \citenamefont {Shankaranarayanan}}]{2022-Johnson.etal-JCAP}%
  \BibitemOpen
  \bibfield  {author} {\bibinfo {author} {\bibfnamefont {J.~P.}\ \bibnamefont {Johnson}}, \bibinfo {author} {\bibfnamefont {A.}~\bibnamefont {Sangwan}}, \ and\ \bibinfo {author} {\bibfnamefont {S.}~\bibnamefont {Shankaranarayanan}},\ }\href {\doibase 10.1088/1475-7516/2022/01/024} {\bibfield  {journal} {\bibinfo  {journal} {JCAP}\ }\textbf {\bibinfo {volume} {01}},\ \bibinfo {pages} {024} (\bibinfo {year} {2022})},\ \Eprint {http://arxiv.org/abs/2102.12367} {arXiv:2102.12367 [astro-ph.CO]} \BibitemShut {NoStop}%
\bibitem [{\citenamefont {Horndeski}(1974)}]{1974-Horndeski-IJTP}%
  \BibitemOpen
  \bibfield  {author} {\bibinfo {author} {\bibfnamefont {G.~W.}\ \bibnamefont {Horndeski}},\ }\href {\doibase 10.1007/BF01807638} {\bibfield  {journal} {\bibinfo  {journal} {Int. J. Theor. Phys.}\ }\textbf {\bibinfo {volume} {10}},\ \bibinfo {pages} {363} (\bibinfo {year} {1974})}\BibitemShut {NoStop}%
\bibitem [{\citenamefont {Kobayashi}(2019)}]{2019-Kobayashi-Rept.Prog.Phys.}%
  \BibitemOpen
  \bibfield  {author} {\bibinfo {author} {\bibfnamefont {T.}~\bibnamefont {Kobayashi}},\ }\href {\doibase 10.1088/1361-6633/ab2429} {\bibfield  {journal} {\bibinfo  {journal} {Rept. Prog. Phys.}\ }\textbf {\bibinfo {volume} {82}},\ \bibinfo {pages} {086901} (\bibinfo {year} {2019})},\ \Eprint {http://arxiv.org/abs/1901.07183} {arXiv:1901.07183 [gr-qc]} \BibitemShut {NoStop}%
\bibitem [{\citenamefont {De~Felice}\ and\ \citenamefont {Tsujikawa}(2010)}]{2010-DeFelice-Phys.Rev.Lett.}%
  \BibitemOpen
  \bibfield  {author} {\bibinfo {author} {\bibfnamefont {A.}~\bibnamefont {De~Felice}}\ and\ \bibinfo {author} {\bibfnamefont {S.}~\bibnamefont {Tsujikawa}},\ }\href {\doibase 10.1103/PhysRevLett.105.111301} {\bibfield  {journal} {\bibinfo  {journal} {Phys. Rev. Lett.}\ }\textbf {\bibinfo {volume} {105}},\ \bibinfo {pages} {111301} (\bibinfo {year} {2010})},\ \Eprint {http://arxiv.org/abs/1007.2700} {arXiv:1007.2700 [astro-ph.CO]} \BibitemShut {NoStop}%
\bibitem [{\citenamefont {Deffayet}\ \emph {et~al.}(2009)\citenamefont {Deffayet}, \citenamefont {Esposito-Farese},\ and\ \citenamefont {Vikman}}]{Deffayet:2009wt}%
  \BibitemOpen
  \bibfield  {author} {\bibinfo {author} {\bibfnamefont {C.}~\bibnamefont {Deffayet}}, \bibinfo {author} {\bibfnamefont {G.}~\bibnamefont {Esposito-Farese}}, \ and\ \bibinfo {author} {\bibfnamefont {A.}~\bibnamefont {Vikman}},\ }\href {\doibase 10.1103/PhysRevD.79.084003} {\bibfield  {journal} {\bibinfo  {journal} {Phys. Rev. D}\ }\textbf {\bibinfo {volume} {79}},\ \bibinfo {pages} {084003} (\bibinfo {year} {2009})},\ \Eprint {http://arxiv.org/abs/0901.1314} {arXiv:0901.1314 [hep-th]} \BibitemShut {NoStop}%
\bibitem [{\citenamefont {Bekenstein}(1993)}]{1993-Bekenstein-Phys.Rev.D}%
  \BibitemOpen
  \bibfield  {author} {\bibinfo {author} {\bibfnamefont {J.~D.}\ \bibnamefont {Bekenstein}},\ }\href {\doibase 10.1103/PhysRevD.48.3641} {\bibfield  {journal} {\bibinfo  {journal} {Phys. Rev. D}\ }\textbf {\bibinfo {volume} {48}},\ \bibinfo {pages} {3641} (\bibinfo {year} {1993})},\ \Eprint {http://arxiv.org/abs/gr-qc/9211017} {arXiv:gr-qc/9211017} \BibitemShut {NoStop}%
\bibitem [{\citenamefont {Zumalacarregui}\ \emph {et~al.}(2013)\citenamefont {Zumalacarregui}, \citenamefont {Koivisto},\ and\ \citenamefont {Mota}}]{2013-Zumalacarregui.etal-PRD}%
  \BibitemOpen
  \bibfield  {author} {\bibinfo {author} {\bibfnamefont {M.}~\bibnamefont {Zumalacarregui}}, \bibinfo {author} {\bibfnamefont {T.~S.}\ \bibnamefont {Koivisto}}, \ and\ \bibinfo {author} {\bibfnamefont {D.~F.}\ \bibnamefont {Mota}},\ }\href {\doibase 10.1103/PhysRevD.87.083010} {\bibfield  {journal} {\bibinfo  {journal} {Phys. Rev. D}\ }\textbf {\bibinfo {volume} {87}},\ \bibinfo {pages} {083010} (\bibinfo {year} {2013})},\ \Eprint {http://arxiv.org/abs/1210.8016} {arXiv:1210.8016 [astro-ph.CO]} \BibitemShut {NoStop}%
\bibitem [{\citenamefont {van~de Bruck}\ and\ \citenamefont {Morrice}(2015)}]{2015-vandeBruck.etal-JCAP}%
  \BibitemOpen
  \bibfield  {author} {\bibinfo {author} {\bibfnamefont {C.}~\bibnamefont {van~de Bruck}}\ and\ \bibinfo {author} {\bibfnamefont {J.}~\bibnamefont {Morrice}},\ }\href {\doibase 10.1088/1475-7516/2015/04/036} {\bibfield  {journal} {\bibinfo  {journal} {JCAP}\ }\textbf {\bibinfo {volume} {04}},\ \bibinfo {pages} {036} (\bibinfo {year} {2015})},\ \Eprint {http://arxiv.org/abs/1501.03073} {arXiv:1501.03073 [gr-qc]} \BibitemShut {NoStop}%
\bibitem [{\citenamefont {Chamings}\ \emph {et~al.}(2020)\citenamefont {Chamings}, \citenamefont {Avgoustidis}, \citenamefont {Copeland}, \citenamefont {Green},\ and\ \citenamefont {Pourtsidou}}]{Chamings:2019kcl}%
  \BibitemOpen
  \bibfield  {author} {\bibinfo {author} {\bibfnamefont {F.~N.}\ \bibnamefont {Chamings}}, \bibinfo {author} {\bibfnamefont {A.}~\bibnamefont {Avgoustidis}}, \bibinfo {author} {\bibfnamefont {E.~J.}\ \bibnamefont {Copeland}}, \bibinfo {author} {\bibfnamefont {A.~M.}\ \bibnamefont {Green}}, \ and\ \bibinfo {author} {\bibfnamefont {A.}~\bibnamefont {Pourtsidou}},\ }\href {\doibase 10.1103/PhysRevD.101.043531} {\bibfield  {journal} {\bibinfo  {journal} {Phys. Rev. D}\ }\textbf {\bibinfo {volume} {101}},\ \bibinfo {pages} {043531} (\bibinfo {year} {2020})},\ \Eprint {http://arxiv.org/abs/1912.09858} {arXiv:1912.09858 [astro-ph.CO]} \BibitemShut {NoStop}%
\bibitem [{\citenamefont {Amendola}\ and\ \citenamefont {Tsujikawa}(2020)}]{2020-Amendola.Tsujikawa-JCAP}%
  \BibitemOpen
  \bibfield  {author} {\bibinfo {author} {\bibfnamefont {L.}~\bibnamefont {Amendola}}\ and\ \bibinfo {author} {\bibfnamefont {S.}~\bibnamefont {Tsujikawa}},\ }\href {\doibase 10.1088/1475-7516/2020/06/020} {\bibfield  {journal} {\bibinfo  {journal} {JCAP}\ }\textbf {\bibinfo {volume} {06}},\ \bibinfo {pages} {020} (\bibinfo {year} {2020})},\ \Eprint {http://arxiv.org/abs/2003.02686} {arXiv:2003.02686 [gr-qc]} \BibitemShut {NoStop}%
\bibitem [{\citenamefont {Bertschinger}(2006)}]{2006-Bertschinger-Astrophys.J.}%
  \BibitemOpen
  \bibfield  {author} {\bibinfo {author} {\bibfnamefont {E.}~\bibnamefont {Bertschinger}},\ }\href {\doibase 10.1086/506021} {\bibfield  {journal} {\bibinfo  {journal} {Astrophys. J.}\ }\textbf {\bibinfo {volume} {648}},\ \bibinfo {pages} {797} (\bibinfo {year} {2006})},\ \Eprint {http://arxiv.org/abs/astro-ph/0604485} {arXiv:astro-ph/0604485} \BibitemShut {NoStop}%
\bibitem [{\citenamefont {Caldwell}\ \emph {et~al.}(2007)\citenamefont {Caldwell}, \citenamefont {Cooray},\ and\ \citenamefont {Melchiorri}}]{2007-Caldwell.etal-Phys.Rev.D}%
  \BibitemOpen
  \bibfield  {author} {\bibinfo {author} {\bibfnamefont {R.}~\bibnamefont {Caldwell}}, \bibinfo {author} {\bibfnamefont {A.}~\bibnamefont {Cooray}}, \ and\ \bibinfo {author} {\bibfnamefont {A.}~\bibnamefont {Melchiorri}},\ }\href {\doibase 10.1103/PhysRevD.76.023507} {\bibfield  {journal} {\bibinfo  {journal} {Phys. Rev. D}\ }\textbf {\bibinfo {volume} {76}},\ \bibinfo {pages} {023507} (\bibinfo {year} {2007})},\ \Eprint {http://arxiv.org/abs/astro-ph/0703375} {arXiv:astro-ph/0703375} \BibitemShut {NoStop}%
\bibitem [{\citenamefont {Johnson}\ and\ \citenamefont {Shankaranarayanan}(2019)}]{2019-Joseph.Shanki-PRD}%
  \BibitemOpen
  \bibfield  {author} {\bibinfo {author} {\bibfnamefont {J.~P.}\ \bibnamefont {Johnson}}\ and\ \bibinfo {author} {\bibfnamefont {S.}~\bibnamefont {Shankaranarayanan}},\ }\href {\doibase 10.1103/PhysRevD.100.083526} {\bibfield  {journal} {\bibinfo  {journal} {Phys. Rev. D}\ }\textbf {\bibinfo {volume} {100}},\ \bibinfo {pages} {083526} (\bibinfo {year} {2019})},\ \Eprint {http://arxiv.org/abs/1904.07608} {arXiv:1904.07608 [astro-ph.CO]} \BibitemShut {NoStop}%
\bibitem [{\citenamefont {Kobayashi}\ \emph {et~al.}(2010)\citenamefont {Kobayashi}, \citenamefont {Yamaguchi},\ and\ \citenamefont {Yokoyama}}]{2010-Kobayashi-Phys.Rev.Lett.}%
  \BibitemOpen
  \bibfield  {author} {\bibinfo {author} {\bibfnamefont {T.}~\bibnamefont {Kobayashi}}, \bibinfo {author} {\bibfnamefont {M.}~\bibnamefont {Yamaguchi}}, \ and\ \bibinfo {author} {\bibfnamefont {J.}~\bibnamefont {Yokoyama}},\ }\href {\doibase 10.1103/PhysRevLett.105.231302} {\bibfield  {journal} {\bibinfo  {journal} {Phys. Rev. Lett.}\ }\textbf {\bibinfo {volume} {105}},\ \bibinfo {pages} {231302} (\bibinfo {year} {2010})},\ \Eprint {http://arxiv.org/abs/1008.0603} {arXiv:1008.0603 [hep-th]} \BibitemShut {NoStop}%
\bibitem [{\citenamefont {Unnikrishnan}\ and\ \citenamefont {Shankaranarayanan}(2014)}]{Unnikrishnan:2013rka}%
  \BibitemOpen
  \bibfield  {author} {\bibinfo {author} {\bibfnamefont {S.}~\bibnamefont {Unnikrishnan}}\ and\ \bibinfo {author} {\bibfnamefont {S.}~\bibnamefont {Shankaranarayanan}},\ }\href {\doibase 10.1088/1475-7516/2014/07/003} {\bibfield  {journal} {\bibinfo  {journal} {JCAP}\ }\textbf {\bibinfo {volume} {07}},\ \bibinfo {pages} {003} (\bibinfo {year} {2014})},\ \Eprint {http://arxiv.org/abs/1311.0177} {arXiv:1311.0177 [astro-ph.CO]} \BibitemShut {NoStop}%
\bibitem [{\citenamefont {Creminelli}\ and\ \citenamefont {Vernizzi}(2017)}]{2017-Creminelli.Vernizzi-Phys.Rev.Lett.}%
  \BibitemOpen
  \bibfield  {author} {\bibinfo {author} {\bibfnamefont {P.}~\bibnamefont {Creminelli}}\ and\ \bibinfo {author} {\bibfnamefont {F.}~\bibnamefont {Vernizzi}},\ }\href {\doibase 10.1103/PhysRevLett.119.251302} {\bibfield  {journal} {\bibinfo  {journal} {Phys. Rev. Lett.}\ }\textbf {\bibinfo {volume} {119}},\ \bibinfo {pages} {251302} (\bibinfo {year} {2017})},\ \Eprint {http://arxiv.org/abs/1710.05877} {arXiv:1710.05877 [astro-ph.CO]} \BibitemShut {NoStop}%
\bibitem [{\citenamefont {Baker}\ \emph {et~al.}(2017)\citenamefont {Baker}, \citenamefont {Bellini}, \citenamefont {Ferreira}, \citenamefont {Lagos}, \citenamefont {Noller},\ and\ \citenamefont {Sawicki}}]{2017-Baker.etal-Phys.Rev.Lett.}%
  \BibitemOpen
  \bibfield  {author} {\bibinfo {author} {\bibfnamefont {T.}~\bibnamefont {Baker}}, \bibinfo {author} {\bibfnamefont {E.}~\bibnamefont {Bellini}}, \bibinfo {author} {\bibfnamefont {P.~G.}\ \bibnamefont {Ferreira}}, \bibinfo {author} {\bibfnamefont {M.}~\bibnamefont {Lagos}}, \bibinfo {author} {\bibfnamefont {J.}~\bibnamefont {Noller}}, \ and\ \bibinfo {author} {\bibfnamefont {I.}~\bibnamefont {Sawicki}},\ }\href {\doibase 10.1103/PhysRevLett.119.251301} {\bibfield  {journal} {\bibinfo  {journal} {Phys. Rev. Lett.}\ }\textbf {\bibinfo {volume} {119}},\ \bibinfo {pages} {251301} (\bibinfo {year} {2017})},\ \Eprint {http://arxiv.org/abs/1710.06394} {arXiv:1710.06394 [astro-ph.CO]} \BibitemShut {NoStop}%
\bibitem [{\citenamefont {Ezquiaga}\ and\ \citenamefont {Zumalac\'arregui}(2017)}]{2017-Ezquiaga.Zumalacarregui-Phys.Rev.Lett.}%
  \BibitemOpen
  \bibfield  {author} {\bibinfo {author} {\bibfnamefont {J.~M.}\ \bibnamefont {Ezquiaga}}\ and\ \bibinfo {author} {\bibfnamefont {M.}~\bibnamefont {Zumalac\'arregui}},\ }\href {\doibase 10.1103/PhysRevLett.119.251304} {\bibfield  {journal} {\bibinfo  {journal} {Phys. Rev. Lett.}\ }\textbf {\bibinfo {volume} {119}},\ \bibinfo {pages} {251304} (\bibinfo {year} {2017})},\ \Eprint {http://arxiv.org/abs/1710.05901} {arXiv:1710.05901 [astro-ph.CO]} \BibitemShut {NoStop}%
\bibitem [{\citenamefont {Amendola}\ \emph {et~al.}(2018)\citenamefont {Amendola}, \citenamefont {Bettoni}, \citenamefont {Dom\`enech},\ and\ \citenamefont {Gomes}}]{Amendola:2018ltt}%
  \BibitemOpen
  \bibfield  {author} {\bibinfo {author} {\bibfnamefont {L.}~\bibnamefont {Amendola}}, \bibinfo {author} {\bibfnamefont {D.}~\bibnamefont {Bettoni}}, \bibinfo {author} {\bibfnamefont {G.}~\bibnamefont {Dom\`enech}}, \ and\ \bibinfo {author} {\bibfnamefont {A.~R.}\ \bibnamefont {Gomes}},\ }\href {\doibase 10.1088/1475-7516/2018/06/029} {\bibfield  {journal} {\bibinfo  {journal} {JCAP}\ }\textbf {\bibinfo {volume} {06}},\ \bibinfo {pages} {029} (\bibinfo {year} {2018})},\ \Eprint {http://arxiv.org/abs/1803.06368} {arXiv:1803.06368 [gr-qc]} \BibitemShut {NoStop}%
\bibitem [{\citenamefont {Terente~D\'\i{}az}\ \emph {et~al.}(2023)\citenamefont {Terente~D\'\i{}az}, \citenamefont {Dimopoulos}, \citenamefont {Kar\v{c}iauskas},\ and\ \citenamefont {Racioppi}}]{TerenteDiaz:2023iqk}%
  \BibitemOpen
  \bibfield  {author} {\bibinfo {author} {\bibfnamefont {J.~J.}\ \bibnamefont {Terente~D\'\i{}az}}, \bibinfo {author} {\bibfnamefont {K.}~\bibnamefont {Dimopoulos}}, \bibinfo {author} {\bibfnamefont {M.}~\bibnamefont {Kar\v{c}iauskas}}, \ and\ \bibinfo {author} {\bibfnamefont {A.}~\bibnamefont {Racioppi}},\ }\href {\doibase 10.1088/1475-7516/2023/10/031} {\bibfield  {journal} {\bibinfo  {journal} {JCAP}\ }\textbf {\bibinfo {volume} {10}},\ \bibinfo {pages} {031} (\bibinfo {year} {2023})},\ \Eprint {http://arxiv.org/abs/2307.06163} {arXiv:2307.06163 [astro-ph.CO]} \BibitemShut {NoStop}%
\bibitem [{\citenamefont {Zumalac\'arregui}\ and\ \citenamefont {Garc\'\i{}a-Bellido}(2014)}]{2014-Zumalacarregui-Phys.Rev.D}%
  \BibitemOpen
  \bibfield  {author} {\bibinfo {author} {\bibfnamefont {M.}~\bibnamefont {Zumalac\'arregui}}\ and\ \bibinfo {author} {\bibfnamefont {J.}~\bibnamefont {Garc\'\i{}a-Bellido}},\ }\href {\doibase 10.1103/PhysRevD.89.064046} {\bibfield  {journal} {\bibinfo  {journal} {Phys. Rev. D}\ }\textbf {\bibinfo {volume} {89}},\ \bibinfo {pages} {064046} (\bibinfo {year} {2014})},\ \Eprint {http://arxiv.org/abs/1308.4685} {arXiv:1308.4685 [gr-qc]} \BibitemShut {NoStop}%
\bibitem [{\citenamefont {Gleyzes}\ \emph {et~al.}(2015)\citenamefont {Gleyzes}, \citenamefont {Langlois}, \citenamefont {Piazza},\ and\ \citenamefont {Vernizzi}}]{2015-Gleyzes-Phys.Rev.Lett.}%
  \BibitemOpen
  \bibfield  {author} {\bibinfo {author} {\bibfnamefont {J.}~\bibnamefont {Gleyzes}}, \bibinfo {author} {\bibfnamefont {D.}~\bibnamefont {Langlois}}, \bibinfo {author} {\bibfnamefont {F.}~\bibnamefont {Piazza}}, \ and\ \bibinfo {author} {\bibfnamefont {F.}~\bibnamefont {Vernizzi}},\ }\href {\doibase 10.1103/PhysRevLett.114.211101} {\bibfield  {journal} {\bibinfo  {journal} {Phys. Rev. Lett.}\ }\textbf {\bibinfo {volume} {114}},\ \bibinfo {pages} {211101} (\bibinfo {year} {2015})},\ \Eprint {http://arxiv.org/abs/1404.6495} {arXiv:1404.6495 [hep-th]} \BibitemShut {NoStop}%
\bibitem [{\citenamefont {Langlois}(2019)}]{Langlois:2018dxi}%
  \BibitemOpen
  \bibfield  {author} {\bibinfo {author} {\bibfnamefont {D.}~\bibnamefont {Langlois}},\ }\href {\doibase 10.1142/S0218271819420069} {\bibfield  {journal} {\bibinfo  {journal} {Int. J. Mod. Phys. D}\ }\textbf {\bibinfo {volume} {28}},\ \bibinfo {pages} {1942006} (\bibinfo {year} {2019})},\ \Eprint {http://arxiv.org/abs/1811.06271} {arXiv:1811.06271 [gr-qc]} \BibitemShut {NoStop}%
\bibitem [{\citenamefont {Langlois}(2017)}]{Langlois:2017mdk}%
  \BibitemOpen
  \bibfield  {author} {\bibinfo {author} {\bibfnamefont {D.}~\bibnamefont {Langlois}},\ }in\ \href@noop {} {\emph {\bibinfo {booktitle} {{52nd Rencontres de Moriond on Gravitation}}}}\ (\bibinfo {year} {2017})\ pp.\ \bibinfo {pages} {221--228},\ \Eprint {http://arxiv.org/abs/1707.03625} {arXiv:1707.03625 [gr-qc]} \BibitemShut {NoStop}%
\bibitem [{\citenamefont {Langlois}\ and\ \citenamefont {Noui}(2016)}]{Langlois:2015cwa}%
  \BibitemOpen
  \bibfield  {author} {\bibinfo {author} {\bibfnamefont {D.}~\bibnamefont {Langlois}}\ and\ \bibinfo {author} {\bibfnamefont {K.}~\bibnamefont {Noui}},\ }\href {\doibase 10.1088/1475-7516/2016/02/034} {\bibfield  {journal} {\bibinfo  {journal} {JCAP}\ }\textbf {\bibinfo {volume} {02}},\ \bibinfo {pages} {034} (\bibinfo {year} {2016})},\ \Eprint {http://arxiv.org/abs/1510.06930} {arXiv:1510.06930 [gr-qc]} \BibitemShut {NoStop}%
\bibitem [{\citenamefont {Ben~Achour}\ \emph {et~al.}(2016)\citenamefont {Ben~Achour}, \citenamefont {Langlois},\ and\ \citenamefont {Noui}}]{BenAchour:2016cay}%
  \BibitemOpen
  \bibfield  {author} {\bibinfo {author} {\bibfnamefont {J.}~\bibnamefont {Ben~Achour}}, \bibinfo {author} {\bibfnamefont {D.}~\bibnamefont {Langlois}}, \ and\ \bibinfo {author} {\bibfnamefont {K.}~\bibnamefont {Noui}},\ }\href {\doibase 10.1103/PhysRevD.93.124005} {\bibfield  {journal} {\bibinfo  {journal} {Phys. Rev. D}\ }\textbf {\bibinfo {volume} {93}},\ \bibinfo {pages} {124005} (\bibinfo {year} {2016})},\ \Eprint {http://arxiv.org/abs/1602.08398} {arXiv:1602.08398 [gr-qc]} \BibitemShut {NoStop}%
\bibitem [{\citenamefont {Crisostomi}\ \emph {et~al.}(2016)\citenamefont {Crisostomi}, \citenamefont {Koyama},\ and\ \citenamefont {Tasinato}}]{Crisostomi:2016czh}%
  \BibitemOpen
  \bibfield  {author} {\bibinfo {author} {\bibfnamefont {M.}~\bibnamefont {Crisostomi}}, \bibinfo {author} {\bibfnamefont {K.}~\bibnamefont {Koyama}}, \ and\ \bibinfo {author} {\bibfnamefont {G.}~\bibnamefont {Tasinato}},\ }\href {\doibase 10.1088/1475-7516/2016/04/044} {\bibfield  {journal} {\bibinfo  {journal} {JCAP}\ }\textbf {\bibinfo {volume} {04}},\ \bibinfo {pages} {044} (\bibinfo {year} {2016})},\ \Eprint {http://arxiv.org/abs/1602.03119} {arXiv:1602.03119 [hep-th]} \BibitemShut {NoStop}%
\bibitem [{\citenamefont {Crisostomi}\ and\ \citenamefont {Koyama}(2018)}]{Crisostomi:2017pjs}%
  \BibitemOpen
  \bibfield  {author} {\bibinfo {author} {\bibfnamefont {M.}~\bibnamefont {Crisostomi}}\ and\ \bibinfo {author} {\bibfnamefont {K.}~\bibnamefont {Koyama}},\ }\href {\doibase 10.1103/PhysRevD.97.084004} {\bibfield  {journal} {\bibinfo  {journal} {Phys. Rev. D}\ }\textbf {\bibinfo {volume} {97}},\ \bibinfo {pages} {084004} (\bibinfo {year} {2018})},\ \Eprint {http://arxiv.org/abs/1712.06556} {arXiv:1712.06556 [astro-ph.CO]} \BibitemShut {NoStop}%
\bibitem [{\citenamefont {Crisostomi}\ \emph {et~al.}(2019)\citenamefont {Crisostomi}, \citenamefont {Koyama}, \citenamefont {Langlois}, \citenamefont {Noui},\ and\ \citenamefont {Steer}}]{Crisostomi:2018bsp}%
  \BibitemOpen
  \bibfield  {author} {\bibinfo {author} {\bibfnamefont {M.}~\bibnamefont {Crisostomi}}, \bibinfo {author} {\bibfnamefont {K.}~\bibnamefont {Koyama}}, \bibinfo {author} {\bibfnamefont {D.}~\bibnamefont {Langlois}}, \bibinfo {author} {\bibfnamefont {K.}~\bibnamefont {Noui}}, \ and\ \bibinfo {author} {\bibfnamefont {D.~A.}\ \bibnamefont {Steer}},\ }\href {\doibase 10.1088/1475-7516/2019/01/030} {\bibfield  {journal} {\bibinfo  {journal} {JCAP}\ }\textbf {\bibinfo {volume} {01}},\ \bibinfo {pages} {030} (\bibinfo {year} {2019})},\ \Eprint {http://arxiv.org/abs/1810.12070} {arXiv:1810.12070 [hep-th]} \BibitemShut {NoStop}%
\bibitem [{\citenamefont {Frusciante}\ \emph {et~al.}(2019)\citenamefont {Frusciante}, \citenamefont {Kase}, \citenamefont {Koyama}, \citenamefont {Tsujikawa},\ and\ \citenamefont {Vernieri}}]{2019-Frusciante-Phys.Lett.B}%
  \BibitemOpen
  \bibfield  {author} {\bibinfo {author} {\bibfnamefont {N.}~\bibnamefont {Frusciante}}, \bibinfo {author} {\bibfnamefont {R.}~\bibnamefont {Kase}}, \bibinfo {author} {\bibfnamefont {K.}~\bibnamefont {Koyama}}, \bibinfo {author} {\bibfnamefont {S.}~\bibnamefont {Tsujikawa}}, \ and\ \bibinfo {author} {\bibfnamefont {D.}~\bibnamefont {Vernieri}},\ }\href {\doibase 10.1016/j.physletb.2019.01.009} {\bibfield  {journal} {\bibinfo  {journal} {Phys. Lett. B}\ }\textbf {\bibinfo {volume} {790}},\ \bibinfo {pages} {167} (\bibinfo {year} {2019})},\ \Eprint {http://arxiv.org/abs/1812.05204} {arXiv:1812.05204 [gr-qc]} \BibitemShut {NoStop}%
\bibitem [{\citenamefont {Lazanu}(2024)}]{Lazanu:2024mzj}%
  \BibitemOpen
  \bibfield  {author} {\bibinfo {author} {\bibfnamefont {A.}~\bibnamefont {Lazanu}},\ }\href@noop {} {\  (\bibinfo {year} {2024})},\ \Eprint {http://arxiv.org/abs/2407.18234} {arXiv:2407.18234 [astro-ph.CO]} \BibitemShut {NoStop}%
\bibitem [{\citenamefont {Archidiacono}\ and\ \citenamefont {Gariazzo}(2022)}]{Archidiacono:2022ich}%
  \BibitemOpen
  \bibfield  {author} {\bibinfo {author} {\bibfnamefont {M.}~\bibnamefont {Archidiacono}}\ and\ \bibinfo {author} {\bibfnamefont {S.}~\bibnamefont {Gariazzo}},\ }\href {\doibase 10.3390/universe8030175} {\bibfield  {journal} {\bibinfo  {journal} {Universe}\ }\textbf {\bibinfo {volume} {8}},\ \bibinfo {pages} {175} (\bibinfo {year} {2022})},\ \Eprint {http://arxiv.org/abs/2201.10319} {arXiv:2201.10319 [hep-ph]} \BibitemShut {NoStop}%
\bibitem [{\citenamefont {Kase}\ and\ \citenamefont {Tsujikawa}(2020)}]{2020-Kase.Tsujikawa-JCAP}%
  \BibitemOpen
  \bibfield  {author} {\bibinfo {author} {\bibfnamefont {R.}~\bibnamefont {Kase}}\ and\ \bibinfo {author} {\bibfnamefont {S.}~\bibnamefont {Tsujikawa}},\ }\href {\doibase 10.1088/1475-7516/2020/11/032} {\bibfield  {journal} {\bibinfo  {journal} {JCAP}\ }\textbf {\bibinfo {volume} {11}},\ \bibinfo {pages} {032} (\bibinfo {year} {2020})},\ \Eprint {http://arxiv.org/abs/2005.13809} {arXiv:2005.13809 [gr-qc]} \BibitemShut {NoStop}%
\bibitem [{\citenamefont {Woodard}(2015)}]{Woodard:2015zca}%
  \BibitemOpen
  \bibfield  {author} {\bibinfo {author} {\bibfnamefont {R.~P.}\ \bibnamefont {Woodard}},\ }\href {\doibase 10.4249/scholarpedia.32243} {\bibfield  {journal} {\bibinfo  {journal} {Scholarpedia}\ }\textbf {\bibinfo {volume} {10}},\ \bibinfo {pages} {32243} (\bibinfo {year} {2015})},\ \Eprint {http://arxiv.org/abs/1506.02210} {arXiv:1506.02210 [hep-th]} \BibitemShut {NoStop}%
\bibitem [{\citenamefont {Jorge}\ \emph {et~al.}(2007)\citenamefont {Jorge}, \citenamefont {Mimoso},\ and\ \citenamefont {Wands}}]{Pedro:2007}%
  \BibitemOpen
  \bibfield  {author} {\bibinfo {author} {\bibfnamefont {P.}~\bibnamefont {Jorge}}, \bibinfo {author} {\bibfnamefont {J.}~\bibnamefont {Mimoso}}, \ and\ \bibinfo {author} {\bibfnamefont {D.}~\bibnamefont {Wands}},\ }\href {\doibase 10.1088/1742-6596/66/1/012031} {\bibfield  {journal} {\bibinfo  {journal} {Journal of Physics: Conference Series}\ }\textbf {\bibinfo {volume} {66}},\ \bibinfo {pages} {012031} (\bibinfo {year} {2007})}\BibitemShut {NoStop}%
\bibitem [{\citenamefont {Mukhanov}\ \emph {et~al.}(1992)\citenamefont {Mukhanov}, \citenamefont {Feldman},\ and\ \citenamefont {Brandenberger}}]{1992-Mukhanov.etal-PRep}%
  \BibitemOpen
  \bibfield  {author} {\bibinfo {author} {\bibfnamefont {V.~F.}\ \bibnamefont {Mukhanov}}, \bibinfo {author} {\bibfnamefont {H.~A.}\ \bibnamefont {Feldman}}, \ and\ \bibinfo {author} {\bibfnamefont {R.~H.}\ \bibnamefont {Brandenberger}},\ }\href {\doibase 10.1016/0370-1573(92)90044-Z} {\bibfield  {journal} {\bibinfo  {journal} {Phys. Rept.}\ }\textbf {\bibinfo {volume} {215}},\ \bibinfo {pages} {203} (\bibinfo {year} {1992})}\BibitemShut {NoStop}%
\bibitem [{\citenamefont {Tsagas}\ \emph {et~al.}(2008)\citenamefont {Tsagas}, \citenamefont {Challinor},\ and\ \citenamefont {Maartens}}]{2008-Tsagas.etal-PRep}%
  \BibitemOpen
  \bibfield  {author} {\bibinfo {author} {\bibfnamefont {C.~G.}\ \bibnamefont {Tsagas}}, \bibinfo {author} {\bibfnamefont {A.}~\bibnamefont {Challinor}}, \ and\ \bibinfo {author} {\bibfnamefont {R.}~\bibnamefont {Maartens}},\ }\href {\doibase 10.1016/j.physrep.2008.03.003} {\bibfield  {journal} {\bibinfo  {journal} {Phys. Rept.}\ }\textbf {\bibinfo {volume} {465}},\ \bibinfo {pages} {61} (\bibinfo {year} {2008})},\ \Eprint {http://arxiv.org/abs/0705.4397} {arXiv:0705.4397 [astro-ph]} \BibitemShut {NoStop}%
\bibitem [{\citenamefont {Pourtsidou}\ \emph {et~al.}(2013)\citenamefont {Pourtsidou}, \citenamefont {Skordis},\ and\ \citenamefont {Copeland}}]{2013-Pourtsidou.etal-Phys.Rev.D}%
  \BibitemOpen
  \bibfield  {author} {\bibinfo {author} {\bibfnamefont {A.}~\bibnamefont {Pourtsidou}}, \bibinfo {author} {\bibfnamefont {C.}~\bibnamefont {Skordis}}, \ and\ \bibinfo {author} {\bibfnamefont {E.~J.}\ \bibnamefont {Copeland}},\ }\href {\doibase 10.1103/PhysRevD.88.083505} {\bibfield  {journal} {\bibinfo  {journal} {Phys. Rev. D}\ }\textbf {\bibinfo {volume} {88}},\ \bibinfo {pages} {083505} (\bibinfo {year} {2013})},\ \Eprint {http://arxiv.org/abs/1307.0458} {arXiv:1307.0458 [astro-ph.CO]} \BibitemShut {NoStop}%
\bibitem [{\citenamefont {Sobotka}\ \emph {et~al.}(2024)\citenamefont {Sobotka}, \citenamefont {Erickcek},\ and\ \citenamefont {Smith}}]{Sobotka:2023bzr}%
  \BibitemOpen
  \bibfield  {author} {\bibinfo {author} {\bibfnamefont {A.~C.}\ \bibnamefont {Sobotka}}, \bibinfo {author} {\bibfnamefont {A.~L.}\ \bibnamefont {Erickcek}}, \ and\ \bibinfo {author} {\bibfnamefont {T.~L.}\ \bibnamefont {Smith}},\ }\href {\doibase 10.1103/PhysRevD.109.063538} {\bibfield  {journal} {\bibinfo  {journal} {Phys. Rev. D}\ }\textbf {\bibinfo {volume} {109}},\ \bibinfo {pages} {063538} (\bibinfo {year} {2024})},\ \Eprint {http://arxiv.org/abs/2312.13235} {arXiv:2312.13235 [astro-ph.CO]} \BibitemShut {NoStop}%
\bibitem [{\citenamefont {Giar\`e}\ \emph {et~al.}(2024{\natexlab{b}})\citenamefont {Giar\`e}, \citenamefont {Sabogal}, \citenamefont {Nunes},\ and\ \citenamefont {Di~Valentino}}]{Giare:2024smz}%
  \BibitemOpen
  \bibfield  {author} {\bibinfo {author} {\bibfnamefont {W.}~\bibnamefont {Giar\`e}}, \bibinfo {author} {\bibfnamefont {M.~A.}\ \bibnamefont {Sabogal}}, \bibinfo {author} {\bibfnamefont {R.~C.}\ \bibnamefont {Nunes}}, \ and\ \bibinfo {author} {\bibfnamefont {E.}~\bibnamefont {Di~Valentino}},\ }\href@noop {} {\  (\bibinfo {year} {2024}{\natexlab{b}})},\ \Eprint {http://arxiv.org/abs/2404.15232} {arXiv:2404.15232 [astro-ph.CO]} \BibitemShut {NoStop}%
\end{thebibliography}
%

\end{document}